\documentclass[english,british]{aa}
\usepackage{graphicx}
\usepackage{txfonts}
\usepackage{hyperref}
\usepackage{xcolor}
\usepackage{adjustbox} 
\usepackage{booktabs} 
\usepackage{rotating} 
\usepackage{multirow}
\usepackage{geometry}
\usepackage{array}
\usepackage{longtable}
\usepackage{lscape} 
\usepackage{natbib}
\usepackage{float}
\bibpunct{(}{)}{;}{a}{}{,} 
\usepackage{svg}
\newcommand{\orcidicon}[1]{\href{https://orcid.org/#1}{\textcolor{green}{\includegraphics[scale=0.08]{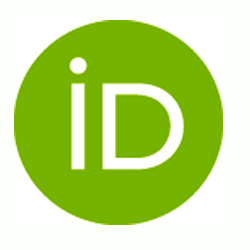}}}}

\begin{document} 

\authorrunning{Guti\'{e}rrez Soto et al.}
\titlerunning{S-PLUS: H{$\alpha$}-Excess Point Sources} 

   \title{Mapping H$\alpha$-Excess Candidate Point Sources in the Southern Hemisphere Using S-PLUS Data}

   \subtitle{}

   \author{L. A. Guti\'{e}rrez-Soto\orcidicon{0000-0002-9891-8017}
          \inst{1,2}\fnmsep\thanks{E-mail: gsotoangel@fcaglp.unlp.edu.ar},
          R. Lopes de Oliveira\orcidicon{0000-0002-6211-7226}\inst{2,3,4},
          S. Akras\orcidicon{0000-0003-1351-7204}\inst{5},
          D. R. Gon\c{c}alves\inst{6},
          L. F. Lomel\'i-N\'u\~nez\orcidicon{0000-0003-2127-2841
}\inst{6},
          C. Mendes de Oliveira\orcidicon{0000-0002-5267-9065}\inst{2},
          E. Telles\orcidicon{0000-0002-8280-4445}\inst{4},
          A. Alvarez-Candal\orcidicon{0000-0002-5045-9675}\inst{7,8},
          M. Borges Fernandes\orcidicon{0000-0001-5740-2914}\inst{4},
          S. Daflon\orcidicon{0000-0001-9205-2307}\inst{4},
          C. E. Ferreira Lopes\orcidicon{0000-0002-8525-7977}\inst{9,10},
          M. Grossi\inst{6},\orcidicon{0000-0003-4675-3246}
          D. Hazarika\orcidicon{0000-0003-4379-6777}\inst{9,10},
          P. K. Humire\orcidicon{0000-0003-3537-4849}\inst{2},
          C. Lima-Dias\orcidicon{0009-0006-0373-8168}\inst{11},
          A. R. Lopes\orcidicon{0000-0002-6164-5051}\inst{1},
          J. L. Nilo Castellón\orcidicon{0000-0002-2370-7784}\inst{11},
          S. Panda\orcidicon{0000-0002-5854-7426}\inst{12}
          A. Kanaan\orcidicon{0000-0002-2484-7551}\inst{13},
          T. Ribeiro\orcidicon{0000-0002-0138-1365}\inst{14},
          W. Schoenell\orcidicon{0000-0002-4064-7234}\inst{15}}

   \institute{Instituto de Astrof\'{i}sica de La Plata (CCT La Plata - CONICET - UNLP), B1900FWA, La Plata, Argentina\\ 
              \email{gsotoangel@fcaglp.unlp.edu.ar}
        \and Departamento de Astronomia, Instituto de Astronomia, Geof\'isica e Ciências Atmosféricas da USP, Cidade Universitária, 05508-900 S\~{a}o Paulo, SP, Brazil
        \and Departamento de F\'isica, Universidade Federal de Sergipe, Av. Marechal Rondon, S/N, 49100-000, S\~ao Crist\'ov\~ao, SE, Brazil
        \and Observat\'orio Nacional, Rua Gal. Jos\'e Cristino 77, 20921-400, Rio~de~Janeiro, RJ, Brazil
        \and Institute for Astronomy, Astrophysics, Space Applications and Remote Sensing, National Observatory of Athens, GR 15236 Penteli, Greece
        \and Universidade Federal do Rio de Janeiro, Observat\'orio do Valongo, Ladeira do Pedro Ant\^onio, 43, Sa\'ude CEP 20080-090 Rio de Janeiro, RJ, Brazil
        \and Instituto de Astrof\'{i}sica de Andaluc\'ia, CSIC, Apt 3004, E18080 Granada, Spain
        \and Instituto de F\'{i}sica Aplicada a las Ciencias y las Tecnolog\'ias, Universidad de Alicante, San Vicent del Raspeig, E03080, Alicante, Spain
        \and Instituto de Astronom\'{i}a y Ciencias Planetarias, Universidad de Atacama, Copayapu 485, Copiap\'{o}, Chile
        \and Millennium Institute of Astrophysics, Nuncio Monse\~{n}or Sotero Sanz 100, Of. 104, Providencia, Santiago, Chile
        \and Departmento de Astronom\'ia, Universidad de La Serena, Avenida Ra\'ul Bitr\'an 1305, La Serena, Chile
        \and International Gemini Observatory/NSF NOIRLab, Casilla 603, La Serena, Chile
        \and Departamento de Física, Universidade Federal de Santa Catarina, Florianópolis, SC, 88040-900, Brazil
        \and Rubin Observatory Project Office, 950 N. Cherry Ave., Tucson, AZ 85719, USA
        \and GMTO Corporation 465 N. Halstead Street, Suite 250 Pasadena, CA 91107}

   \date{Received September 15, 1996; accepted March 16, 1997}


\abstract
{We use the Southern Photometric Local Universe Survey (S-PLUS) Fourth Data Release (DR4) to identify and classify H$\alpha$-excess point source candidates in the Southern Sky. This approach combines photometric data from 12 S-PLUS filters with machine learning techniques to improve source classification and advance our understanding of H$\alpha$-related phenomena.} 
{Our goal is to enhance the classification of H$\alpha$-excess point sources by distinguishing between Galactic and extragalactic objects, particularly those with redshifted emission lines, and to identify sources where the H$\alpha$ excess is associated with variability phenomena, such as short-period RR Lyrae stars.} 
{We selected H$\alpha$-excess candidates using the ($r - J0660$) versus ($r - i$) colour-colour diagram from the S-PLUS main survey (MS) and Galactic Disk Survey (GDS). Dimensionality reduction was achieved using UMAP, followed by HDBSCAN clustering. We refined this by incorporating infrared data, improving the separation of source types. A Random Forest model was then trained on the clustering results to identify key colour features for the classification of H$\alpha$-excess sources. New, effective colour-colour diagrams are constructed by combining data from S-PLUS MS and infrared data. These diagrams, alongside tentative colour criteria, offer a preliminary classification of H$\alpha$-excess sources without the need for complex algorithms.} 
{Combining multiwavelength photometric data with machine learning techniques significantly improved the classification of H$\alpha$-excess sources. We identified 6956 sources with excess in the $J0660$ filter, and cross-matching with SIMBAD allowed us to explore the types of objects present in our catalogue, including emission-line stars, young stellar objects, nebulae, stellar binaries, cataclysmic variables, variable stars, and extragalactic sources such as QSOs, AGNs, and galaxies. The cross-match also revealed X-ray sources, transients, and other peculiar objects. Using S-PLUS colours and machine learning, we successfully separated RR Lyrae stars from both other Galactic stars and extragalactic objects. Additionally, we achieved a clear separation between Galactic and extragalactic sources. However, distinguishing cataclysmic variables from QSOs at specific redshifts remained challenging. Incorporating infrared data refined the classification, enabling us to separate Galactic from extragalactic sources and to distinguish cataclysmic variables from QSOs. The Random Forest model, trained on HDBSCAN results, highlighted key colour features that distinguish the different classes of H$\alpha$-excess sources, providing a robust framework for future studies  such as follow-up spectroscopy.}
{}

   \keywords{surveys -- techniques: photometric -- stars: novae, cataclysmic variables -- quasars: emission lines}

   \maketitle
%

\section{Introduction}

  Hydrogen Balmer emission lines are primarily produced by radiative processes, particularly radiative excitation and ionization, which dominate over collisional excitation under typical nebular conditions. For example, the Einstein A-coefficient for the H$\alpha$ transition ($A_{32} \approx 4.41 \times 10^7 \, \text{s}^{-1}$) is significantly larger than the typical collisional excitation rate coefficient ($\sim\,10^{-9} \text{ to } 10^{-8} \, \text{cm}^3 \, \text{s}^{-1}$) at electron temperatures around 10\,000\,K. While collisional excitation can become more important in shock-heated or very dense environments, it generally remains secondary in the diffuse conditions of most nebulae. Universe being hydrogen-abundant, the observation of those electronic transitions offers an important window into the study of astrophysical objects. Among all the possible electronic transitions, the Balmer series represents an extremely useful tool in Astronomy, because it falls in the commonly used optical spectral range. In particular, the H$\alpha$ emission line, with a rest-frame vacuum wavelength of 6564.614\,\AA, corresponding to the electron transition from $n$ = 3 to $n$ = 2, is the strongest in both emission and absorption. It is the most widely used line for identifying various types of objects, such as star-forming regions, H~{\sc ii} regions, planetary nebulae (PNe), supernovae, novae, young stellar objects (YSO), Herbig-Haro objects, circumstellar disks, post-asymptotic and asymptotic giant stars (AGB), red giant branch (RGB), active late-type dwarfs. Among massive stars, emission lines are observed in Be stars with decretion disks, B[e] supergiants, Luminous Blue Variables (LBVs), Wolf-Rayet (WR) stars, and interacting binary systems that are experiencing mass exchange, like symbiotic stars (SySt), cataclysmic variables (CVs), among others.

In high-redshift sources, such as starburst galaxies and quasi-stellar objects (QSOs), H$\alpha$ emission is present, but redshifted to longer wavelengths. However, when we detect emission near 6563\AA{} from high redshift sources, the recombination of H$\alpha$ is not the cause, instead, it is the outcome of UV emission lines that have shifted towards the visible spectrum.

Most existing databases of the aforementioned classes of objects are not homogeneous and remain far from complete, even in the local universe. Some classes are highly populated, while others are significantly underrepresented. For example, there are $\sim$ 300 known SySts in the Milky Way but only 75 in nearby galaxies \citep{Akras:2019a,Merc:2019}  with constantly new discoveries being made every year \citep[e.g.][]{Merc:2020, Akras:2021, Merc:2021, Merc:2022, Munari:2021, Munari:2022, Akras:2023}.
The number of known PNe in our galaxy is on the order of $\sim$3500 \citep{Parker:2016}, which may represent only 15-30\% of the total population \citep{Frew:2008,Jacoby:2010}.

H$\alpha$ surveys have been conducted with varying angular resolutions, sky coverage, and sensitivity. Some surveys, despite having modest spatial resolutions, have successfully resolved extended nebular emissions, enabling the study of supernova remnants, galaxy groups, and star-forming regions \citep[e.g.][]{1976MmRAS..81...89D,  Blair:2004, Jaiswal:2016, Cook:2019}. 
Others, with higher spatial resolution, 
have revealed compact emission-line sources
in the Milky Way and nearby galaxies.
Examples of them are the INT Photometric H$\alpha$ survey
(IPHAS; \citealt{Drew:2005, Barentsen:2014}), the SuperCOSMOS H$\alpha$ survey with the UK Schmidt Telescope (UKST) of the Anglo-Australian Observatory \citep{2005MNRAS.362..689P}, and the VST Photometric H$\alpha$ Survey (VPHAS+; \citealt{Drew:2014}).

Colour-colour diagrams from photometric surveys are also used to identify possible H$\alpha$ emitters.
For example, the ($r$ - H{$\alpha$}) versus ($r - i$) colour-colour and similar diagrams has been used
to find  CVs \citep{Witham:2006,Witham:2007}, 
YSOs \citep{Vink:2008}, 
SySt \citep{Corradi:2008, Corradi:2010, Corradi:2011, Miszalski:2014, Mikolajewska:2014, Mikolajewska:2017, Akras:2019b}, 
early-type emission-line stars \citep{Drew:2008}, 
and PNe \citep{Miszalski:2009, Viirone:2009, Sabin:2010, Akras:2019c}. Additionally, other combinations of broadband filters have been tailored to distinguish AGNs, QSOs, and compact PNe based on their distinct photometric signatures \citep{Peters:2015, Gutierrez-Soto:2024}.

In particular, the class of Be stars is the more common, nearly 50\%, in the total sample of Ha emitters in IPHAS with only a moderate $r-H\alpha$ excess, well limited near-infrared colours ($J-H, H-K$; \citealp{Corradi:2008, Raddi:2015} and moderate mid-infrared colours (e.g. $W1-W4, W3-W4$; \citealp{Akras:2019b}). Besides the identification of different classes of H$\alpha$ emitters, the $r-H\alpha$ excess derived from H$\alpha$ photometric surveys such as the IPHAS and VPHAS+, it also provides a new  automatic way to derive the accretion rate in large numbers of YSOs \citep{Barentsen:2011, Kalari:2015}.

\citet{Witham:2008} developed a method to select H$\alpha$ emission line sources in the IPHAS survey by implementing the aforementioned colour-colour diagram ($r$ - H{$\alpha$}) versus ($r - i$). Objects with H$\alpha$ excess line were identified by iteratively fitting the stellar locus and considering as candidates those objects that fall several sigma above this stellar locus in the $r$ - H{$\alpha$} colour. This conservative method yields a total of 4853 point sources in the IPHAS catalogue that exhibit strong photometric evidence for H{$\alpha$} emission. They obtained spectra from around 300 sources, confirming more than 95 percent of them as genuine emission-line stars. 

\citet{Monguio:2020} developed the INT Galactic Plane Survey (IGAPS) by merging the IPHAS and UVEX optical surveys. The IGAPS catalogue includes 295.4 million photometric measurements in the $i$, $r$, narrow-band H$\alpha$, $g$, and U$_{\text{RGO}}$ filters. It identifies 8292 candidate emission line stars and over 53\,000 variable stars with confidence greater than $5\sigma$.

More recently, \citet{Fratta:2021} introduced a technique using Gaia data to identify H$\alpha$-bright sources in the IPHAS catalogue. They partitioned the data based on Gaia colour-absolute magnitude and Galactic coordinates to minimize contamination and then applied the strategy from \citet{Witham:2008} to these partitions.

Two ongoing multi-band surveys are observing the sky in a systematic, complementary way, with 5 broad and 7 narrow-band filters, including H$\alpha$: the Javalambre Photometric Local Universe Survey
(J-PLUS\footnote{\url{https://www.j-plus.es}}; \citealp{Cenarro:2019}), covering the Northern celestial hemisphere, and the Southern-Photometric Local Universe Survey
(S-PLUS\footnote{\url{http://www.splus.iag.usp.br}}; \citealp{Mendes:2019}), covering the southern sky with a twin 83\,cm telescope and filter system. The first survey is paving the way for an even more ambitious survey, the Javalambre Physics of the Accelerating Universe Astrophysical Survey (J-PAS; \citealp{Benitez:2014} and miniJ-PAS; \citealp{Bonoli:2021}), which will observe the Northern sky with 56 narrow-band filters. As source hunters, the spectral energy distributions provided by these surveys enable an unprecedented source classification using photometry only. However, in the Big Data era, efficient investigation tools are required to deal with their massive imaging and catalogues production, and machine learning techniques have been increasingly used to explore these data sets \citep[e.g][]{Bom:2021, Yang:2022}.

Here we present a census of H$\alpha$-excess point-like sources from the S-PLUS DR4, identified using the ($r$ - $J$0660) versus ($r$ - $i$) colour-colour diagram. Advanced machine learning techniques are employed to improve the identification and classification of these sources from the S-PLUS DR4 dataset. Specifically, we use Uniform Manifold Approximation and Projection (UMAP; \citealp{Becht:2018, mcinnes:2020}) for dimensionality reduction followed by Hierarchical Density-Based Spatial Clustering of Applications with Noise (HDBSCAN; \citealp{Campello:2013}) clustering to group sources based on their multi-wavelength photometric signatures. This approach allows us to handle high-dimensional data effectively and uncover patterns that traditional methods might overlook. Additionally, we incorporate Wide-Field Infrared Survey Explorer (WISE; \citealp{Wright:2010}) data and apply a Random Forest \citep{Breiman:2001} model to refine our classification and identify key features that distinguish different types of H$\alpha$-excess sources.

Sect.~\ref{sec:Obser} describes the observations related to the S-PLUS project, including important information on the fourth data release, photometry, and data handling. Sect.~\ref{sec:metho} presents the technique implemented to select the H$\alpha$-feature sources. Sect.~\ref{sec:results} includes the analysis of the results. In Sect.~\ref{sec:ML}, we present the machine learning methods used to analyse and make a more accurate classification of the H$\alpha$ sources. Finally, Sect.~\ref{sec:conclu} discusses our main results and conclusions.

\section{S-PLUS Survey Overview}
\label{sec:Obser}


\begin{figure*}
    \includegraphics[width=\linewidth]{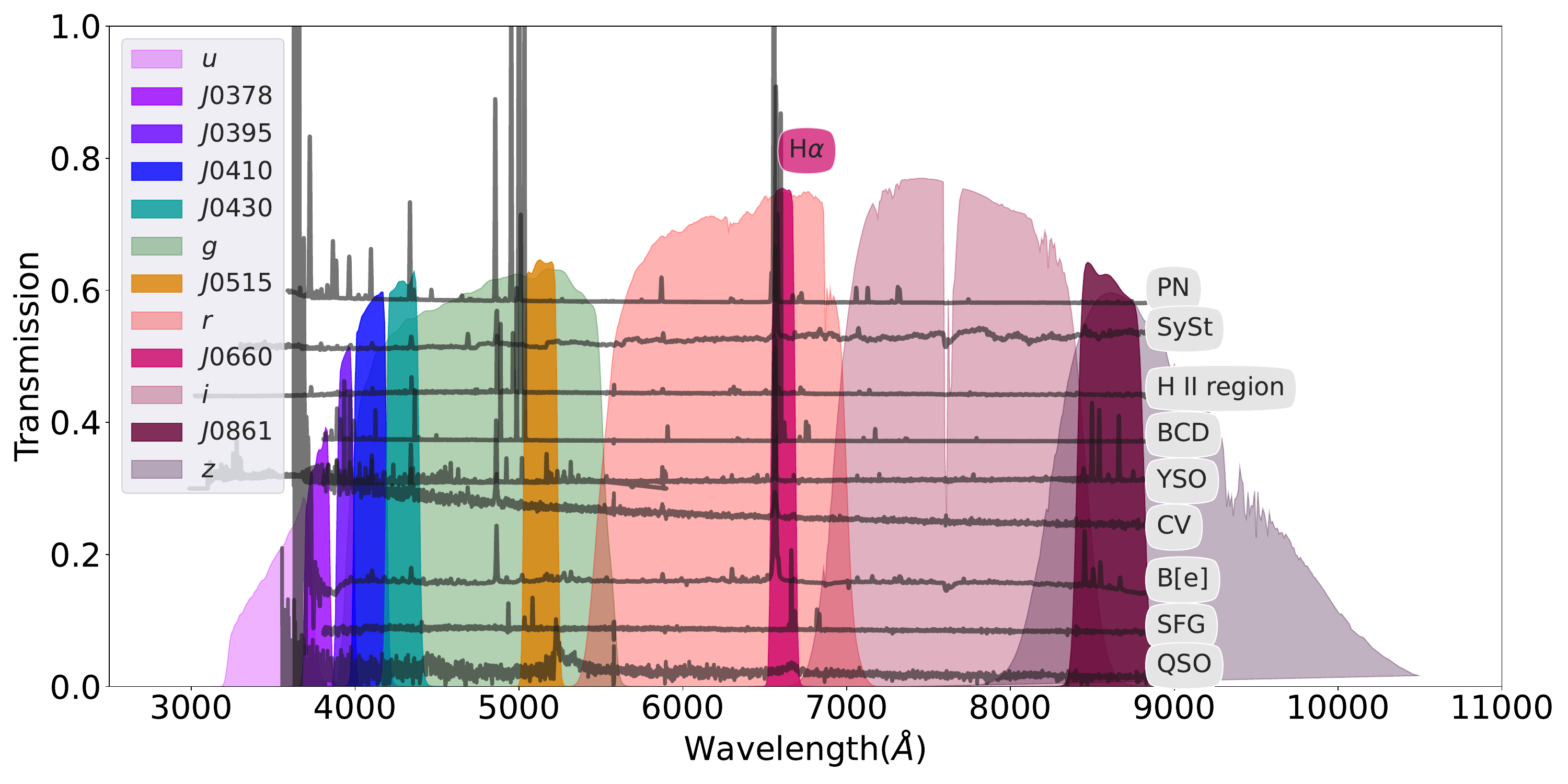}
    \caption{Transmission curves of the S-PLUS filter set. The narrowband filter $J0660$ includes the H$\alpha$ emission line. Over-plotted is spectra of different
      classes of emission line objects. From top to bottom:  a PN, a symbiotic star, an extragalactic H~{\sc ii} region, a blue compact/H~{\sc ii} galaxy, a YSO, a CV star, a B[e] star, a star-forming galaxy and a QSO at a redshift of $\sim$3.31.}
    \label{fig:curves}
\end{figure*}

S-PLUS surveys the southern sky using the 12 filters from the Javalambre filter system \citep{Marin-Franch:2012}, a spanning the wavelength range from 3\,000\AA\ to 10\,000\AA. This system comprises seven narrow-band filters (\textit{J}0378, \textit{J}0395, \textit{J}0410, \textit{J}0430, \textit{J}0515, \textit{J}0660, \textit{J}0861, and five broad-band Sloan-like \citep{Fukugita:1996} filters (see Fig. \ref{fig:curves}). The narrow-band \textit{J}0660 filter used in S-PLUS is centred at $\lambda$ 6614\AA\ and has a width of $\approx$ 147\AA\ (Table 2 of \citealp{Mendes:2019}). Consequently, it covers both the H{$\alpha$} and the [\ion{N}{ii}] doublet $\lambda\lambda$ 6548, 6584 spectral lines for sources up to a redshift of $\approx$ 0.02. S-PLUS is conducted using a dedicated 0.83 m robotic telescope located at Cerro Tololo, Chile \citep{Mendes:2019}.

\begin{figure*}
  \setlength\tabcolsep{0pt}
  \setkeys{Gin}{width=0.5\linewidth}
  \begin{tabular}{ll}
    (a) & (b) \\
    \includegraphics[trim=5 10 10 20, clip]{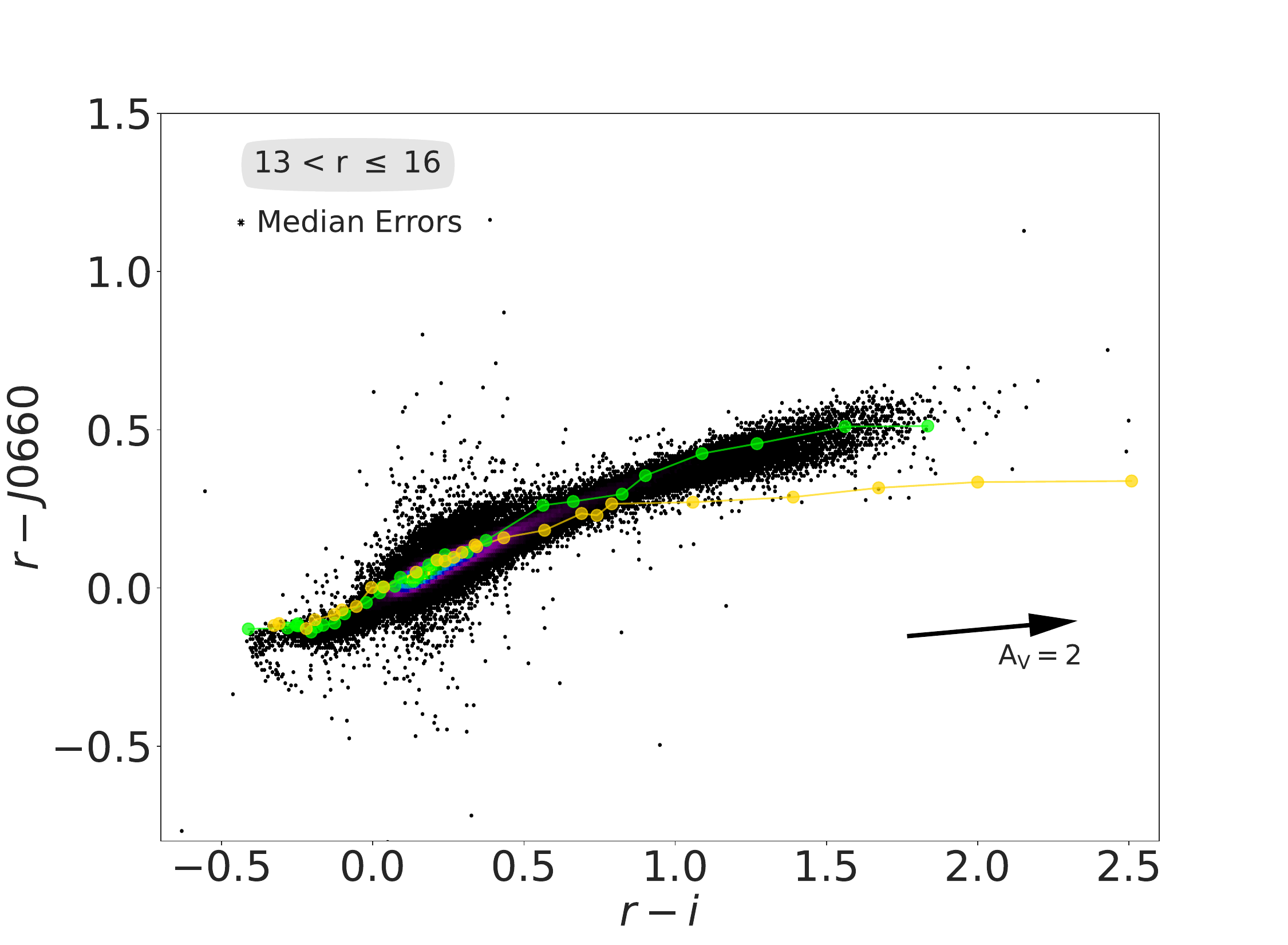}
    & \includegraphics[trim=5 10 10 20, clip]{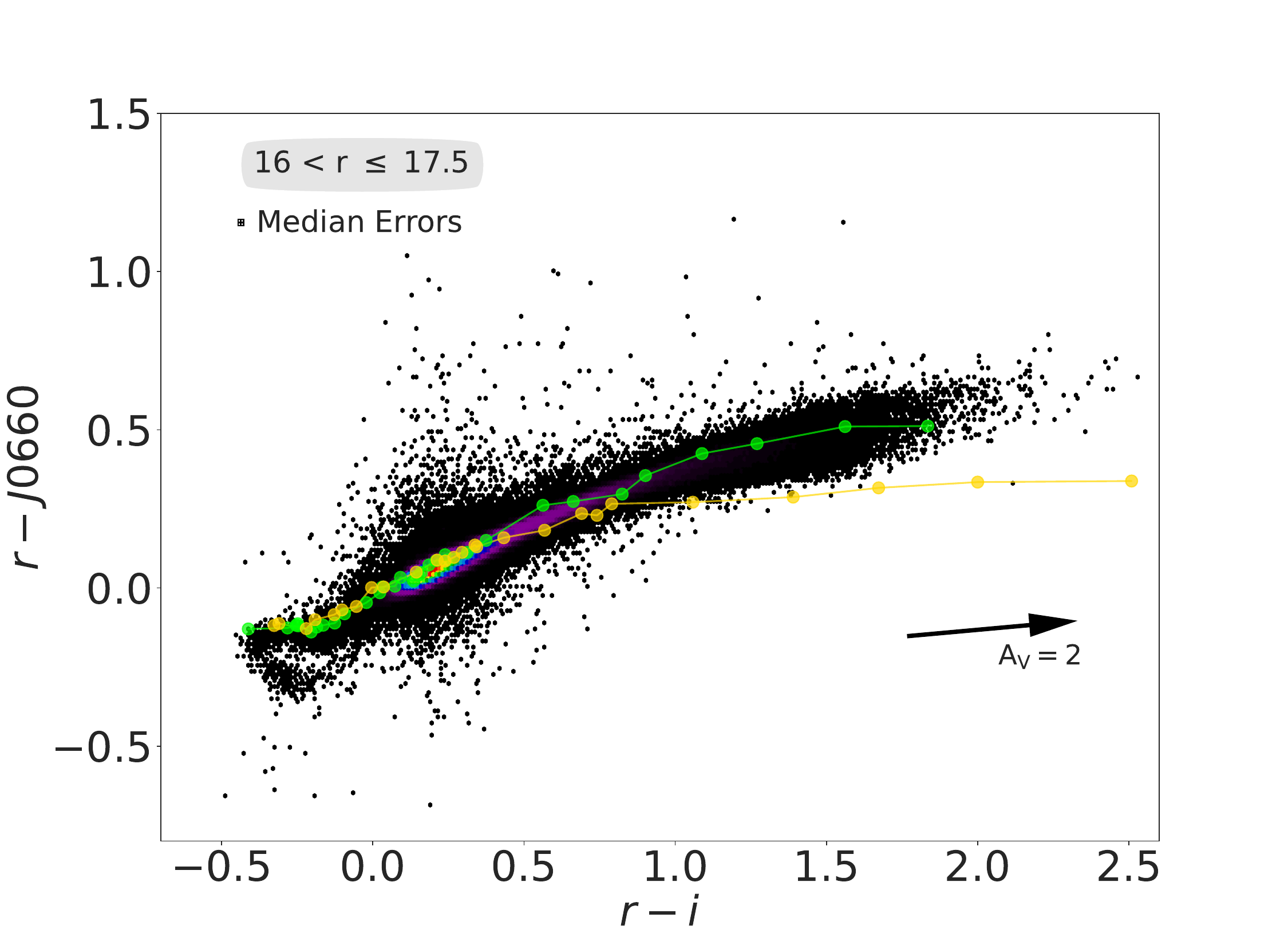}\\
    (c) & (d) \\
    \includegraphics[trim=5 10 10 20, clip]{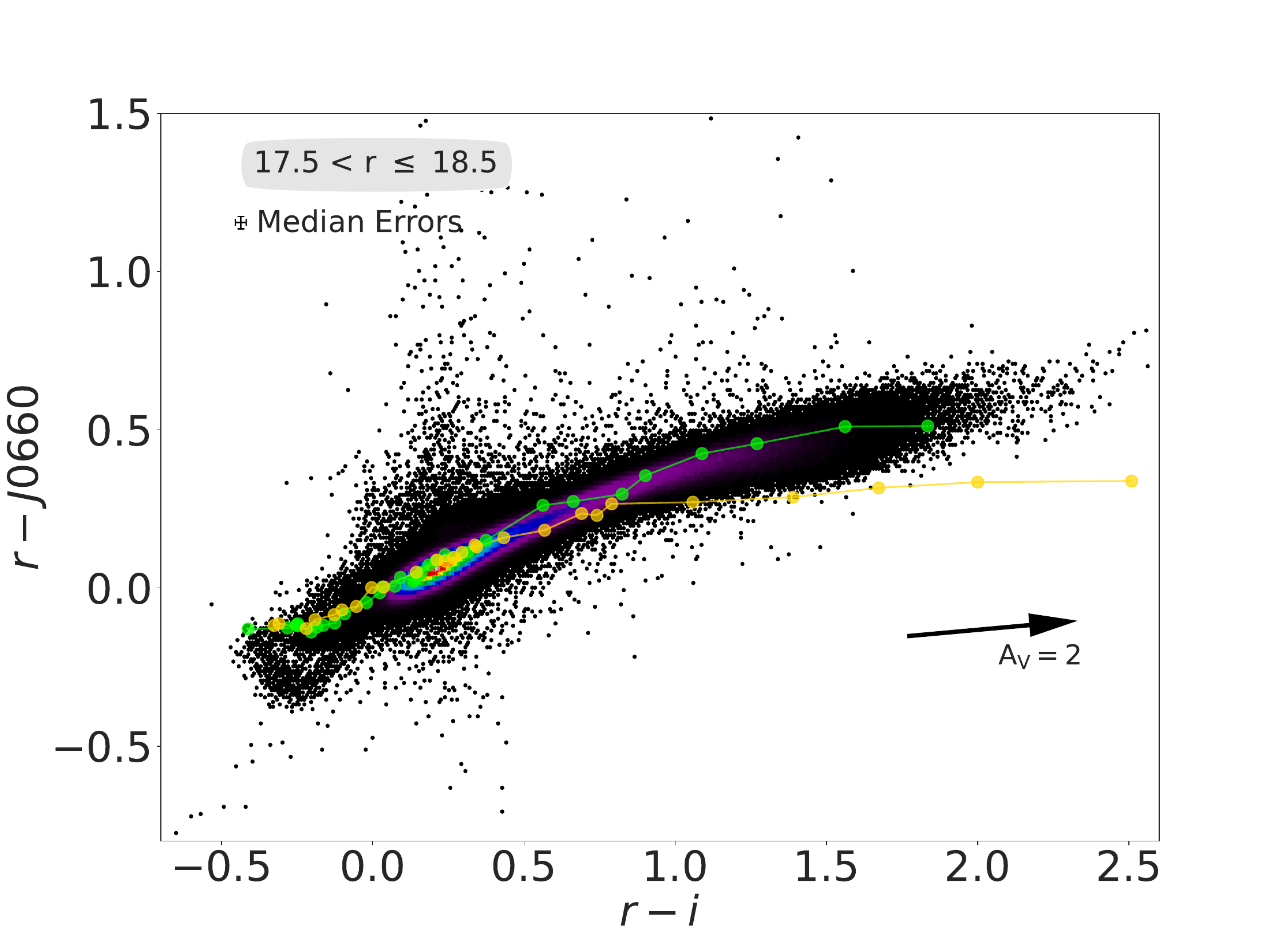}
    & \includegraphics[trim=5 10 10 20, clip]{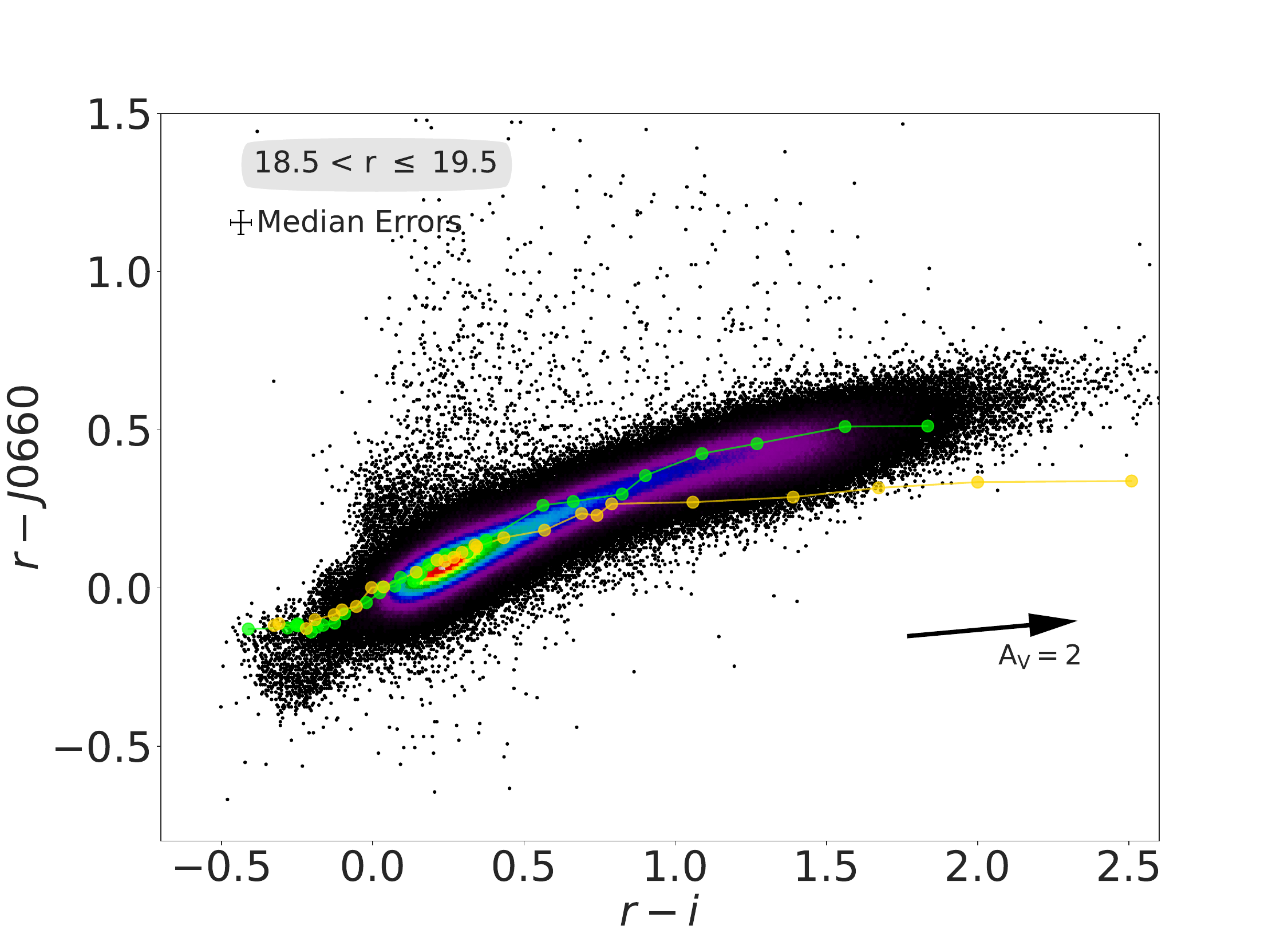}\\
  \end{tabular}
  \caption{The ($r - J0660$) versus ($r - i$) colour-colour plots used to select objects with H$\alpha$ excess. These plots display data for all stars from the S-PLUS DR4 MS, representing the \texttt{PStotal} photometry in these colours. The data are divided into four magnitude bins: \textit{(a)} $13 < r \leq 16$, \textit{(b)} $16 < r \leq 17.5$, \textit{(c)} $17.5 < r \leq 18.5$, and \textit{(d)} $18.5 < r \leq 19.5$. Objects with H$\alpha$ excess are expected to be located towards the top of these diagrams. Lighter green and yellow points connected by lines represent the tracks for main-sequence and giant stars, respectively. These tracks are derived from the synthetic spectra library of \citet{Pickles:1998}. The background colour gradient represents the density of objects, with red indicating the highest concentration of points, followed by green, blue, and purple, which represent progressively lower concentrations.}
  \label{fig:all-DR4}
\end{figure*}

This work uses data from S-PLUS DR4 \citep{Herpich:2024}. DR4 encompasses 171 fields at very low galactic latitudes (|b| < 15°), an additional 341 fields carried over from DR3 spanning the Main Survey (MS) footprint (with |b| > 30°), and 150 fields within the Magellanic Clouds region. This accumulation results in a total of 1629 fields in DR4, covering an expansive area of 3022.7 square degrees. Notably, this coverage includes 347.4 square degrees within the disk regions and 289.5 square degrees within the Magellanic Clouds Survey (MC). Here, we explore the MS and Galactic Disk Survey (GDS), we primary goal of identify objects with H$\alpha$ excess in S-PLUS DR4. 

\subsection{Flux Calibration} \label{sec:calibration}

 The flux calibration of the S-PLUS survey was performed using a combination of external and internal calibration steps to ensure uniformity and accuracy across the entire survey footprint. The calibration process begins with an external calibration, where synthetic photometry is integrated with a reference catalogue to derive the calibrated magnitudes for the different filters. Zero points (ZPs) are determined as the difference between the predicted magnitudes from the synthetic models and the instrumental magnitudes observed in the survey. The synthetic spectral library for this step was constructed by convolving the library of \citet{Coelho:2014} with the transmission curves of multiple reference catalogues and the S-PLUS filter system.

In regions where external calibration data are unavailable, such as for the $u$, $J0410$, and $J0430$ filters, a stellar locus method is applied. This technique calibrates these filters by leveraging the stellar locus, a relationship between specific filter magnitudes observed for a population of stars. This step is crucial when external reference data are insufficient. Once the external calibration is complete, an internal calibration step further refines the ZPs by using pre-calibrated narrowband filters, which better constrain the synthetic models and improve the calibration accuracy to 0.01–0.02 mag. This internal calibration is particularly valuable for cases where external calibration might be lacking or less precise.

Finally, the calibration is aligned to the Gaia system by applying an average offset derived from synthetic photometry, ensuring consistency across the entire survey region. This final alignment guarantees that the flux calibration is homogeneous and compatible with the Gaia photometric system, as outlined by \citet{Herpich:2024}.
 
\subsection{Filter Sequence and Observational Strategy} \label{sec:filters} 

In the S-PLUS survey, the filters are observed in the following fixed sequence: $u$, $J0378$, $J0395$, $J0410$, $J0430$, $J0515$, $g$, $r$, $i$, $z$, $J0660$, $J0861$. Each field is observed for approximately 90 minutes, with three exposures taken per filter, accounting for readout times and filter transitions. This sequence enables the study of variability in sources over different timescales and wavelengths. Furthermore, the time gap between the $r$ and $J0660$ filters, caused by the intervening filters, plays a crucial role in detecting variability in sources with periods that are captured within this observational sequence.

\section{Selection Procedure}
\label{sec:strategy}

The selection of H$\alpha$ candidates is based on applying a series of restrictions to catalogues provided by S-PLUS DR4 in the $r$, $i$, and $J$0660 bands. We have reinforced this information in the Galactic disk region by generating catalogues with PSF photometry using SExtractor + PSFEx.

\begin{figure*}
  \setlength\tabcolsep{0pt}
  \setkeys{Gin}{width=0.5\linewidth}
  \begin{tabular}{ll}
    (a) & (b) \\
    \includegraphics[trim=5 10 10 20, clip]{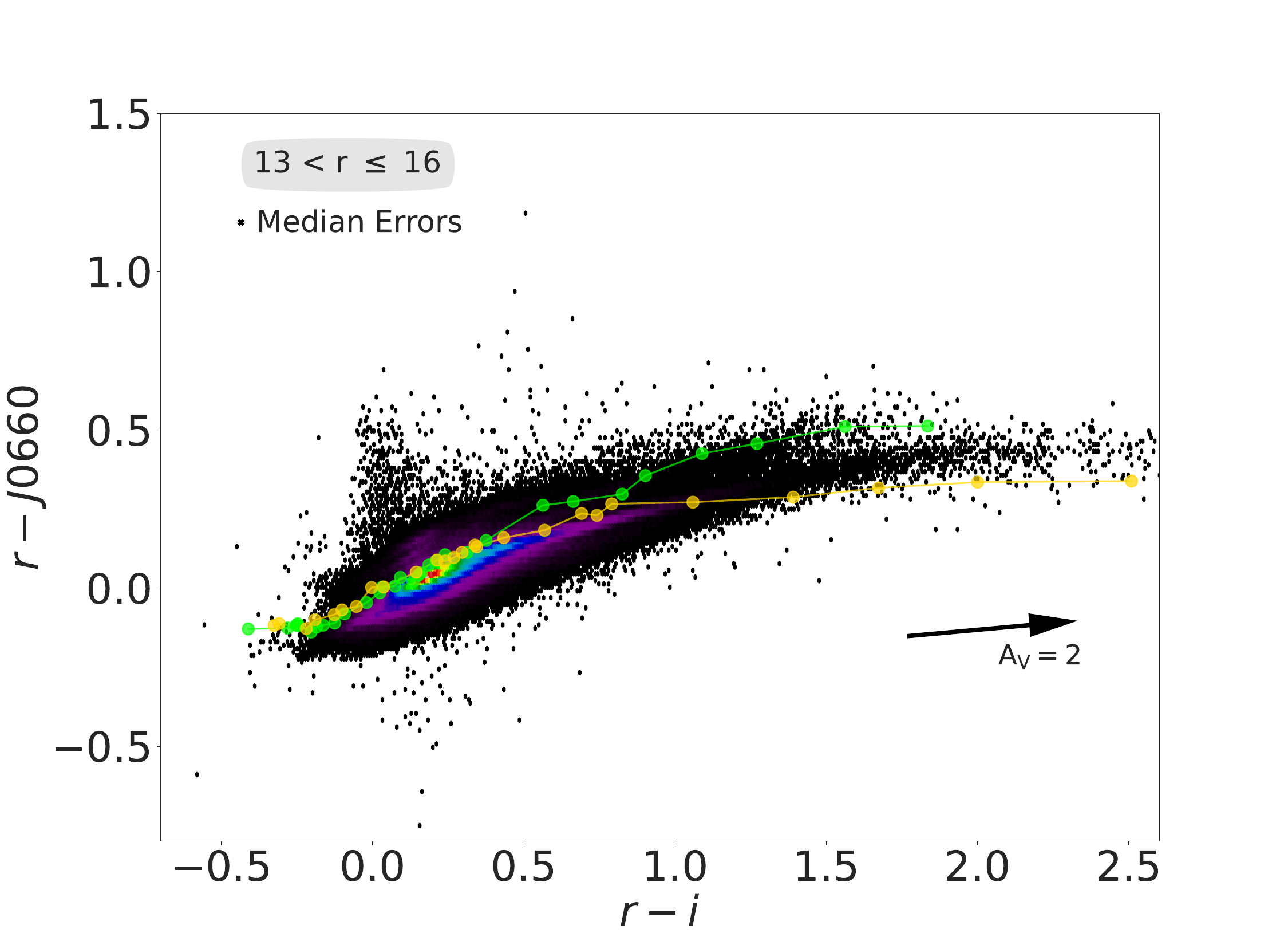}
    & \includegraphics[trim=5 10 10 20, clip]{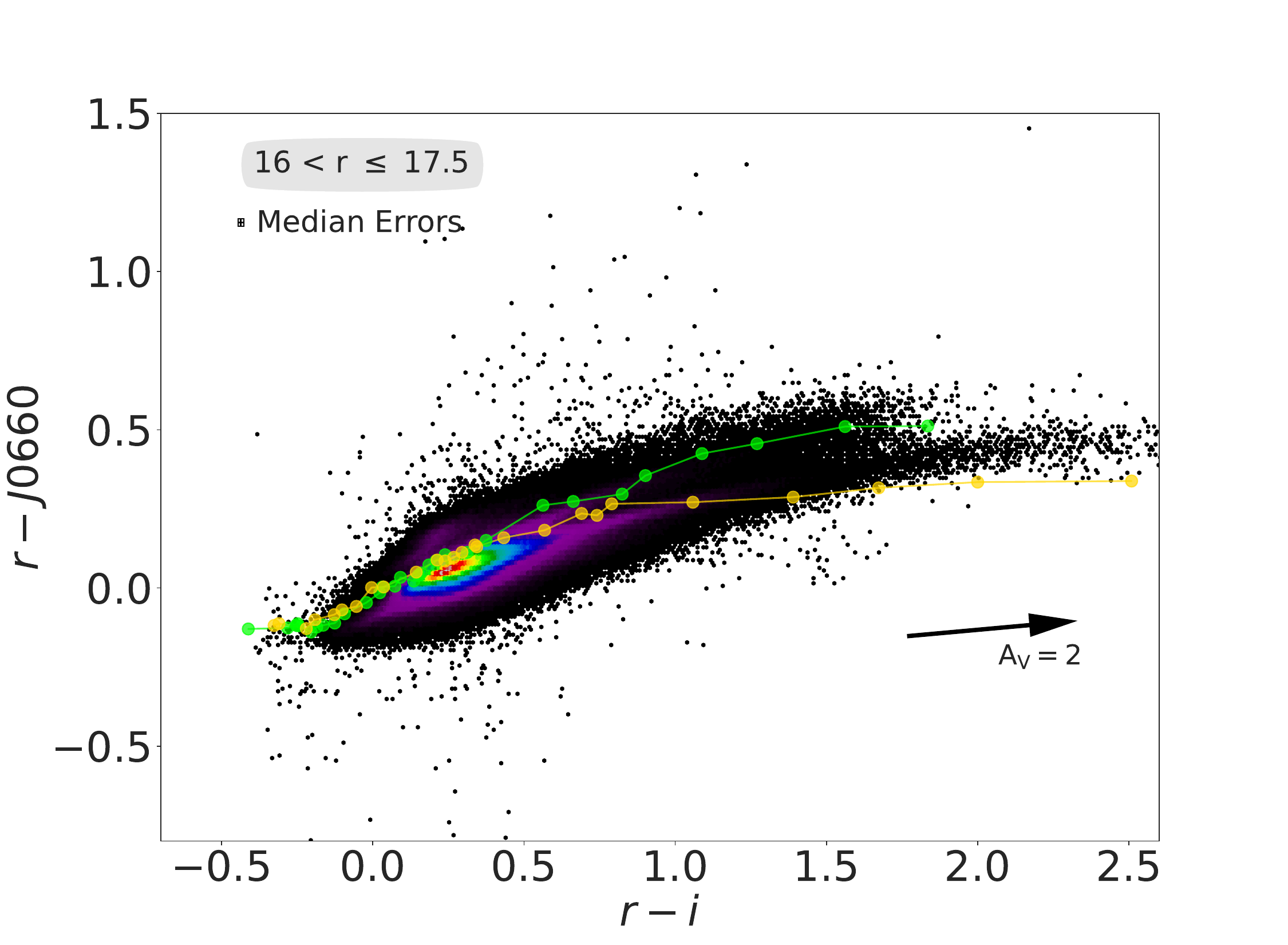}\\
    (c) & (d) \\
    \includegraphics[trim=5 10 10 20, clip]{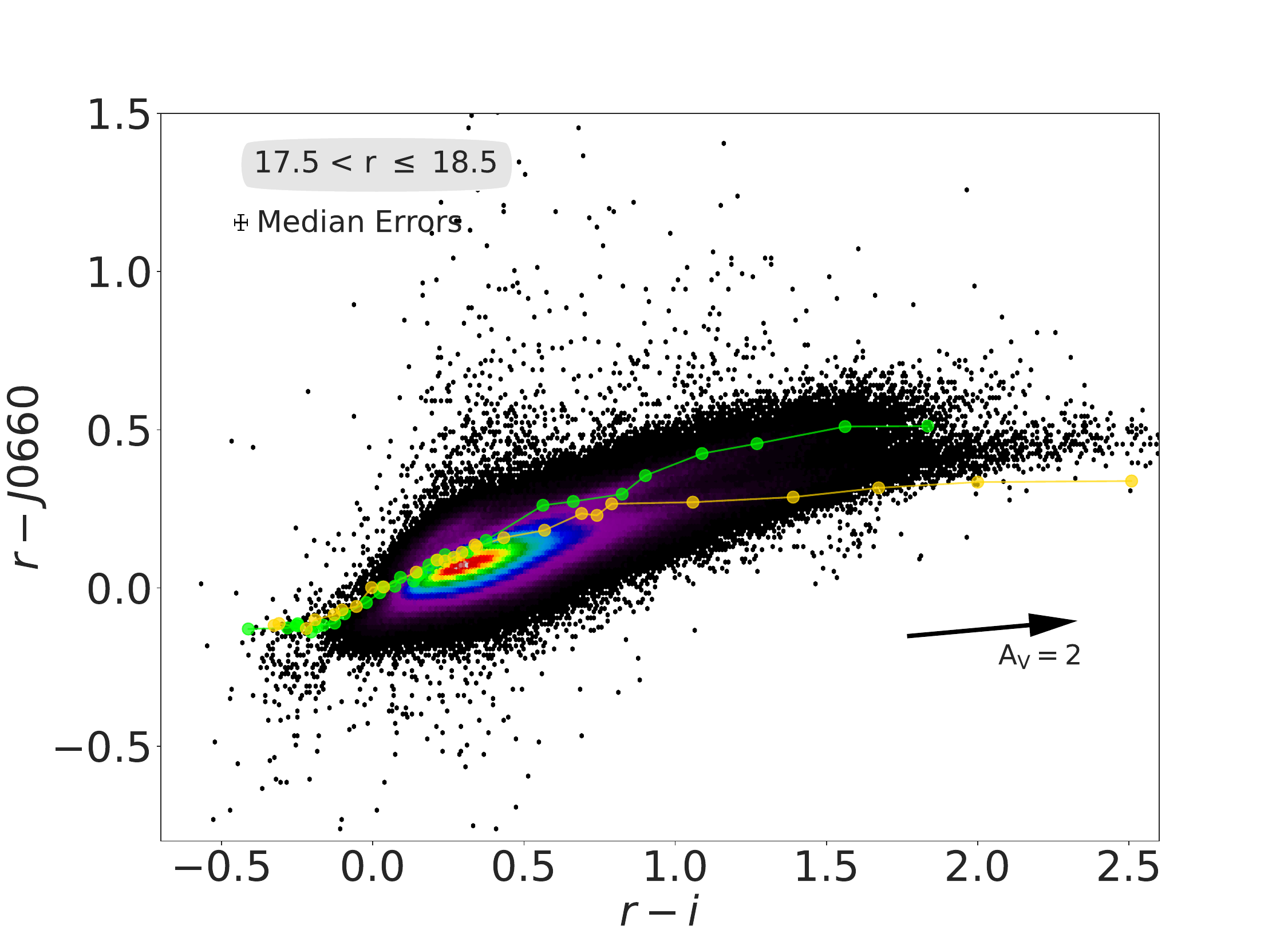}
    & \includegraphics[trim=5 10 10 20, clip]{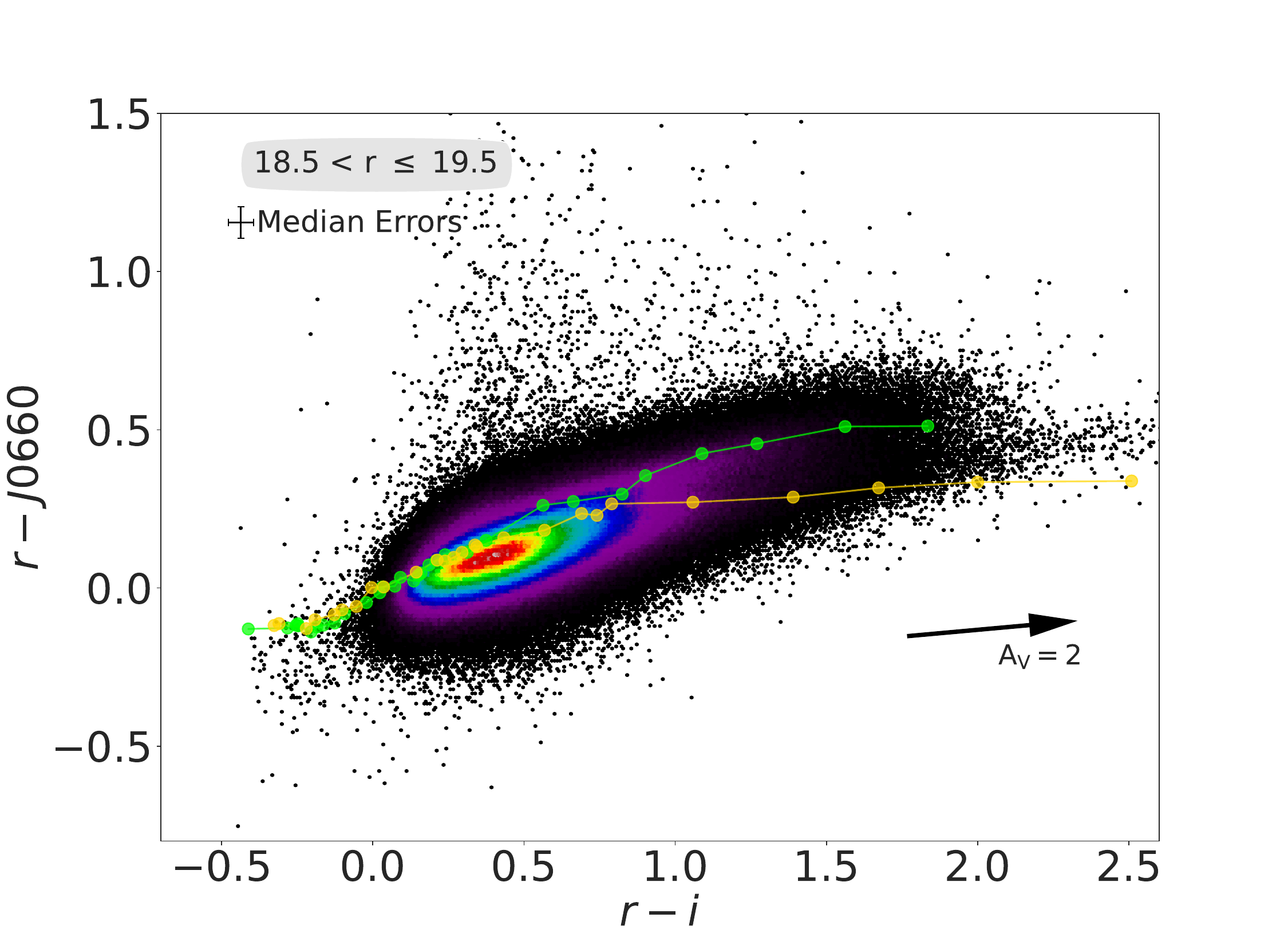}\\
  \end{tabular}
  \caption{Same as Fig. \ref{fig:all-DR4}, but for the GDS and using PSF photometry.}
  \label{fig:disk-DR4}
\end{figure*}

\subsection{Main Survey (MS) Data}
\label{sec:MS}

Amongst the different aperture photometry available in the S-PLUS DR4 catalogue, the \texttt{PStotal}\footnote{\texttt{PStotal} refers to photometry obtained using a 3-arcsecond circular aperture, with corrections applied to account for the fraction of flux that falls outside this aperture. This method is intended to provide the best estimate of the total magnitude of point sources.} photometry is used \citep[][]{Fernandes:2022}. To acquire data with high-quality photometry and identify compact objects in the MS, several criteria were applied:

\begin{itemize}
\item Objects must exhibit an $r$ magnitude within the range of $13 < r \leq 19.5$.
\item $J$0660 magnitude < 19.4 and $i$ magnitude < 19.2, which are average photometric depths for a S/N >10 threshold (see Table 4 of \citealp{Fernandes:2022}).
\item Errors less than 0.2 mag in the $r$, $J$0660, and $i$ filters.
\item The signal-to-noise ratio (S/N) in the respective filter should be higher than 10.
\item Objects should have \texttt{SEX\_FLAGS\_DET} < 4. The \texttt{SEX\_FLAGS\_DET} parameter is a bit-flag generated by SExtractor, indicating potential issues during photometry. The value corresponds to the sum of all flags, each represented by an integer. A value lower than four indicates that the flags can only sum up to: 0 (no flag), 1 (a minor issue: blending with another object), 2 (another minor issue: the object was originally blended with another one), or 3 (a combination of minor issues). Therefore, at most, we are selecting data affected by a combination of two minor issues, such as contamination from a nearby source and previous blending among sources. For more information, we refer the reader to \url{https://splus.cloud/documentation/DR4}.
\item Objects must satisfy \texttt{CLASS\_STAR\_r} = 1 and \texttt{CLASS\_STAR\_i} = 1, corresponding to the binary classification in the $r$ and $i$ filters, where a value of 1 indicates that the source is classified as a point source (star) in each filter, and a value of 0 denotes non-stellar or extended sources. The \texttt{CLASS\_STAR} parameter in SExtractor represents a probability value ranging from 0 to 1, with higher values indicating a greater likelihood that the source is a point source. In our selection, we applied this binary classification for \texttt{CLASS\_STAR\_r} and \texttt{CLASS\_STAR\_i} to ensure a higher likelihood that the sources are stars.

\end{itemize}

Additional criteria were implemented. These criteria are systematically chosen to ensure the robustness and reliability of the selected sample, considering various photometric and morphological properties of the sources.

\begin{itemize}

\item We consider the morphological properties of the sources by imposing a threshold on ellipticity. Sources with ellipticity values greater than 0.2 are likely to have non-galactic types (e.g., AGN, QSOs, galaxies, radio sources) or irregular shapes and are therefore excluded.
\item We select sources with compact morphology by constraining the radius enclosing 50\% of the total flux, setting \texttt{FLUX\_RADIUS\_50} < 3. Sources with a flux radius exceeding 3 pixels are likely to have extended morphology and are thus excluded from the sample.

\end{itemize}

These constraints led to the selection of 6\,655\,139 sources. The data were obtained by querying the project's database using the \texttt{splusdata} Python package, accessible via \texttt{S-PLUS Cloud}\footnote{\url{https://splus.cloud/}}. 

\subsection{Galactic Disk Survey (GDS)}
\label{sec:GDS}

We used a combination of  SExtractor\footnote{\url{https://www.astromatic.net/software/sextractor}} \citep{Sextractor:1996} and PSFex\footnote{\url{https://www.astromatic.net/software/psfex}} \citep{Psfex:2011} for source detection and posterior photometric measurements. We performed a serie of proofs with different SExtractor (e.g. {\sc detect\_minarea, detect\_thresh, phot\_apertures}) and PSFex (e.g. {\sc psf\_size, photflux\_key, psfvar\_degrees}) parameters plus test images (e.g. {\sc background}, {\sc background\_rms}, {\sc -background}, {\sc apertures}) to detect the largest number of objects with the best measurement possible of  PSF-magnitude, {\sc mag\_psf}. The crucial parameters for PSF photometry are listed in Table\ref{table:sexpara}.
The detection was performed on images from which their median-filtered versions was subtracted; faint sources are detected more easily in a median-subtracted image \citep{Lomeli:2017}. All median images were produced with a 11$\times$11 pix$^{2}$ median-filter.

\begin{table}
    \setlength\tabcolsep{3.7pt}
     \centering
	\caption{SExtractor and PSFex input parameters.}
	\label{table:sexpara}
	\begin{tabular}{l c c } 
\hline
\multicolumn{2}{c}{SExtractor} & \\
\hline
Parameter        &  Value &    \\ 
\hline
\hline
{\sc detect\_minarea}   & 3  \\
{\sc detect\_thresh}    & 1.5 \\
{\sc analysis\_thresh}  & 1.5 \\
{\sc pixel\_scale}      & 0.55 \\
{\sc back\_size}        & 64 \\
{\sc back\_filtersize}  & 3   \\
\hline
\multicolumn{2}{c}{PSFex} & \\ 
\hline
{\sc psf\_size}           &  18 \\
{\sc psfvar\_degrees}     & 3 \\
\hline
	\end{tabular}
\end{table}

The PSF photometry method is described in \citet{Lomeli:2017}, \citet{Lomeli:2022}, \citet{Lomeli:2022cfht} and Lomel\'i-N\'u\~nez in prep. A brief description of the photometric method is given below.
a) {\it First run of SExtractor}: we run SExtractor for the first time for the detection and selection of point sources based on their brightness versus compactness, as measured by the parameters of SExtractor {\sc mag\_auto} (a Kron-like elliptical aperture magnitude; \citealt{Kron:1980}) and {\sc flux\_radius} (similar to the effective radius). For the creation of the PSF, we selected sources in the space {\sc mag\_auto} vs {\sc flux\_radius}, in  a range to: 12$\,\lesssim$\,{\sc mag\_auto}$\,\lesssim$\,21.5 and 1\,$\lesssim$\,{\sc flux\_radius}$\,\lesssim$\,3.5. Because we are observing towards the Galactic disk, the number of sources to creation each PSF can reach $\sim$20000 sources, which was not possible in the previous works because they were focused on extragalactic sources far away from the Galactic disk.
b) {\it PSF creation:} we used PSFex to create the PSF using the point sources selected in the last step. The spatial variations of the PSF were modelled using third-degree polynomials as a function of the pixel coordinates (X, Y). For PSF creation, the ﬂux of each star was measured in an aperture of 9 pixels of radius in all bands (equivalent to 4\arcsec.95$\times$4\arcsec.95); such aperture, determined through the growth-curve method for each passband, is large enough to measure the total ﬂux of the stars, but small enough to reduce the likelihood of contamination by external sources. 
c) {\it Second run of SExtractor:} we run SExtractor again this time using the PSF created in the last step as an input parameter to measure the magnitude of the PSF ({\sc mag\_psf}). In this work we always used the {\sc mag\_psf}, for simplicity only the name of each band is written. 

The constraints described in Sect. \ref{sec:MS} were applied with adjustments specific to the GDS to ensure high-quality data, resulting in the selection of 7\,007\,778 sources. To ensure the reliability of the data, five fields from the GDS were excluded from the analysis due to apparent calibration issues. These fields showed systematic offsets in the $(r - J0660)$ colour when compared to other fields, indicating potential zero-point calibration problems. Excluding these fields minimizes the impact of systematic errors and enhances the robustness of the results. Furthermore, potential edge effects were mitigated by carefully handling sources near the CCD boundaries, ensuring consistent photometric quality across the dataset.

\subsection{Selection of H$\alpha$ Excess Sources}
\label{sec:metho}

\begin{figure*}
  \setlength\tabcolsep{0pt}
  \setkeys{Gin}{width=0.5\linewidth}
  \begin{tabular}{ll}
    (a) & (b) \\
    \includegraphics[trim=5 10 10 20, clip]{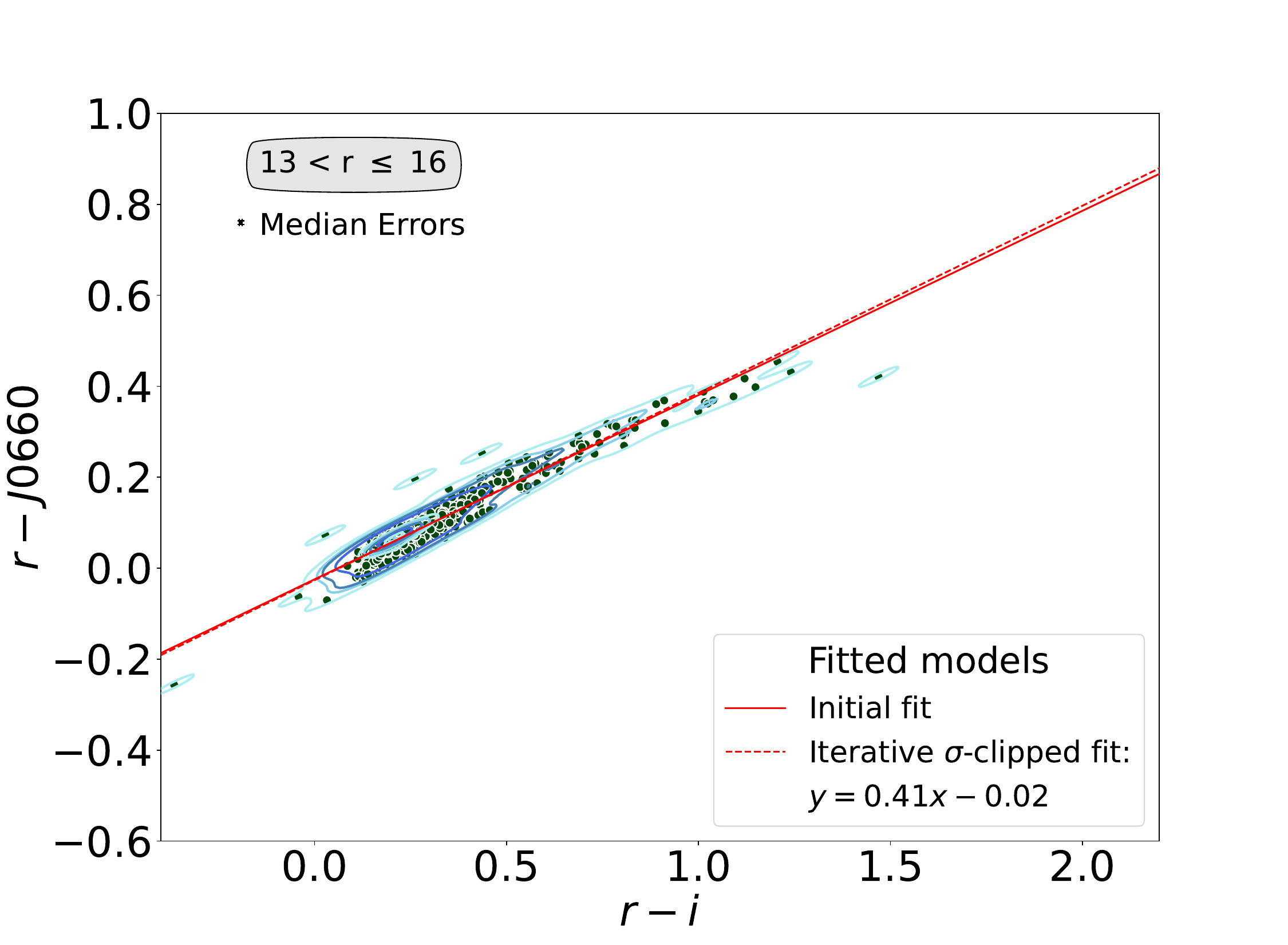}
    & \includegraphics[trim=5 10 10 20, clip]{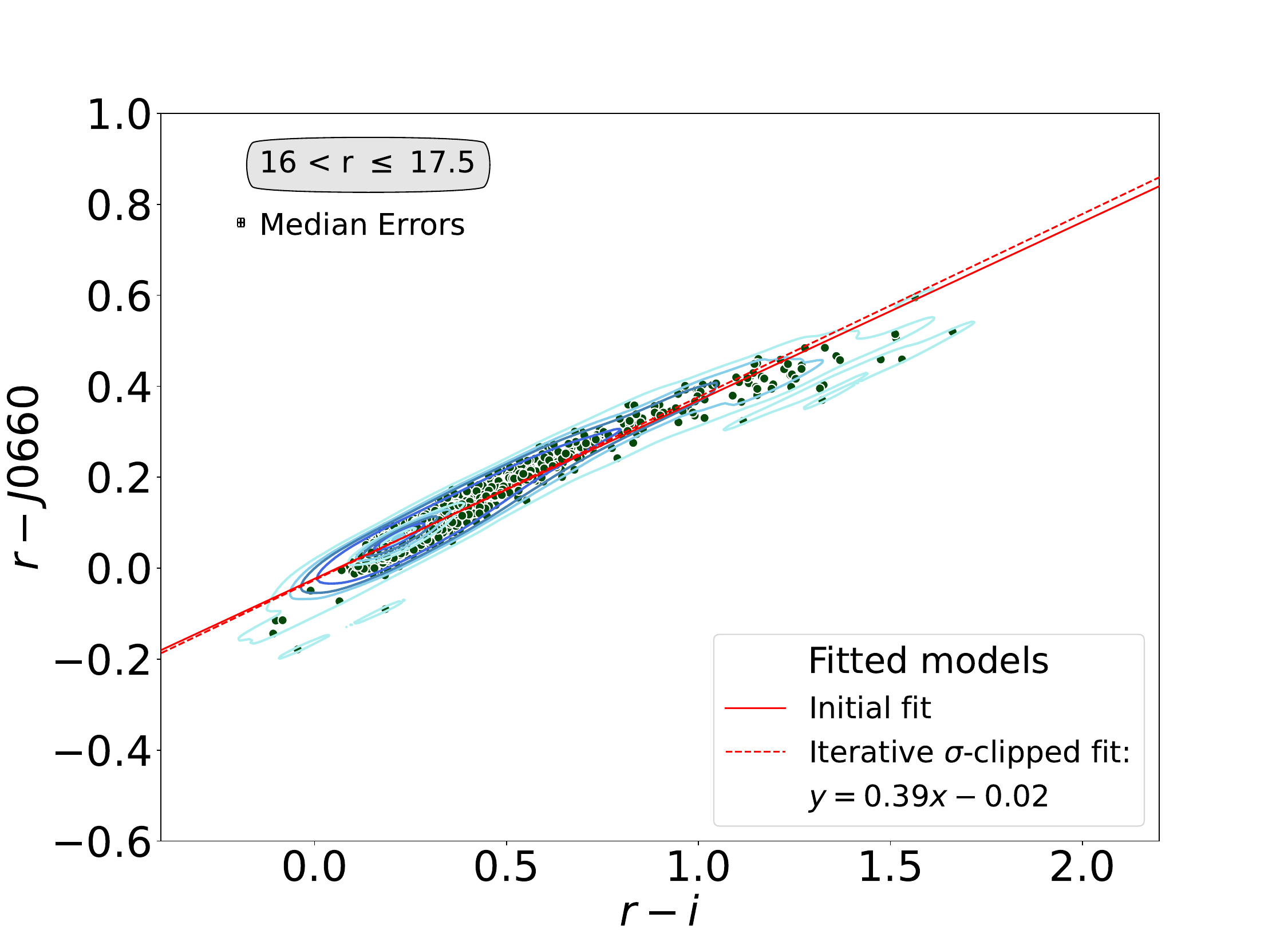}\\
    (c) & (d) \\
    \includegraphics[trim=5 10 10 20, clip]{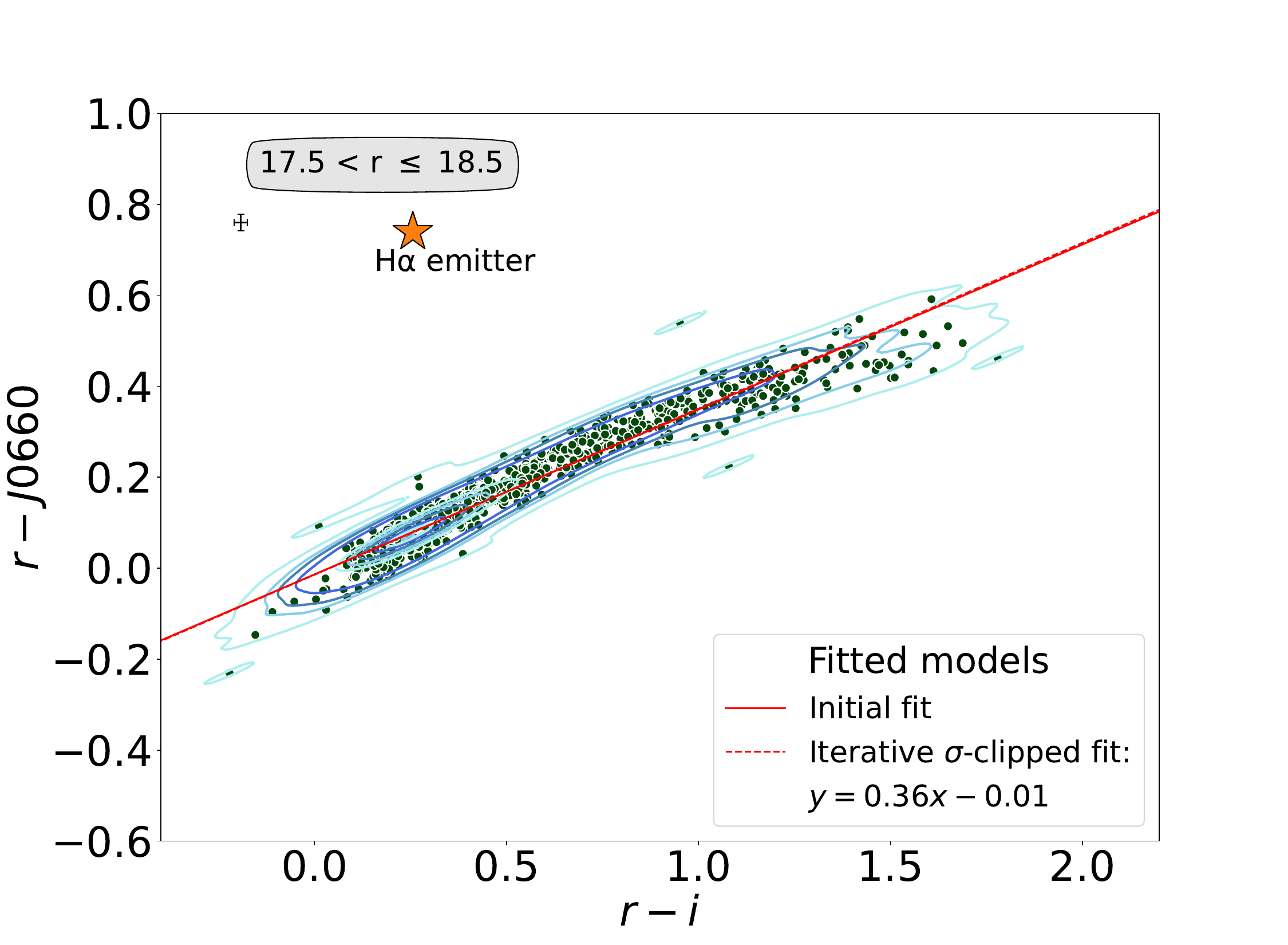}
    & \includegraphics[trim=5 10 10 20, clip]{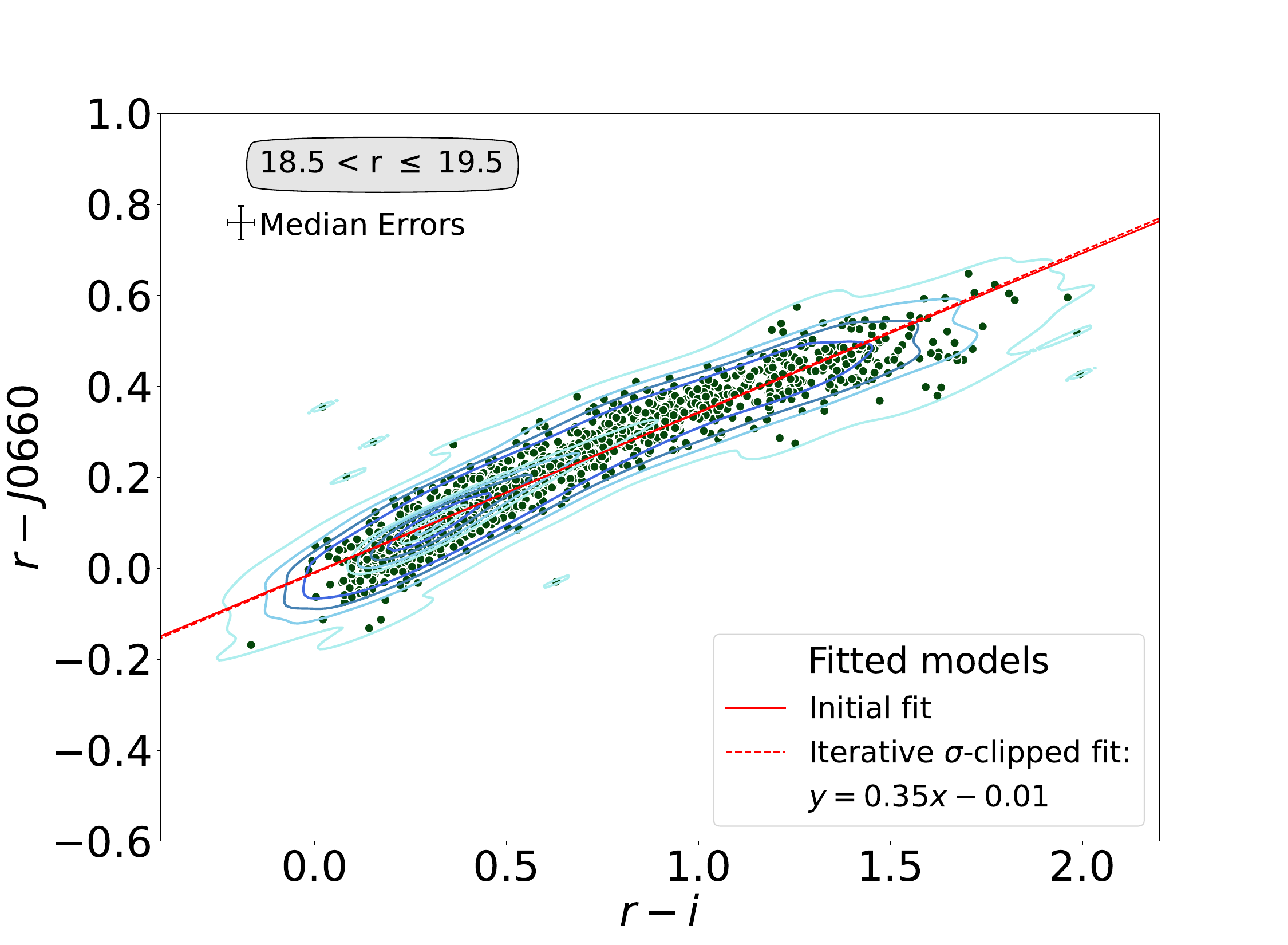}\\
  \end{tabular}
  \caption{Illustration of the selection criteria used to identify strong emission-line objects via colour-colour plots. The data shown here are from the S-PLUS field STRIPE82-0142, split into four magnitude bins, as displayed in the four panels. The thin red continuous lines show the initial linear fit to all data points (in green), while the dashed red line is the fit after applying iterative $\sigma$-clipping. Objects selected as H$\alpha$ emitters are located above the dashed line. The orange star in panel (\textit{c}) represents the cataclysmic variable (CV) FASTT 1560 (S-PLUS ID: DR4\_3\_STRIPE82-0142\_0021237), highlighted here as an example of an H$\alpha$ emitter identified by our criteria.}
 
  \label{fig:criteria-color-plot}
\end{figure*}

Before searching for potential sources of H$\alpha$ excess sources hidden in the S-PLUS DR4 footprint, we first divided our sample into four subsamples based on their magnitudes in the $r$ band: (i) $13  \leq r  < 16$, (ii) $16 \leq r < 17.5$, (iii) $17.5 \leq r < 18.5$, and (iv) $18.5 \leq r < 19.5$. This way, we avoided mixing up bright and faint sources with low and high uncertainties, respectively. Otherwise, the selection criteria could be affected by the intrinsic scatter in the measurement of faint objects. Figures~\ref{fig:all-DR4} and~\ref{fig:disk-DR4} display the ($r$-$J$0660) versus ($r$-$i$) colour-colour diagrams for the sources
from the MS of S-PLUS and the sub-survey of the GDS, respectively. The lighter green and yellow points connected by lines represent the tracks of main sequence and giant stars, respectively. These loci for main sequence and giant stars were derived from the synthetic spectra library by \citet{Pickles:1998}, convolved with the S-PLUS transmission curves in the AB magnitude system \citep{Oke:1983}. It is important to note that in these diagrams, the magnitudes for the MS correspond to \texttt{PStotal}, while for the GDS sources they correspond to PSF photometry. 

\begin{figure*}
  \setlength\tabcolsep{0pt}
  \setkeys{Gin}{width=0.5\linewidth}
  \begin{tabular}{ll}
    (a) & (b) \\
    \includegraphics[trim=5 10 10 20, clip]{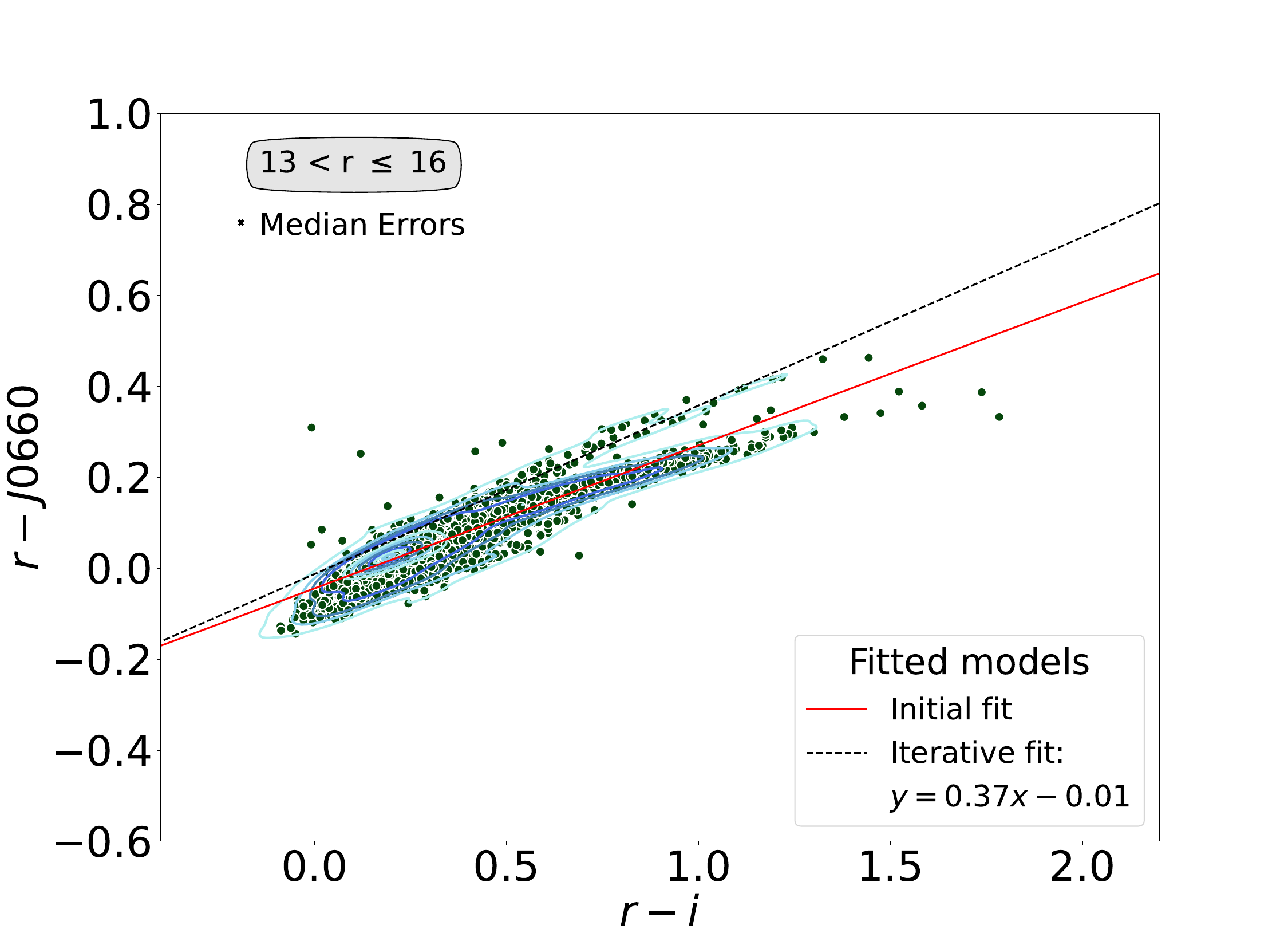}
    & \includegraphics[trim=5 10 10 20, clip]{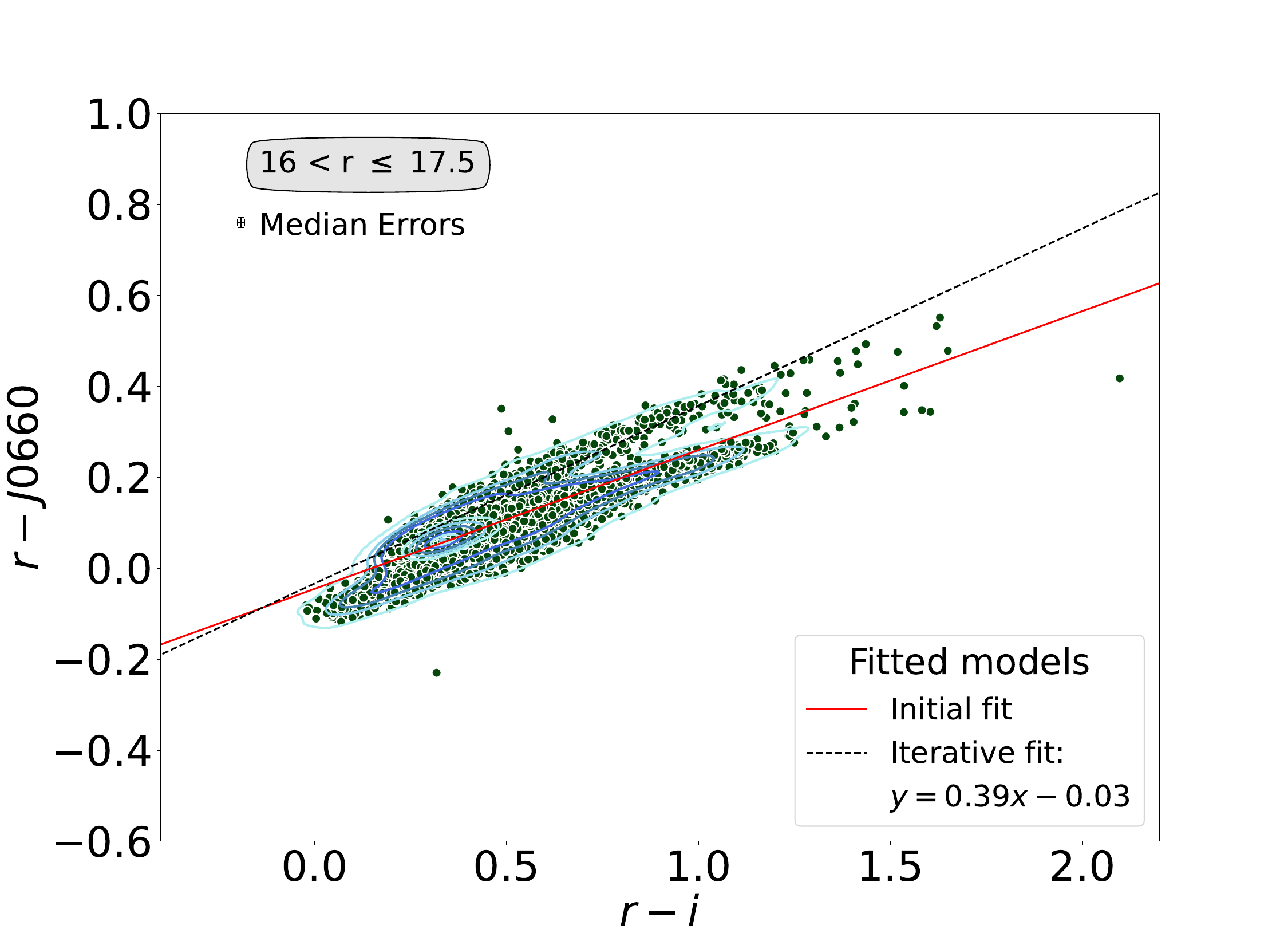}\\
    (c) & (d) \\
    \includegraphics[trim=5 10 10 20, clip]{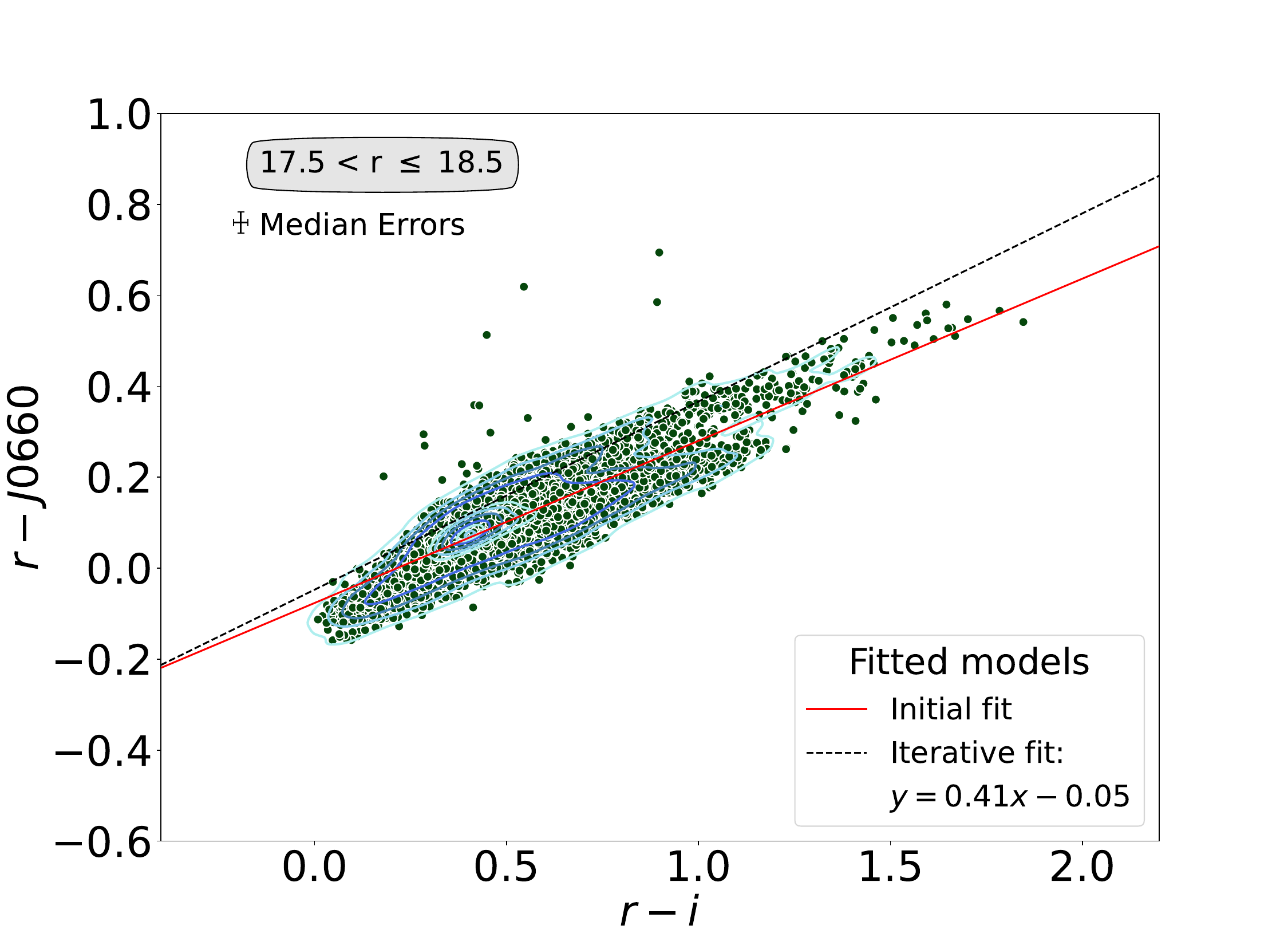}
    & \includegraphics[trim=5 10 10 20, clip]{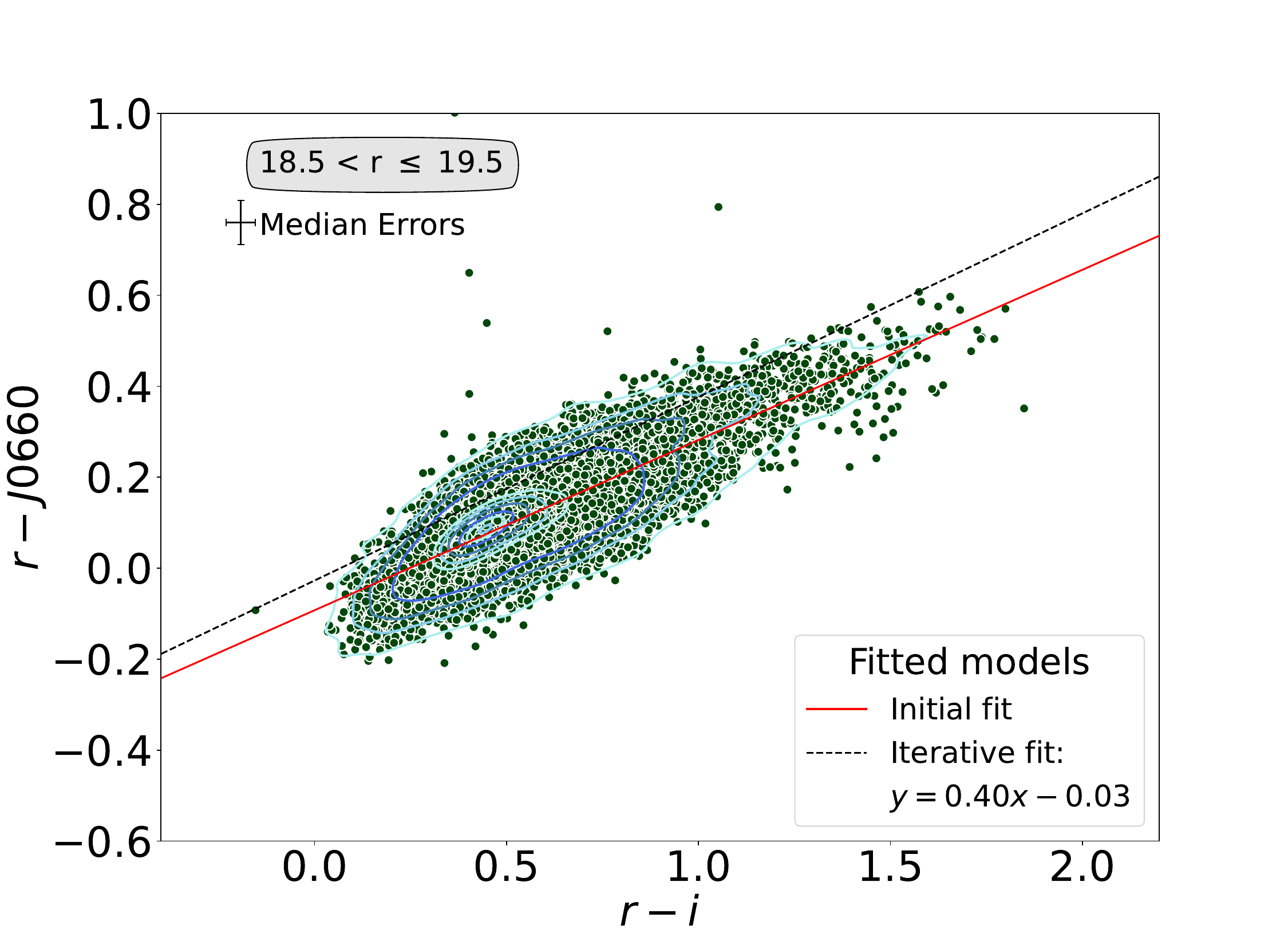}\\
  \end{tabular}
  \caption{Colour-colour diagram of the stars locate at field SPLUS-d288 like those found in Fig. \ref{fig:criteria-color-plot}. The red lines represent the original fit to all data, while the black dashed lines represent the final fits to the upper locus of points, obtained by applying an iterative fitting process to the initial fit.}
  \label{fig:criteria-color-plot-disk}
\end{figure*}

The identification of objects is based on the method successfully applied by \citet[][]{Witham:2006, Witham:2008} to the IPHAS catalogue, since similar filters are also available in S-PLUS: $r$, $J$0660, and $i$. Similar technique was also used by \citet{Scaringi:2013, Wevers:2017, Monguio:2020, Fratta:2021} to reveal H{$\alpha$} excess sources.

We first generated ($r$ - $J0660$) versus ($r$ - $i$) diagrams for each magnitude bin in each field and then attempted to fit the regions predominantly occupied by main-sequence and giant stars using a linear regression model. After this, we applied an iterative $\sigma$-clipping technique, where data points more than several $\sigma$ away from the fitted line were excluded in successive iterations to refine the fit. This process primarily aimed to remove outliers, ensuring that the final fit closely follows the bulk of the non-emitting stars, and was applied to the MS fields. 
Objects with H$\alpha$ emission typically exhibit an excess in ($r$ - $J0660$), causing them to appear above the main stellar loci in these plots. Therefore, it is expected that objects with H$\alpha$ signatures will be located above these fitted lines. For fields in the MS with low stellar density, mostly those outside the Galactic plane, this  fit often works well (as illustrated in Fig. \ref{fig:criteria-color-plot}). However, many fields of the GDS display (at least) two distinct stellar loci in the colour–colour plane, resulting from differential reddening and/or contributions from both main-sequence stars and giants, where the fit is likely to align with the reddened locus (also illustrated in Fig. \ref{fig:criteria-color-plot-disk}).

To address this aspect in the GDS, we followed the procedure implemented by \citet{Witham:2008}: we selected the objects above the initially fitted line and iteratively adjusted the fit, moving it upwards towards the uppermost locus of points in the colour–colour diagram. As shown in Fig. \ref{fig:criteria-color-plot-disk}, this upper locus generally corresponds to the unreddened main sequence. In cases where the final fit is poorer than the initial one (e.g., in fields containing only a single stellar locus), we reverted to the initial fit. Once the appropriate fit for each magnitude bin was established, we identified objects significantly above the fit as likely H$\alpha$ excess candidates. During this process, we examined the colour–colour diagram for each field and bin to ensure the fit was suitable, and found that, in general, 2 to 3 iterations were sufficient to locate the upper locus. This method ensures that objects exhibiting excess in H$\alpha$ emission should adhere to the specified criterion:

\begin{equation}
(r - J0660)_{\mathrm{obs}} - (r - J0660)_{\mathrm{fit}} \geq C \times \sigma_{\text{est}},
\label{eq:criterion}
\end{equation}

\noindent where $(r - J0660)_{\mathrm{obs}}$ denotes the observed colour difference between the $r$ and $J0660$ bands, $(r - J0660)_{\mathrm{fit}}$ represents the colour difference predicted by the linear regression fit, $C$ is a constant parameter set to 5, and $\sigma_{\text{est}}$ is the estimated standard deviation of the residuals around the fit, defined as:

\begin{equation}
\sigma_{\text{est}} = \sqrt{\sigma^2_{\mathrm{s}} + (1 - m)^2 \times \sigma^2_{(r - J0660)} + m^2 \times \sigma^2_{(r - i)}}, 
\label{eq:sigma}
\end{equation}

 \noindent where $\sigma_{\mathrm{s}}$ represents the root mean squared value of the residuals around the fit, $\sigma_{(r - i)}$ denotes the error in the colour index between the $r$ and $i$ bands, $\sigma_{(r - \mathrm{J0660})}$ denotes the error in the colour index between the $r$ and $J0660$ bands, and $m$ represents the slope of the linear regression fit. The fits were performed using the \texttt{astropy.modeling} library
\footnote{\url{https://docs.astropy.org/en/stable/modeling/index.html}}.

Figure~\ref{fig:criteria-color-plot} illustrates the procedure applied to one field in the MS (STRIPE82-0142). The iterative approach was used for each individual field, with solid red lines indicating the initial fit. Sources showing $J$0660 excess or lying significantly above the stellar locus were identified as deviations from these fitted lines. The large orange star in panel {\sc c} of Fig.~\ref{fig:criteria-color-plot} represents a known H$\alpha$ emitter (CV, FASTT 1560, \citealp{Abril:2020}) that lies significantly above the stellar locus, with $(r - J0660) > 0.5$. Figure~\ref{fig:criteria-color-plot-disk} shows the same procedure applied to the GDS. The red lines indicate the initial fit, while the black dashed lines represent the final iterative fits.

\section{Results and Analysis}
\label{sec:results}

\begin{figure}
    \includegraphics[width=\linewidth]{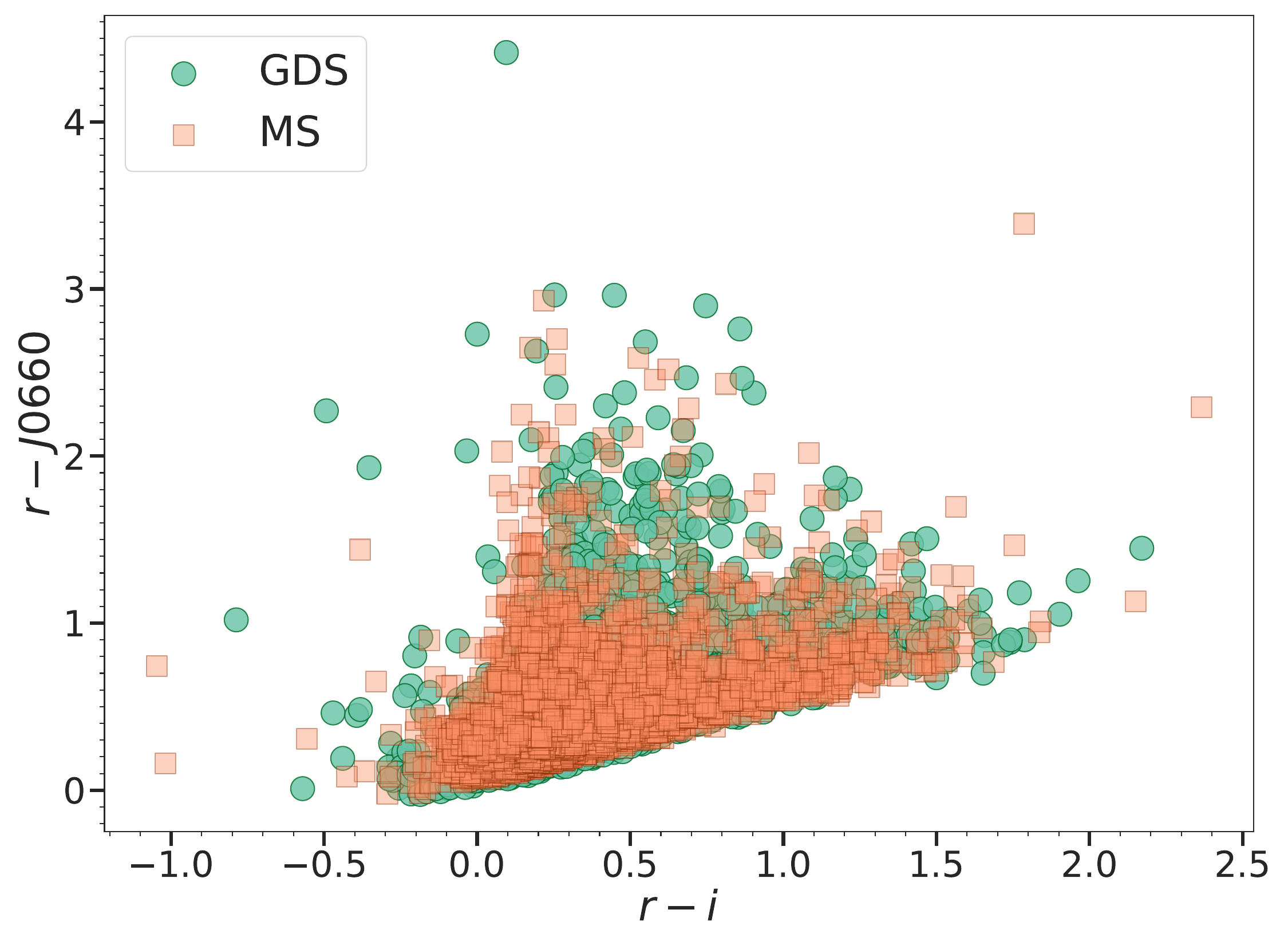}
    \caption{The colour-colour diagram shows the distribution of H$\alpha$-feature sources in the ($r - J0660$) versus ($r - i$) colour-colour space. The data is divided into two populations: GDS and MS representing distinct galactic components. The GDS population, depicted by filled circles in greenish colour, corresponds to H$\alpha$ excess sources associated with disk structures. In contrast, the MS population, represented by square symbols in light orange, includes H$\alpha$ excess sources primarily located in the direction of the Galactic halo and extragalactic sources.} 
    \label{fig:Halpha-excess}
\end{figure}

Our objective is the identification of H{$\alpha$} excess sources within the S-PLUS footprint, leveraging the unique filter system of the survey. This effort resulted in 3\,637 outliers for the MS and 3\,319 for the GDS. The distribution of the sources with excess H{$\alpha$} emission in the ($r - J0600$) versus ($r - i$) colour-colour plane is depicted in Fig.~\ref{fig:Halpha-excess}. Square light orange symbols represent objects with H$\alpha$ excess identified in the MS, while greenish circle symbols denote those found in the GDS. All the sources placed above the locus of the main and giant stars exhibit an excess in the $J0600$ filter, attributed to the H$\alpha$ excess. The broad distribution of sources on the colour-colour diagram of ($r - J0660$) and ($r - i$) indicates the selection of several types of H{$\alpha$} sources. These sources are likely associated with PNe, CVs, SySt, YSOs, Be stars, as well as extragalactic compact objects like QSOs and galaxies, among others (see Fig. 2 of \citealp{Gutierrez:2020}).

The fractional contribution of different classes of sources to the overall sample was evaluated by cross-matching the objects' list with the \texttt{SIMBAD} database\footnote{\url{http://simbad.u-strasbg.fr/simbad/}}. Optical spectra available in the Sloan Digital Sky Survey (SDSS; \citealp{York:2000}) and in the Large Sky Area Multi-Object Fiber Spectroscopic Telescope (LAMOST; \citealp{Wu:2011}) were also explored. 
In all cases, positive matches between the different catalogues were considered for those sources that have an angular distance on the sky-plane within a given limit ($d_{max,proj}$). Verification of the photometry and assessment of H$\alpha$ excess in the selected objects within the disk area were conducted by cross-matching the H$\alpha$ source list identified in S-PLUS with photometric data from VPHAS+ DR2.

\subsection{Matches with SIMBAD sources}

\begin{table*}[h]
\centering
\caption{Summary of the positional cross-match results between the S-PLUS list of H$\alpha$ source and the \texttt{SIMBAD} database. A search radius of 2 arcsec was used for the MS, while 1 arcsec was used for the GDS. The first column indicates main object categories, the second column lists \texttt{SIMBAD} object types, and the third column indicates the number of objects in each category.}
\label{tab:simbad-sources}
\begin{adjustbox}{max width=\textwidth}
\begin{tabular}{llc}
\hline\hline
\textbf{Main Type}          & \textbf{Associated \texttt{SIMBAD} Types}                               & \textbf{Number of S-PLUS Objects} \\
                            &                                                                         & \textbf{with \texttt{SIMBAD} Match} \\
\hline
\multicolumn{3}{c}{\textbf{Main Survey}} \\
\hline
Stellar Binary System       & CataclyV*, CV*\_Candidate, RSCVn, EB*, EB*\_Candidate, SB*\_Candidate   & 353 \\
Variable Star               & PulsV*, V*, PulsV*delSct, RotV*,  RRLyr                                 & 139 \\
Star                        & Star, Blue, low-mass*, WD*, WD*\_Candidate, PM*, BlueStraggler          & 47  \\
Radio Source                & Radio, Radio(cm), RadioG                                                & 9  \\
Active Galactic Nucleus (AGN) & AGN, AGN\_Candidate, Seyfert\_1                                       & 23  \\
Quasar                      & QSO, QSO\_Candidate                                                     & 143 \\
Galaxy                      & Galaxy                                                                  & 9   \\
Other                       & Hsd\_Candidate, Pec*, AGB*, MIR                                         & 8   \\
\textbf{Total}&                                                                                       & \textbf{731} \\
\hline
\multicolumn{3}{c}{\textbf{Disk}} \\
\hline
Emission-line star          & Em*, Be*                                                               & 125 \\
Young stellar object        & YSO, YSO\_Candidate, Orion\_V*, TTau*\_Candidate       & 102 \\
Stellar Binary System       & CataclyV*, CV*\_Candidate, RSCVn, EB*, EB*\_Candidate, SB*             & 146 \\
Variable star               & PulsV*delSct, PulsV*, LPV*, LP*\_Candidate, Mira, RRLyr, V*, V*?\_Candidate, BYDra   & 43 \\
Star                        & Star, **, RGB*, C*, WD*\_Candidate                                     & 104 \\
Nebula                      & PN?\_Candidate, RfNeb, Nova                                     & 3  \\
Other                       & EmObj, Hsd\_Candidate, deltaCep, Cepheid\_Candidate, Transient, X     & 9 \\
\textbf{Total}              &                                                                        & \textbf{532} \\
\hline\hline
\end{tabular}
\end{adjustbox}
\end{table*}

We identified a total of 1\,263 positive matches between our catalogues of H$\alpha$ compact excess sources and the \texttt{SIMBAD} database, assuming a search radius of $d_{\text{max,proj}}$\,=\,2,arcsec for the MS and 1\,arcsec for the GDS. In the MS, the identified objects primarily fall into categories such as variable stars, predominantly cataclysmic variables and/or candidates (CataclyV*), eclipsing binaries and/or candidates (EB*), and RR Lyrae Variables (RRLyr), as well as various kinds of stars including normal stars, white dwarfs, and/or candidates (WD*). Additionally, extragalactic compact sources that exhibit redshifted lines coinciding with the $J0660$ filter, simulating the H$\alpha$ emission line, are also present. These include AGNs, Seyfert galaxies, QSOs, and other objects. It is important to note that the presence of redshifted emission lines in extragalactic sources may contribute to the identification of some of these objects as H$\alpha$-excess candidates (see Table \ref{tab:simbad-sources} for details).

For the GDS, the identified categories include emission-line stars (Em*), young stellar objects (YSO) and candidates, which encompass T Tauri (TTau*) and Herbig Ae/Be (Ae*) star candidates. Additionally, variable stars such as cataclysmic variables (CataclyV*), eclipsing binaries (EB*), and RR Lyrae variables (RRLyr) are found, along with objects exhibiting nebular components, such as planetary nebula (PN) candidates, novae, and reflection nebulae (RfNeb), among others. As shown in Table \ref{tab:simbad-sources}, the highest number of sources in the disk belong to the Em* and young stellar objects category, reflecting the active star formation processes in the Galactic disk. 

An important consideration regarding the \texttt{SIMBAD} matches is that in the MS, numerous extragalactic sources with emission lines are selected due to the mapping of high latitudes in the southern sky. Conversely, for the GDS, no extragalactic sources have been selected. While the MS emphasizes extragalactic sources and diverse stellar populations, the disk region primarily showcases young stellar objects and variable stars, indicative of ongoing star formation and stellar evolution processes. In both regions, variable stars such as EB*, among others, are also present. The results are described below and listed in Table \ref{tab:simbad-sources}.

In our analysis of H$\alpha$-excess sources, variable stars such as RR Lyrae stars and eclipsing binaries are frequently detected due to their tendency to exhibit significant photometric deviations in the H$\alpha$-related bands. It is important to highlight that RR Lyrae stars, which are known for their characteristic spectral features, often show H$\alpha$ absorption lines. This occasionally causes them to be identified as outliers in our selection, as our criteria are sensitive to any significant deviation from expected stellar colours, whether it involves emission or absorption features. Moreover, the use of the S-PLUS filter system, with its 12 sequential filters, plays a role in detecting these short-period variables. Since both RR Lyrae and eclipsing binaries have short periods (typically hours to days), the sequential observation through S-PLUS's filters can capture these stars at different phases of their variability. This effect can lead to apparent H$\alpha$-excess due to the changes in brightness across different bands during the observation sequence. In particular, eclipsing binaries can display H$\alpha$ emission due to complex interactions between the stellar components and their surrounding material. This phenomenon has been observed in systems such as the eclipsing binary VV Cephei, where periodic variations in H$\alpha$ emission occur during different phases of the eclipse \citep{Pollmann:2018}. 

An important observation is that our selection criteria have predominantly excluded extended sources. In the MS, only 23 AGN and 9 galaxies were identified, making up approximately 3.1\% and 1.2\% of the total 731 SIMBAD matches (see Table~\ref{tab:simbad-sources}), respectively. Additionally, we identified 143 QSOs, representing about 19.6\% of the total matches. These percentages highlight the effectiveness of our selection criteria in isolating compact sources with significant H$\alpha$ excess, while also illustrating the relative proportions of different astrophysical categories identified in our survey.

\subsection{Redshifted Lines Mimicking the H$\alpha$ Emission}

According to the classification in the literature, near 20\% of the H{$\alpha$} sources in our sample are classified as QSOs. It is important to note that the excess observed in the $J$0660 filter is due to QSOs whose emission lines are redshifted to the wavelength range of this filter. For instance, lines such as H$\beta$, Mg \textsc{ii} 2798 \AA, C \textsc{iii}] 1909 \AA, and C \textsc{iv} 1550 \AA~ can contribute to this excess (see \citealp{Gutierrez:2020} and the bottom of Fig. 1 of \citealp{Nakazono:2021}, which shows the main emission lines of a quasar at different redshifts and indicates which of those fall within the $J$0660 filter).

This particular population of apparent H$\alpha$ emitters includes AGNs, Seyfert 1 galaxies, and other emission-line galaxies. In particular, within the redshift range $0.306 < z < 0.376$, lines such as H$\beta$ and [O \textsc{iii}] 4959, 5007 \AA~ are redshifted into the $J$0660 filter.

\subsection{Matches with SDSS and LAMOST}

\begin{figure}
    \includegraphics[width=\linewidth, trim=10 40 5 8, clip]{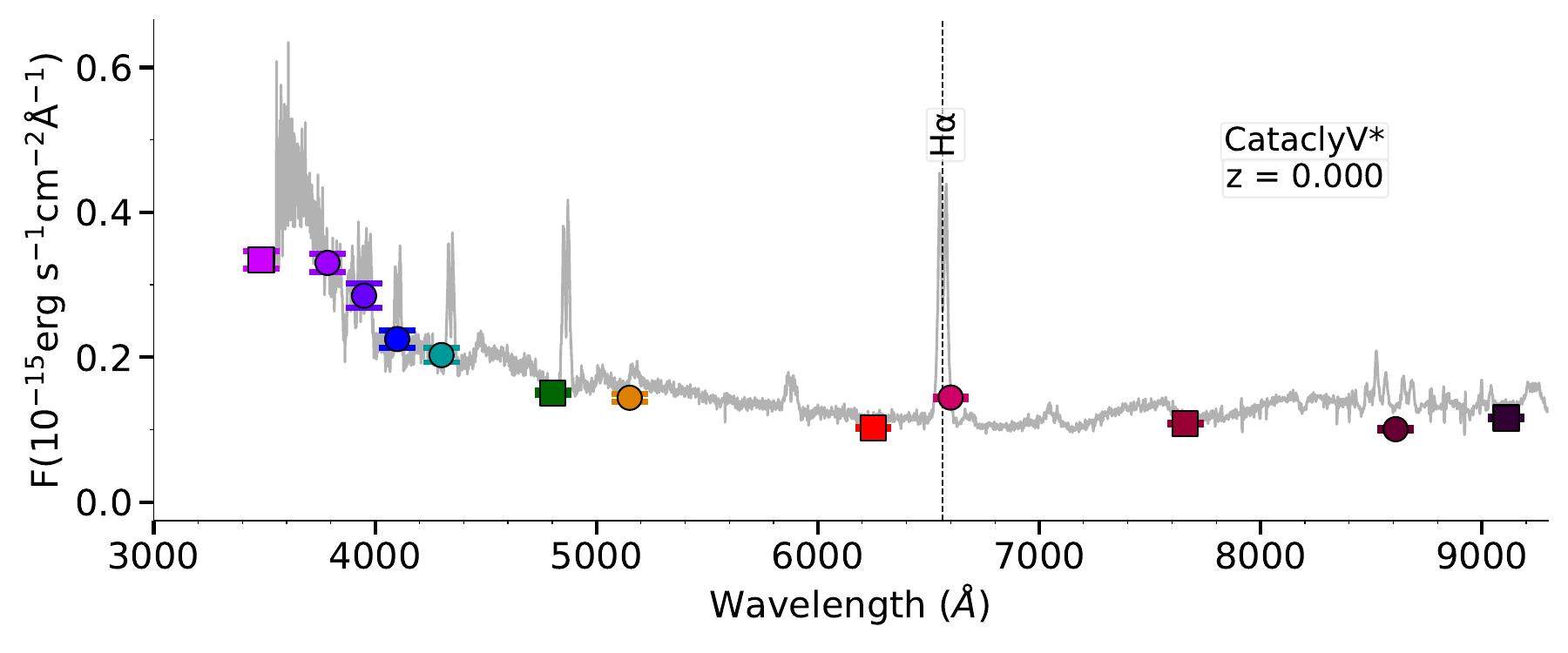} 
    \includegraphics[width=\linewidth, trim=10 10 5 8, clip]{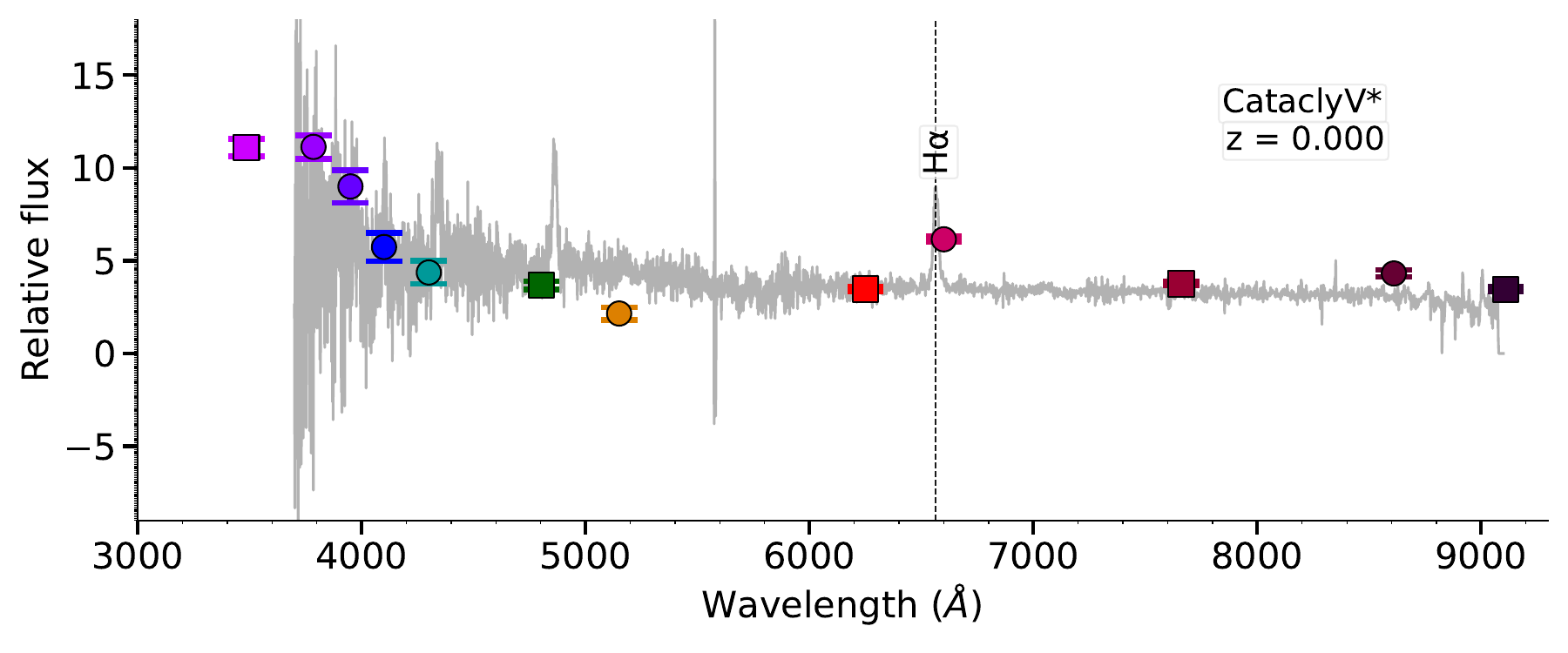}
    \includegraphics[width=\linewidth, trim=10 10 5 8, clip]{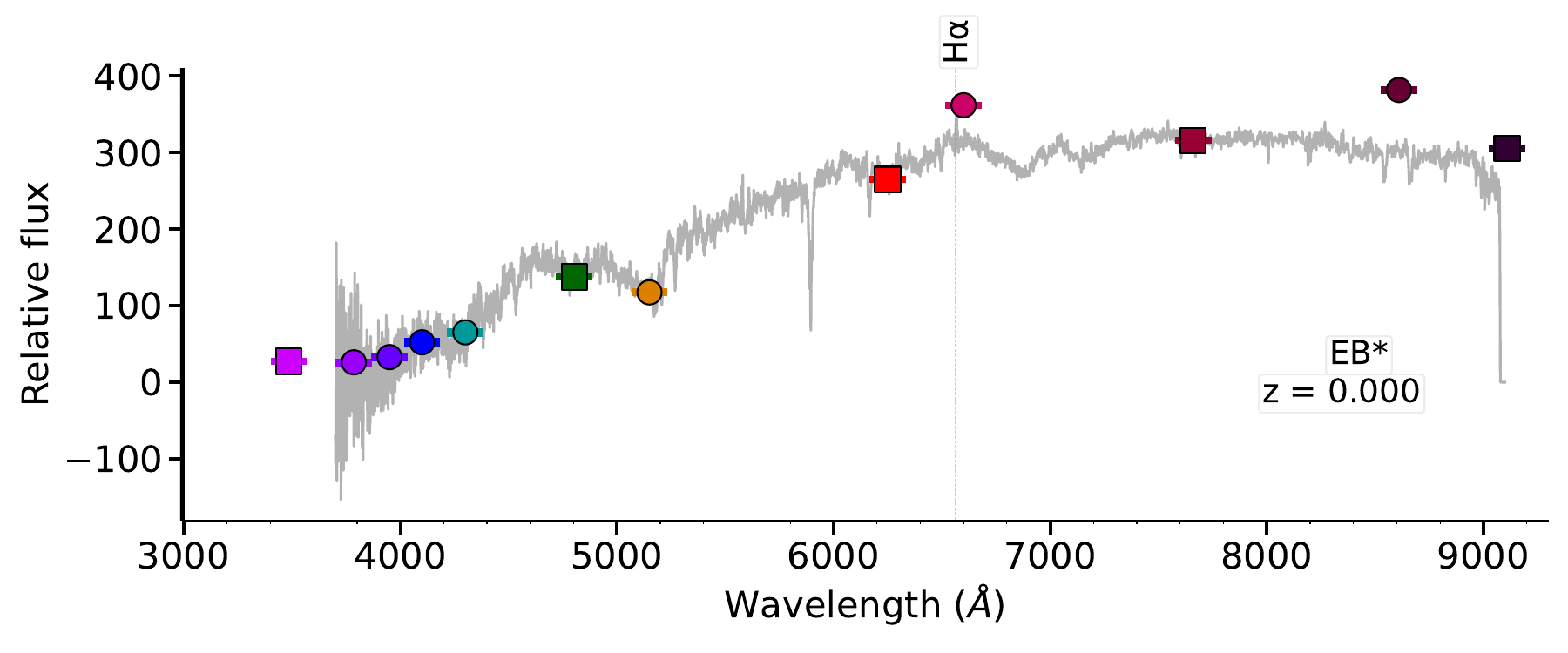}
    \caption{Spectra of three objects identified as H$\alpha$ excess sources using our methodology. The top panel displays the SDSS spectrum, while the middle and bottom panels show LAMOST spectra. The coloured symbols correspond to S-PLUS photometry in flux units for the following filters (from left to right): $u$, $J0378$, $J0395$, $J0410$, $J0430$, $g$, $J0515$, $r$, $J0660$, $i$, $J0861$, and $z$. Square symbols represent broad-band filters, while circle symbols denote narrow-band filters. According to SIMBAD, the objects in the top and middle panels--SDSS ID J113722.24+014858.5 and LAMOST ID J232551.47-014023.5--are classified as cataclysmic variables. The bottom panel shows an eclipsing binary star with weak H$\alpha$ emission (LAMOST ID: J012119.09-001950.0). The dashed line marks the position of the H$\alpha$ wavelength.}
\label{fig:cv}
\end{figure}

\begin{figure}
    \includegraphics[width=\linewidth, trim=10 40 5 8, clip]{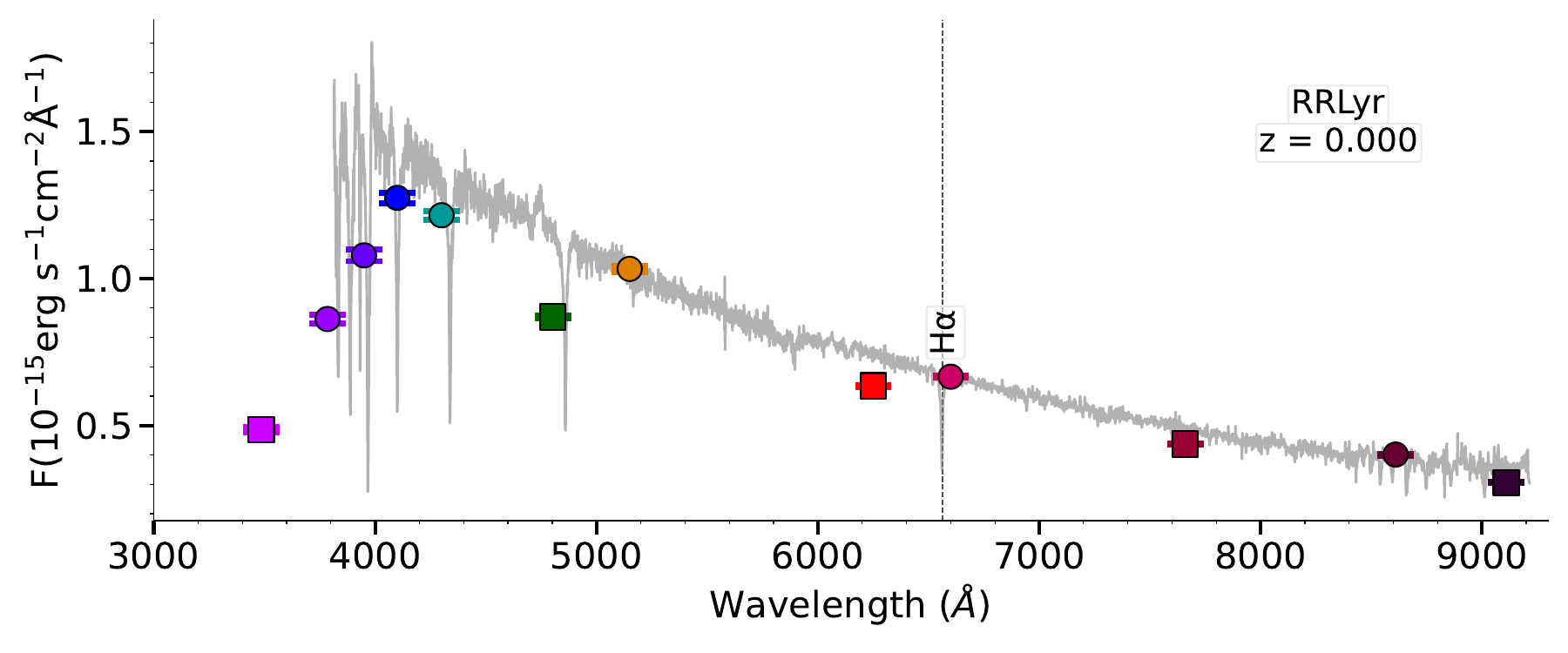} 
    \caption{SDSS spectrum and S-PLUS photometry of the RR Lyrae star SDSS J010045.13-010212.2, showing an H$\alpha$ absorption line.}
\label{fig:RRlyra}
\end{figure}

\begin{figure}
    \includegraphics[width=\linewidth, trim=10 40 5 8, clip]{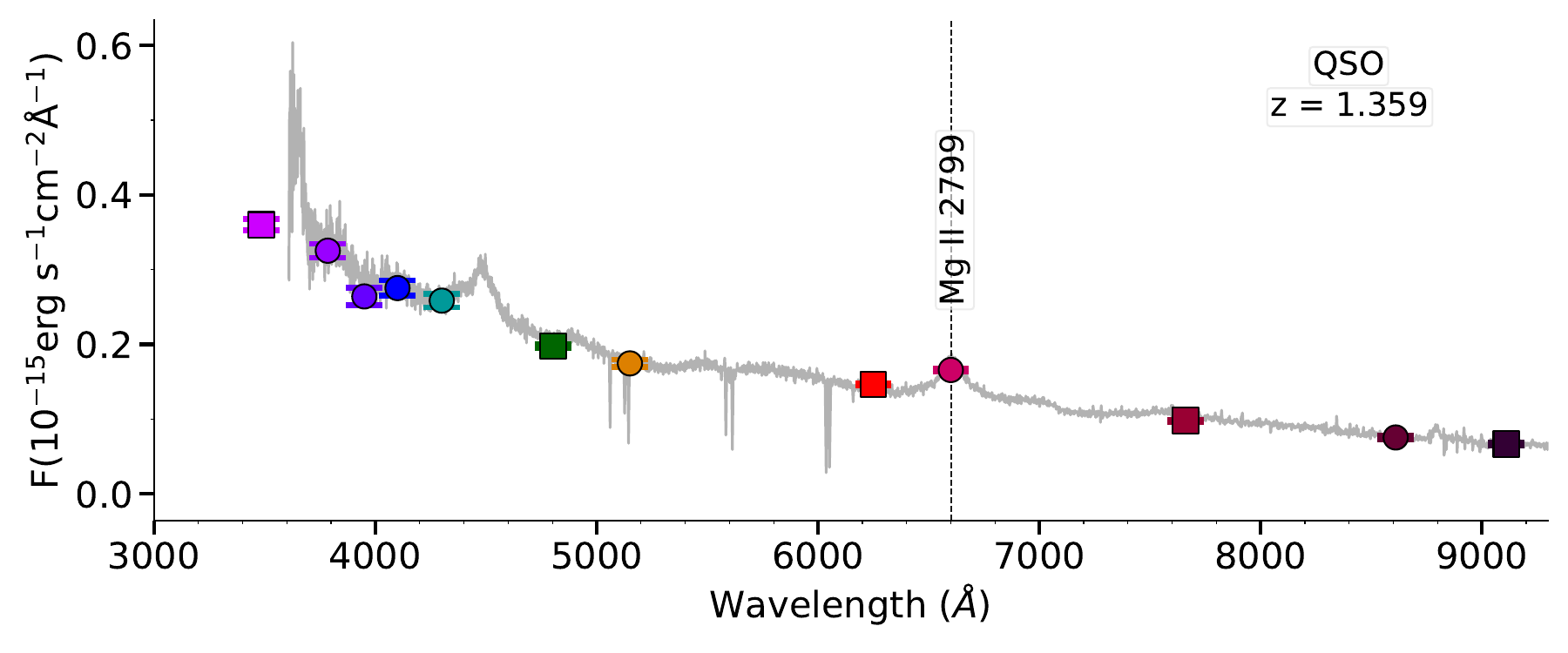} 
    \includegraphics[width=\linewidth, trim=10 40 5 8, clip]{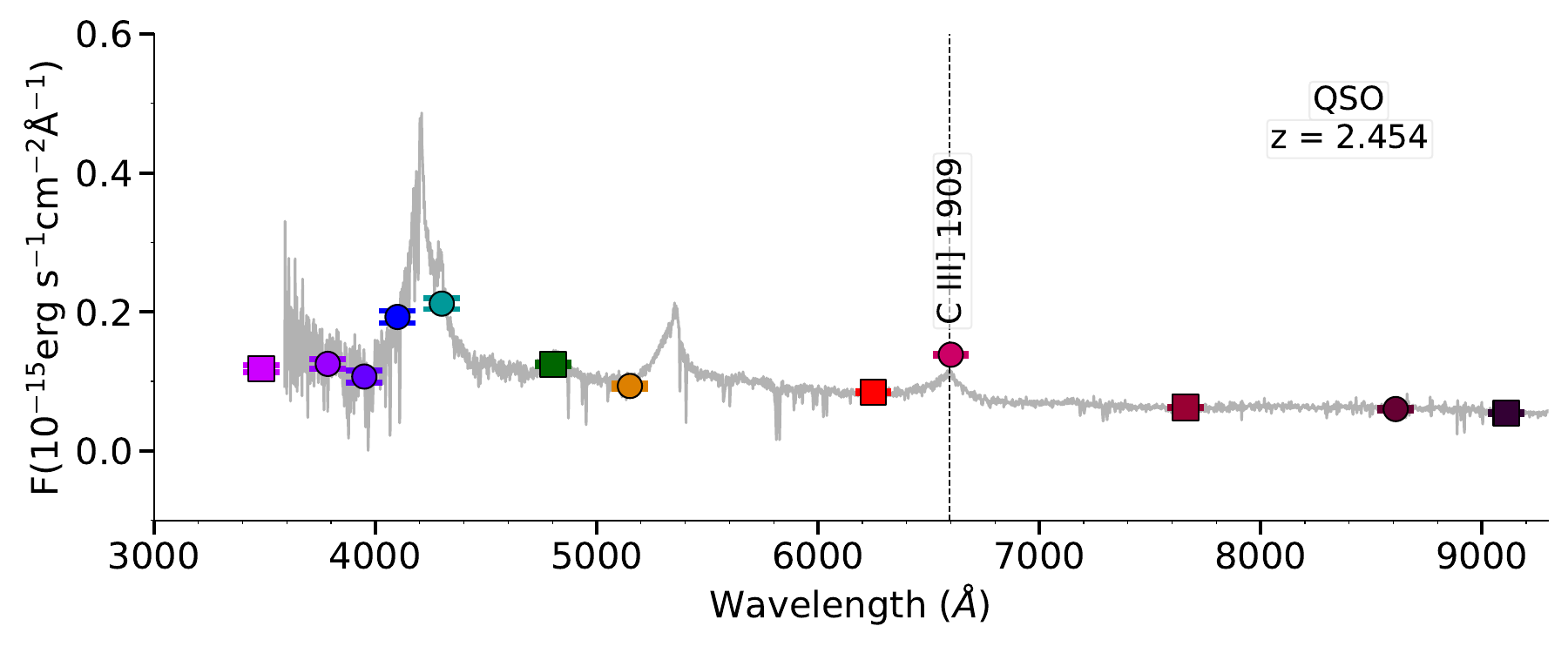} 
    \includegraphics[width=\linewidth, trim=10 10 5 8, clip]{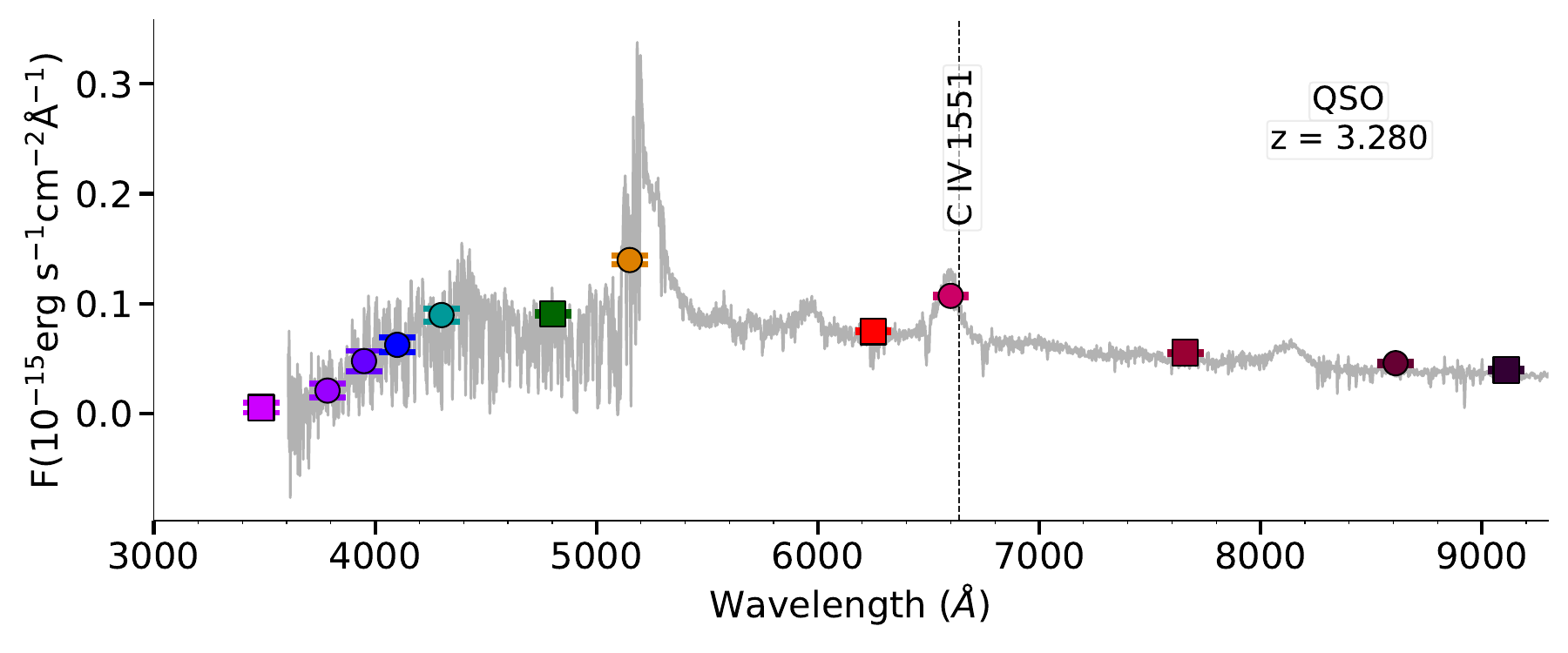}
    \caption{S-PLUS photometry and SDSS spectra of three QSOs with redshifts of 1.359, 2.454 and  3.280 (top to bottom) selected as H$\alpha$ excess sources. At these redshifts, the emission lines Mg II $\lambda$2799, C III] $\lambda$1909 and C IV $\lambda$1551 are detected in the $J0660$ filter. The SDSS IDs of the sources are: J235157.58+003610.5, J220529.34-003110.7, and J224539.94-002419.6.}
\label{fig:qso}
\end{figure}

\begin{figure}
 \includegraphics[width=\linewidth, trim=10 10 5 8, clip]{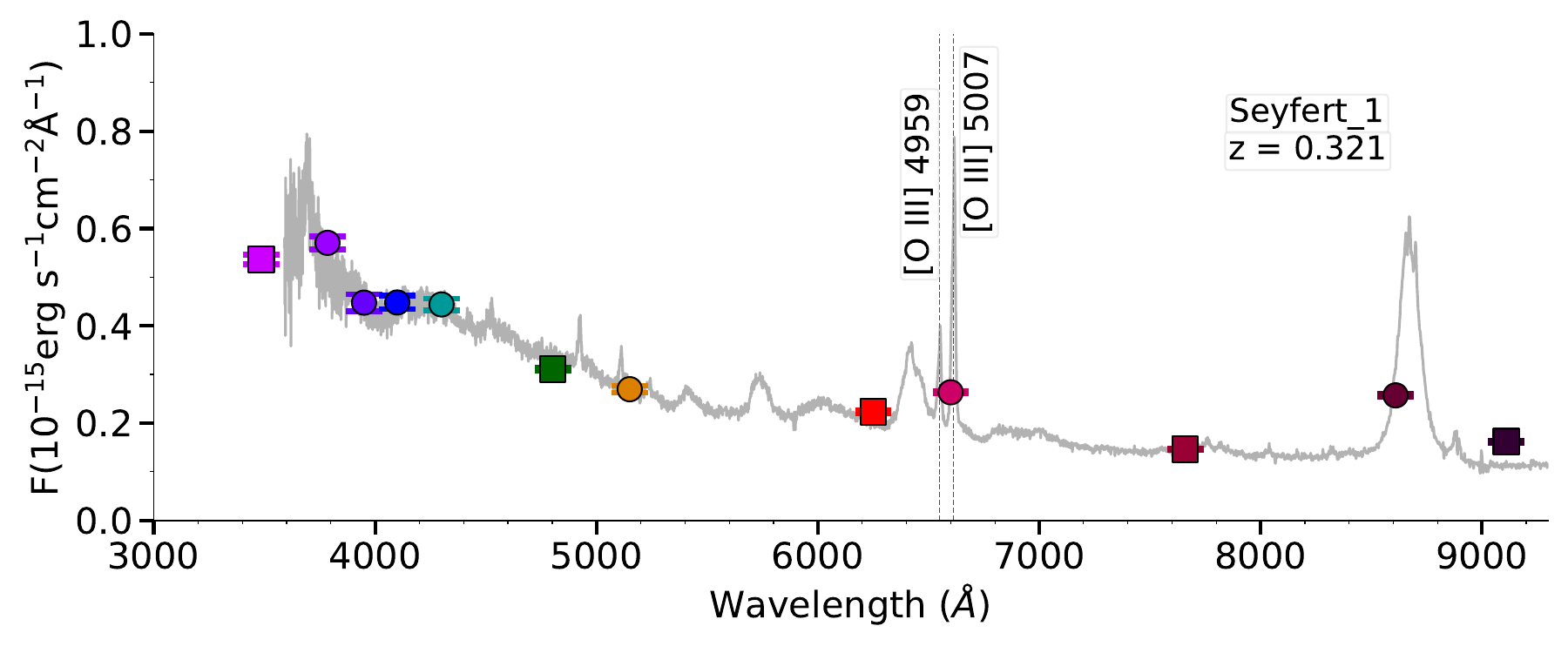}
    \includegraphics[width=\linewidth, trim=10 10 5 8, clip]{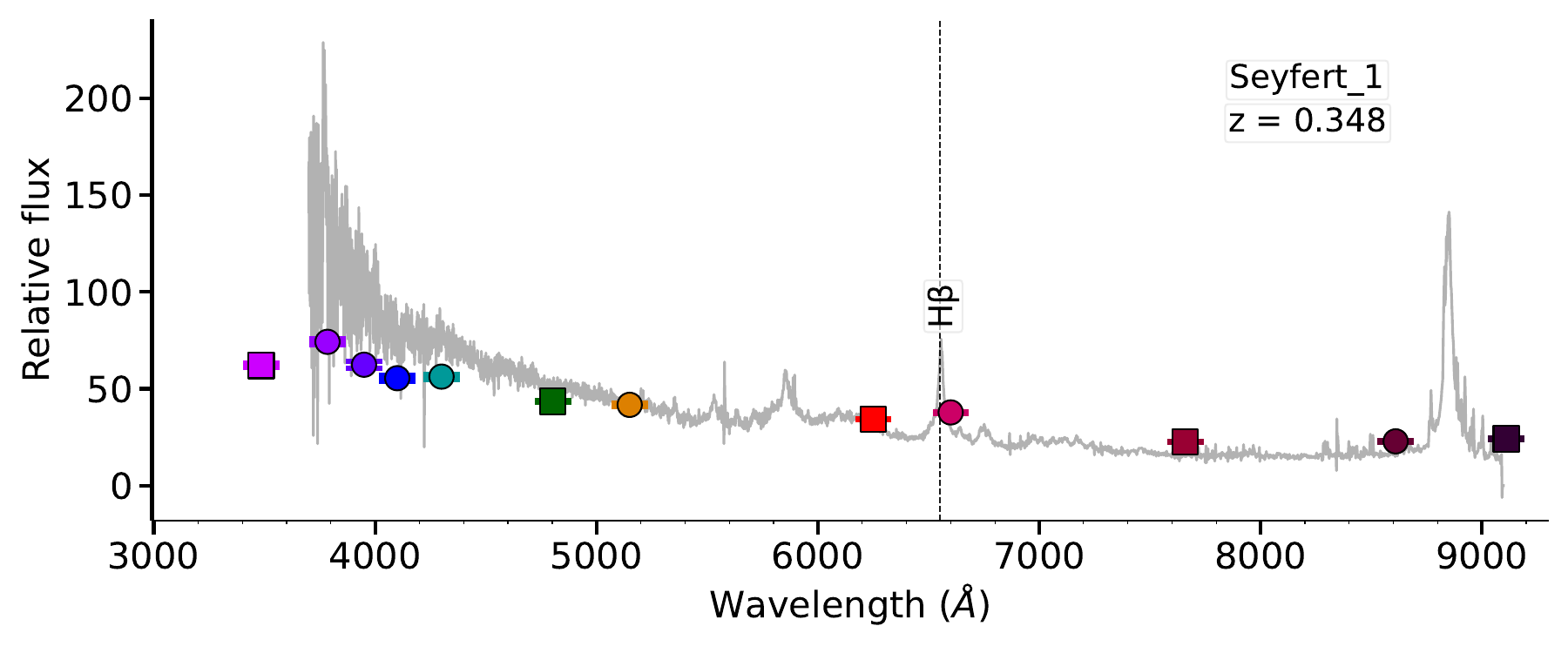}

    \caption{As Fig. \ref{fig:qso} but for sources with redshifts of \(z = 0.321\) and \(z = 0.348\). At these redshifts, the lines [O III] \(\lambda\lambda 4959, 5007\) doublet and H$\beta$ line are detected in the narrow band filter, generating an H$\alpha$ excess. The spectra of the sources are from SDSS (upper) and LAMOST (bottom) with IDs SDSS J231742.60+000535.1 and LAMOST J033429.44+000611.0, respectively.}
    \label{fig:seyfert}
\end{figure}

Our list of H$\alpha$-excess sources identified in the MS was cross-matched with the DR18 SDSS catalog \citep{Ahumada:2020} and the DR9 LAMOST catalogue, using a 2 arcsec radius. These cross-matching identified 212 common sources (138 from SDSS and 74 from LAMOST). The procedure was restricted to the MS due to its overlap with SDSS and LAMOST areas, unlike the S-PLUS Galactic disk survey. It is noteworthy that some H$\alpha$-excess sources detected by our algorithm may exhibit transient behaviour, meaning that H$\alpha$-excess features might be present in spectra from one survey (SDSS or LAMOST) but not in others (S-PLUS), or vice versa. This variability is attributed to differences in observational epochs and conditions across the surveys. Upon spectroscopic examination, approximately 60\% of these sources exhibited emission lines, which might include redshifted lines other than H$\alpha$, while about 30\% showed H$\alpha$-related absorption features.

Most of the objects with available spectroscopic information in SDSS and LAMOST correspond to CVs,
QSOs, AGN, and variable stars. A more detailed spectroscopic characterization of these sources is out the scope of this
paper. Also, it is worth noticing that there is a number of objects without a conclusive classification.

Figure~\ref{fig:cv} presents the SDSS (upper) and LAMOST (lower) spectra, along with the corresponding S-PLUS photometry (coloured symbols) for two known cataclysmic variables (CVs) and one eclipsing binary, respectively. The excess in the $J$0660 filter is evidently produced by the H$\alpha$ line. Note that the bluer emission tends to be more intense, which is consistent with the expected behaviour of CVs showing strong Balmer series emission. Bottom panel of Fig. \ref{fig:cv} displays the LAMOST spectrum and S-PLUS photometry of an eclipsing binary. The spectra exhibit weak H$\alpha$ emission, which is effectively captured by the narrow $J$0660 filter of S-PLUS. 

Figure~\ref{fig:RRlyra} shows the SDSS spectra and S-PLUS photometry of an RR Lyrae star with H$\alpha$ in absorption. The absorption feature in H$\alpha$ affects both the $r$-band and the $J0660$ filter. The apparent H$\alpha$ excess observed in the $(r - J0660)$ colour index for sources with H$\alpha$ absorption is due to the differential effect of the absorption feature on the broadband $r$ filter and the narrowband $J0660$ filter. The H$\alpha$ line lies within the $r$-band, so the absorption feature reduces the total flux detected, making the $r$ band appear fainter. In contrast, the $J$0660 filter, shows a less pronounced reduction in flux. This difference results in a more negative $(r - J0660)$ colour index, creating an apparent H$\alpha$ excess. This photometric effect is important for identifying H$\alpha$ excess sources, as it indicates the presence of H$\alpha$ variations, even in absorption, within various stellar objects \citep{Fratta:2021}. Furthermore, most of the H$\alpha$ absorption line objects in the MS (relatively high latitude) are RR Lyrae stars, as confirmed by SIMBAD, which lists 111 RR Lyrae stars in our sample. We also explore the distribution of RR Lyrae stars in the $(r - J0660)$ versus $(r - i)$ colour diagram, finding that these variable stars span $(r - J0660)$ values between -0.4 and 0.6. This means that a population of these stars have $(r - J0660) > 0$, indicating their selection. For more details, see Fig. \ref{fig:Diagram-RRlyra} in the Appendix \ref{sec:Appenx-A}.

Figure~\ref{fig:qso} presents examples of SDSS spectra for three QSOs, where the $J$0660 filter captures emission from different redshifted lines. For the QSO in the upper panel (redshift $\sim$1.36), the excess corresponds to the Mg II 2798 \AA\ line. In the middle panel (redshift $\sim$2.45), the excess is due to the C III] 1909 \AA\ line. Finally, for the QSO in the bottom panel (redshift $\sim$3.28), the C IV 1550 \AA\ line produces the observed excess. These plots demonstrate how the $J$0660 filter captures redshifted emission lines for QSOs at various redshifts.

Other extragalactic objects for which we found spectra in SDSS and LAMOST include AGNs. For example, Figure~\ref{fig:seyfert} displays the spectra of two nearby AGNs with redshifts of approximately $z \simeq 0.35$ (top) and $z \simeq 0.32$ (bottom). In the first, the H$\beta$ emission line falls within our narrowband filter. For the second source, with $z \simeq 0.32$, the doublet  [O III] 4959, 5007 \AA\ emission lines lie in the $J$0660 filter, resulting in an observed excess.

The analysis of individual spectra reveals distinct H$\alpha$ line features, including both emission and absorption at expected wavelengths, offering valuable insights into the physical characteristics and evolutionary stages of the objects. The spectral confirmation rates we present provide a conservative estimate of the selection purity. This is because our algorithm targets H$\alpha$-excess sources, not strictly H$\alpha$ emitters. Thus, objects with excess in the $J$0660 filter are selected as outliers, even if they lack a prominent H$\alpha$ emission line. By referring to "H$\alpha$-excess" rather than "H$\alpha$-emitters," we highlight that our selection is based on photometric excess in the $J$0660 filter, rather than solely on strong H$\alpha$ emission.

\subsection{Evaluation of Photometric Colour Consistency Between S-PLUS and VPHAS+}

\begin{figure*}
\centering
\begin{tabular}{l l}
    \includegraphics[width=0.5\linewidth, trim=10 10 5 8, clip]{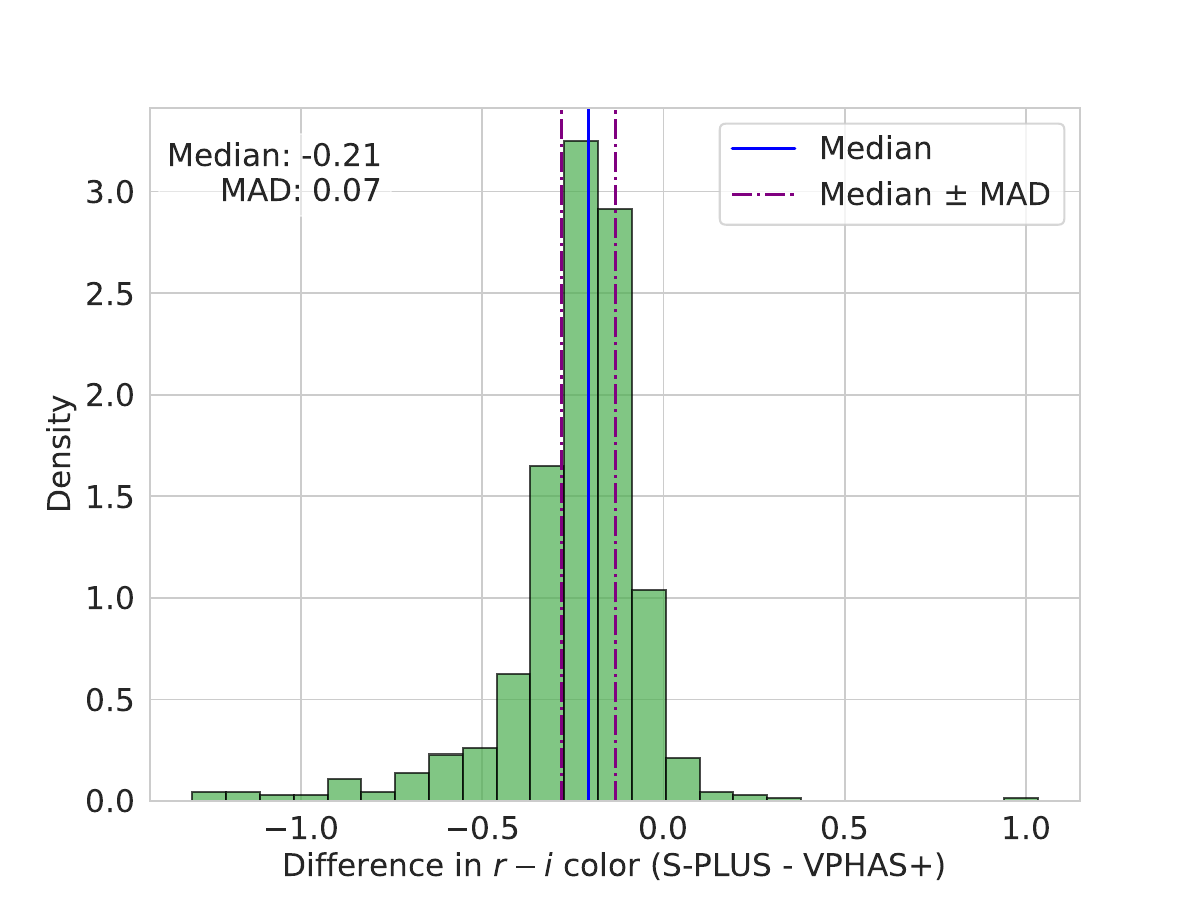} 
    \includegraphics[width=0.5\linewidth, trim=10 10 5 8, clip]{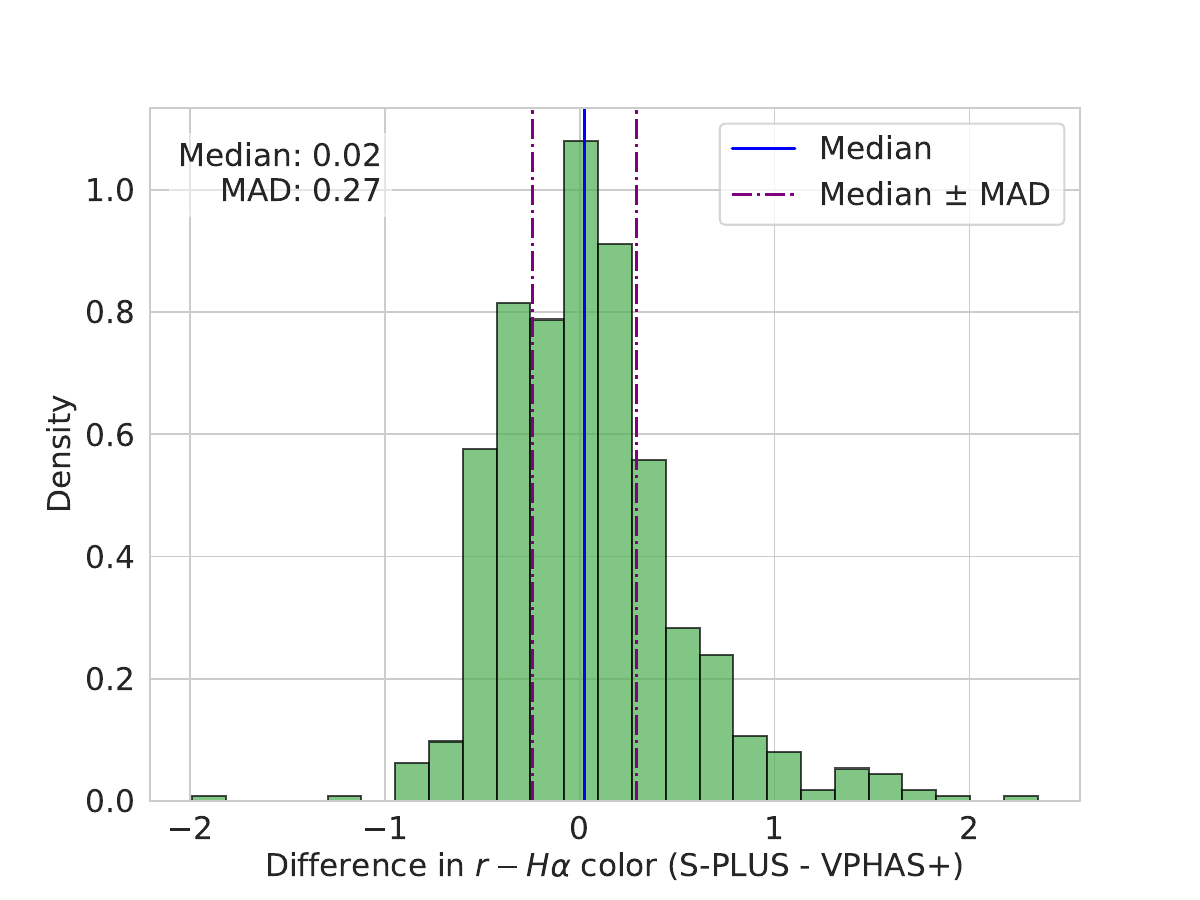}
    \end{tabular}  
    \caption{Histograms illustrating the discrepancies in the photometric colours $r - i$ and $r$ - H$\alpha$ between the S-PLUS and VPHAS+ surveys. The left panel depicts the differences in the $r - i$ colour, while the right panel shows the differences in the $r$ - H$\alpha$ colour. Both histograms reveal significant differences for the stars included in both surveys. The median value and median absolute deviation (MAD) for each colour discrepancy are provided, offering insights into the agreement between the two datasets.}

    \label{fig:VPHAS-SPLUS}
\end{figure*}

We performed a comparative analysis of PSF photometric colours between the S-PLUS data from the GDS and those provided by VPHAS+ DR2\footnote{More detailed information about the VPHAS+ survey can be found at: \url{https://www.vphasplus.org/}}. For the crossmatching, we considered a radius of 1" and ended up with a number of 793 matches. We computed the differences in two key colour indices: $r - i$ and $r - \text{H}\alpha$. Specifically, we investigated the median difference and the median absolute deviation (MAD) of these colours to assess the consistency and agreement between the two surveys. It is worth noting that VPHAS+, like S-PLUS, employs the $r$, $i$, and a narrowband filter (NB-659) designed to detect the H$\alpha$ line, facilitating a meaningful comparison of H$\alpha$ emission.

The comparison of colours reveals important insights into the consistency and reliability of S-PLUS photometry (see Fig.~\ref{fig:VPHAS-SPLUS}). The median difference in the $r - i$ colour between S-PLUS and VPHAS+ was $-0.21$, with a MAD of $0.07$. For the $r - \text{H}\alpha$ colour, the median difference was $0.02$ with a MAD of $0.27$. These results indicate a systematic offset between the photometric colours of the two surveys, which is within the expected range considering differences in instrumentation and filter systems.

A key factor contributing to the differences in the $r - \text{H}\alpha$ colour index is the distinct characteristics of the H$\alpha$ filters used in S-PLUS and VPHAS+. The S-PLUS H$\alpha$ filter ($J$0660) has an effective wavelength of 6614 \AA\ and a width of 147 \AA\, whereas the VPHAS+ NB-659 filter has an effective wavelength of 6588 \AA\ and a width of 107 \AA. These differences can significantly affect the measurement of H$\alpha$ excess, as the narrower VPHAS+ filter captures a more restricted range of wavelengths, potentially leading to higher precision. The broader S-PLUS filter, on the other hand, may include additional continuum emission, affecting the photometric measurement. Additionally, the exposure times in the two surveys differ, with VPHAS+ using a 120-second exposure and S-PLUS using a 290-second exposure. The longer exposure time in S-PLUS allows for greater sensitivity to faint sources and potentially higher signal-to-noise ratios (SNR), contributing to the observed differences in photometric colours. 

Despite the observed systematic differences, the MAD values suggest that the photometric measurements from both surveys exhibit good agreement. This consistency is crucial for cross-referencing and integrating datasets from different surveys for comprehensive astrophysical studies. The observed differences in photometric colours may result from various factors, including differences in filter characteristics, photometric calibration, and data processing techniques. Further investigations are warranted to better understand these factors' contributions to the observed discrepancies.

\subsection{H$\alpha$ Excess Source Distributions}

\begin{figure}
    \includegraphics[width=\linewidth]{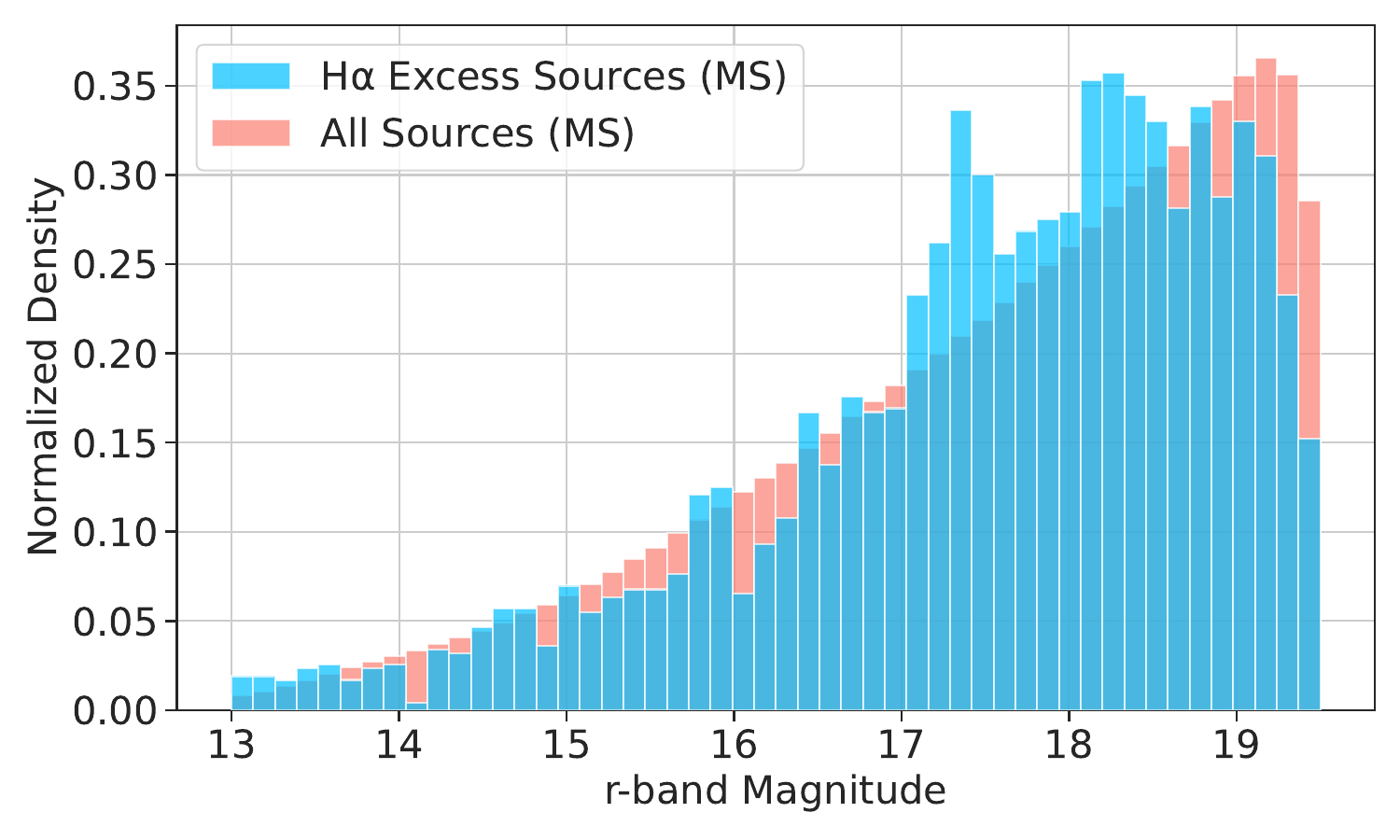}
    \includegraphics[width=\linewidth]{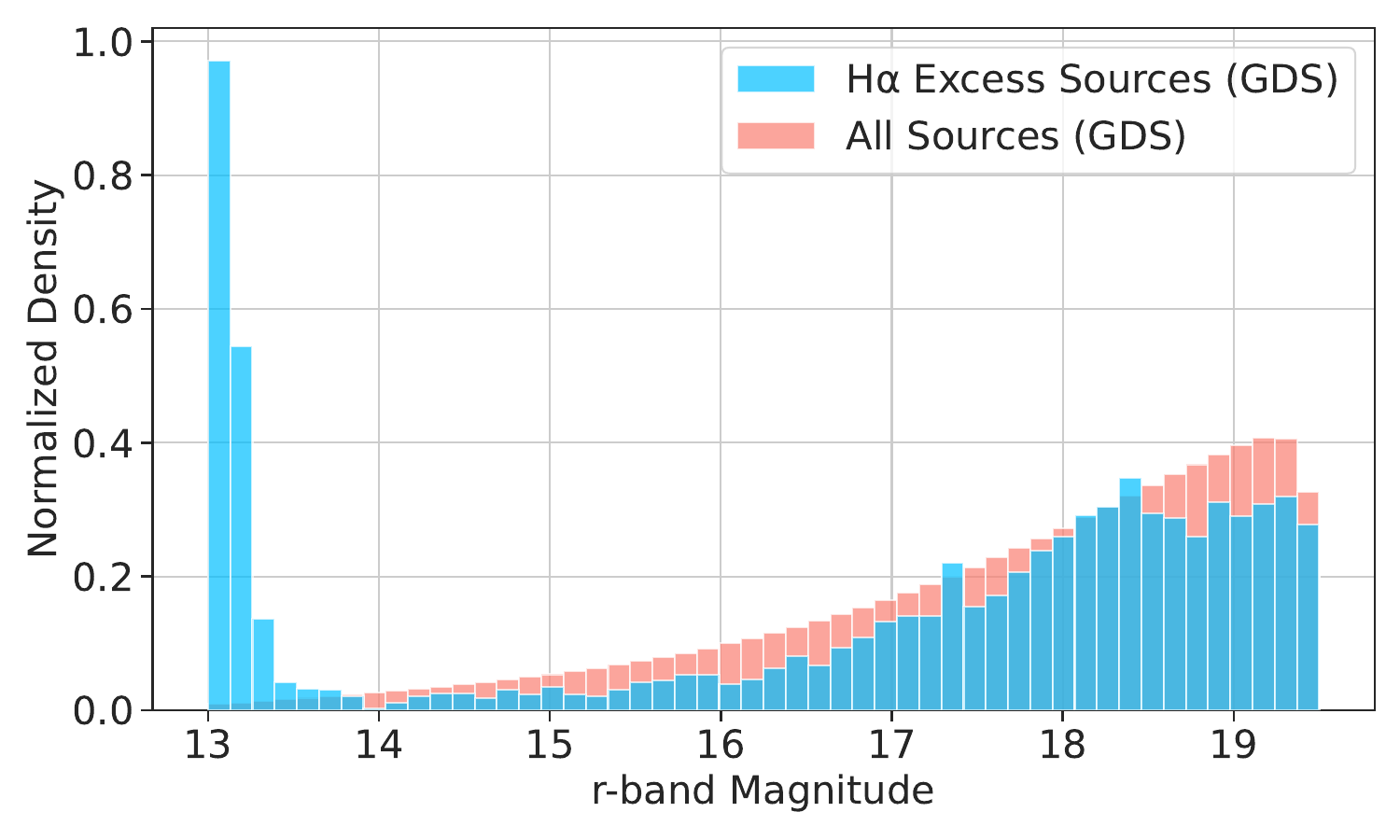}
    \caption{\textit{Upper panel}: Distribution of $r$-band magnitudes for H$\alpha$ excess sources (blue curve) compared to all the stars (salmon curve) in the MS. \textit{Lower panel}: Distribution of $r$-band magnitudes for H$\alpha$ excess sources (blue curve) in the GDS compared to all stars (salmon curve).}
    \label{fig:r-distribution}
\end{figure}

The upper panel of Fig.~\ref{fig:r-distribution} presents a histogram of the $r$-band magnitude distribution for all objects in our study from the MS. The normalized density facilitates comparison between different subsets. The blue curve represents H$\alpha$ excess objects, while the salmon curve represents all stars from the MS. The magnitude distribution for H$\alpha$ excess sources shows a higher concentration at intermediate magnitudes. The lower panel of Fig.~\ref{fig:r-distribution} focuses on the $r$-band magnitude distribution sources for the subset of H$\alpha$ excess objects in the disk. A noticeable large number of sources with H$\alpha$ excess have magnitudes in the $r$-band
between 13 and 13.5, something that we do not see in the stars of GDS. This implies that H$\alpha$ excess objects could be intrinsically more luminous or closer to us than the general population of all stars. However, these stars are closer to the saturation limit. Therefore, we recommend exercising caution with all sources in our sample that have an $r$-band magnitude less than 13.5. 

\begin{figure}
    \includegraphics[width=\linewidth]{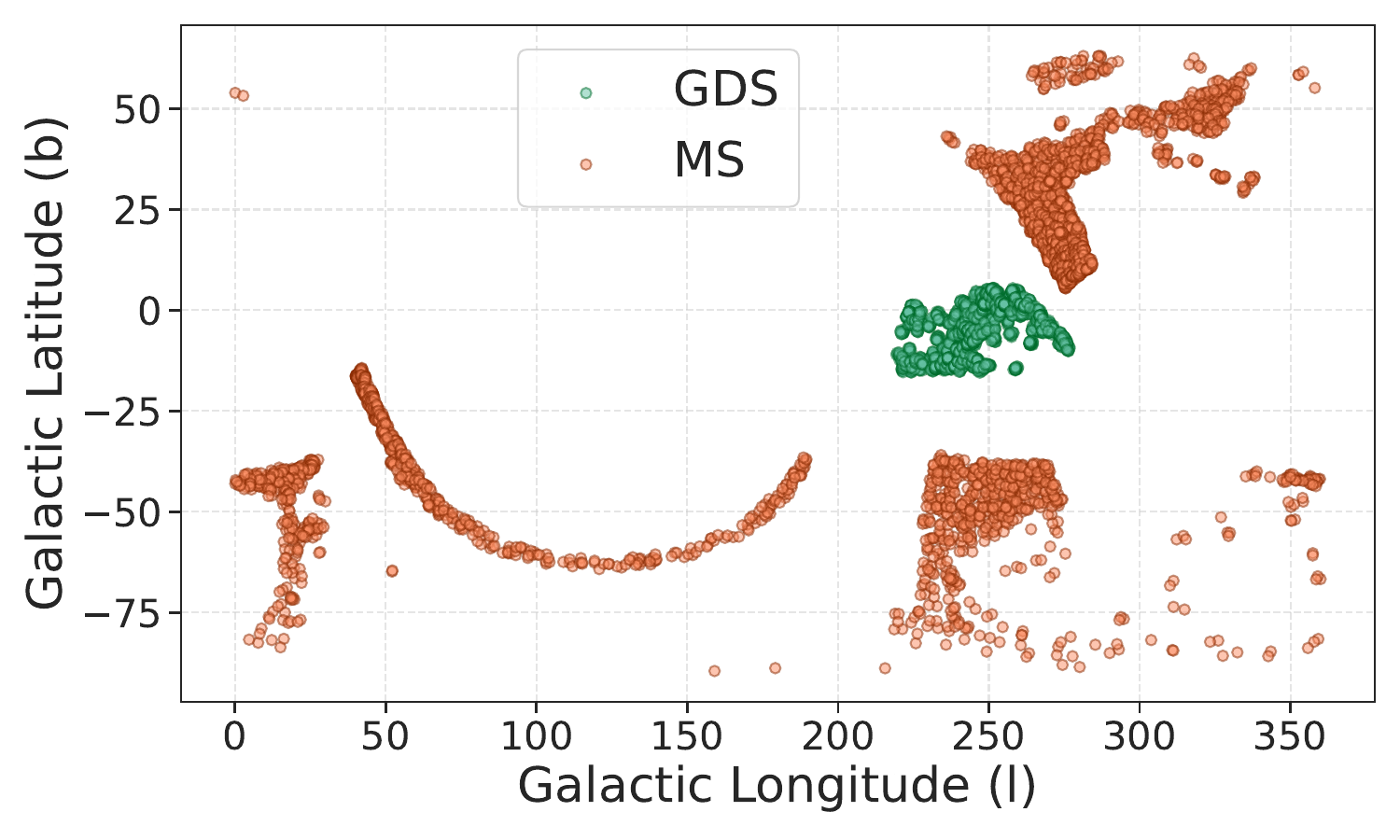}
    \includegraphics[width=\linewidth]{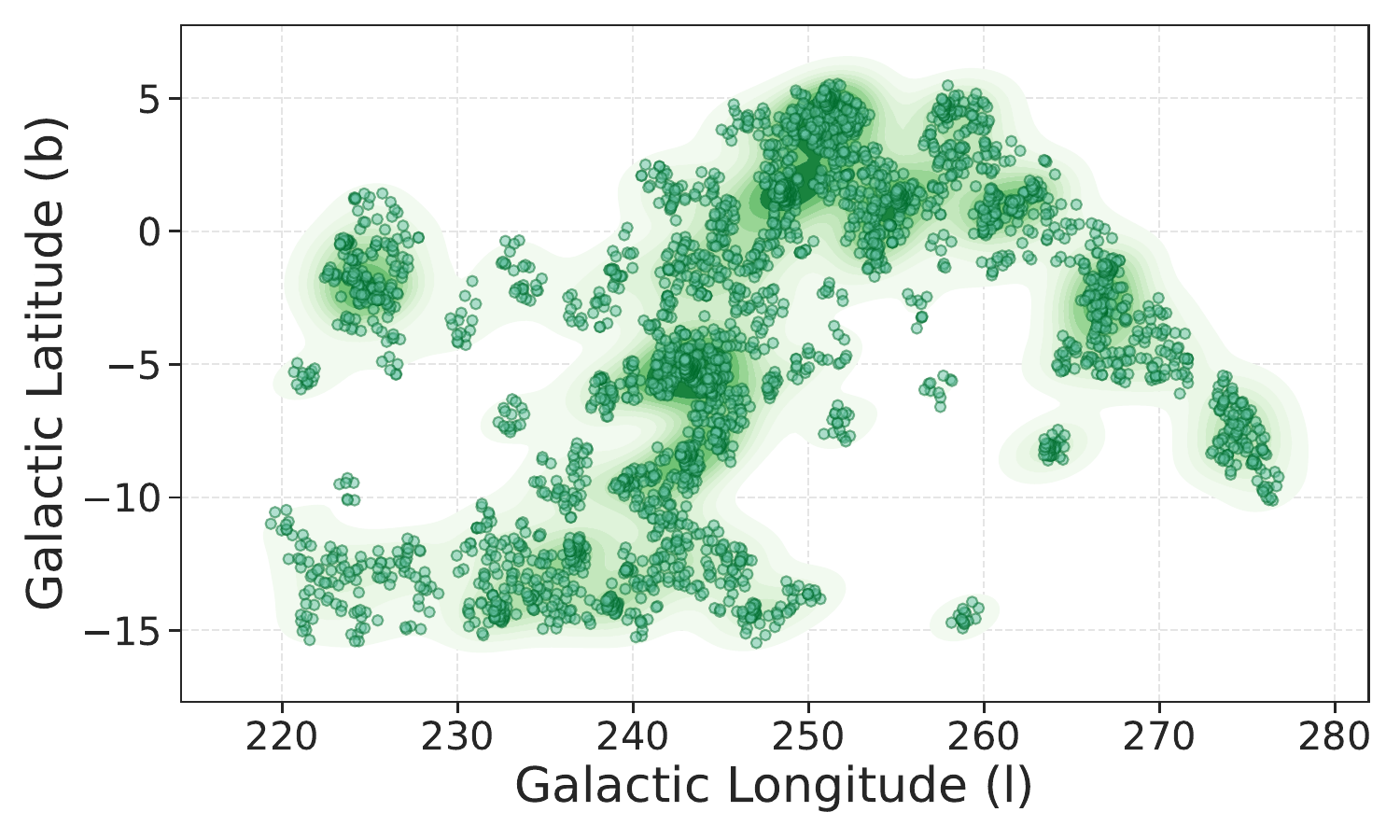}
    \caption{Distribution of H$\alpha$ excess objects in Galactic longitude and latitude coordinates. The \textit{upper panel} shows all the H$\alpha$ sources selected, and the \textit{lower panel} is a zoomed-in view of the GDS.}
    \label{fig:Halpha-excess-galactic}
\end{figure}

\begin{figure*}
\centering
\begin{tabular}{l l}
    \includegraphics[width=0.5\linewidth, trim=10 10 5 8, clip]{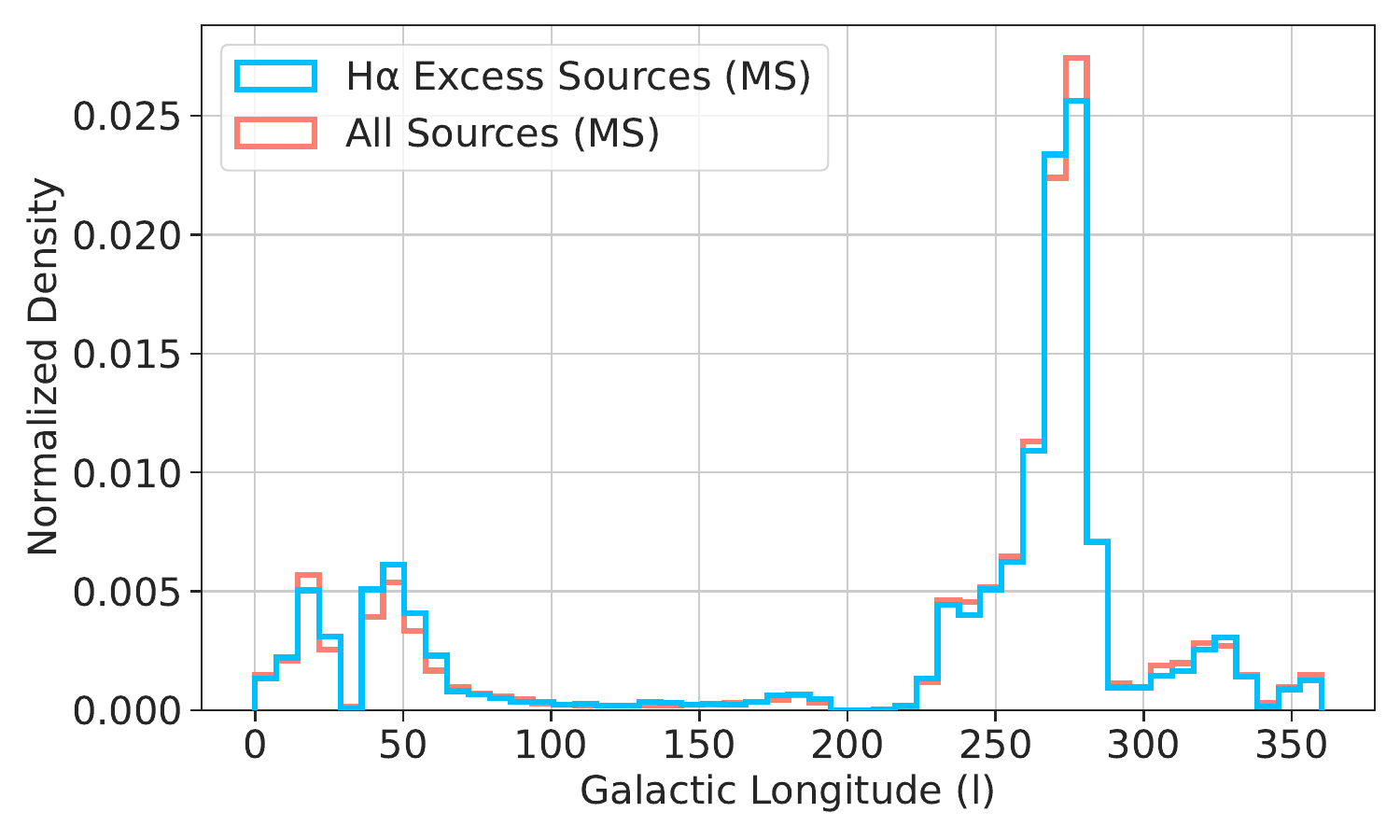}
    \includegraphics[width=0.5\linewidth, trim=10 10 5 8, clip]{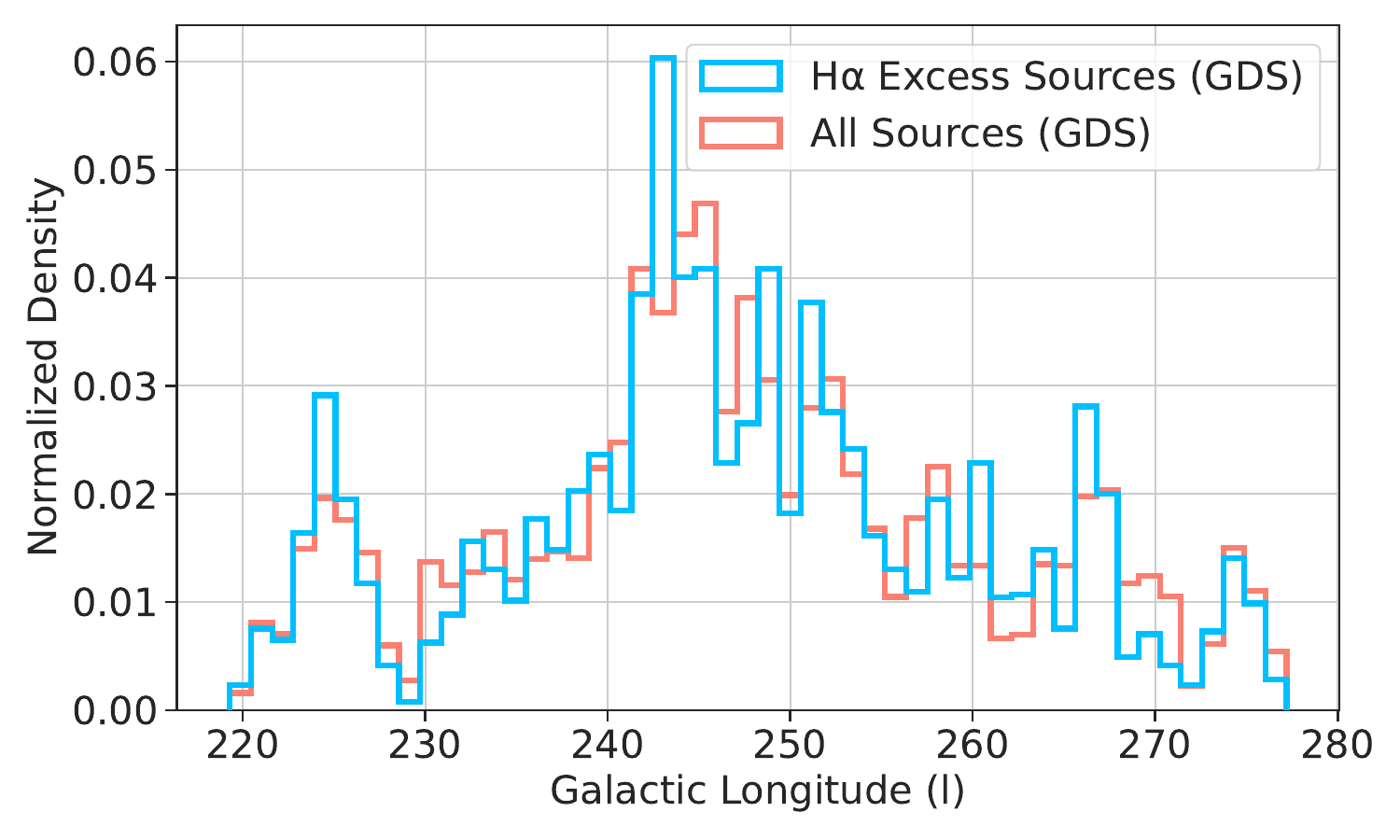} 
    \end{tabular}  
    \caption{Distribution of the objects in galactic longitude for H$\alpha$ excess sources (blue bars) and all stars (salmon bars) for the MS (\textit{left panel}) and the GDS (\textit{right panel}).}

    \label{fig:longitude}
\end{figure*}

Figure \ref{fig:Halpha-excess-galactic} shows the distribution of all H$\alpha$ excess sources in Galactic latitude and longitude, along with a zoomed-in view of the GDS in the bottom panel. The distribution of objects in Galactic longitude for the MS (left panel of Fig. \ref{fig:longitude}) indicates that the blue bars, representing H$\alpha$ excess sources, are relatively evenly spread across the Galactic longitude, similar to the general population of stars from the MS (pink bars). Peaks are observed around Galactic longitudes of 15$^{\circ}$, 50$^{\circ}$, and 270$^{\circ}$, which are also present in the general star population of the MS.

The bottom panel of Fig. \ref{fig:Halpha-excess-galactic} and the right panel of Fig. \ref{fig:longitude} show the distribution of objects in Galactic longitude specifically within the Galactic disk. There is a noticeable concentration of H$\alpha$ excess sources at specific longitudes, particularly around 243$^{\circ}$. Additionally, there are small peaks around 225$^{\circ}$ and 268$^{\circ}$ in Galactic longitude. While H$\alpha$ excess sources follow a distribution similar to that of all stars, the peaks are more pronounced for H$\alpha$ excess sources.

It should be noted that the observed concentrations of H$\alpha$ excess sources in certain Galactic regions may be influenced by the uneven sky coverage of the S-PLUS survey (see Fig. 6 in \citealp{Herpich:2024}). In the MS, this effect may be more pronounced due to the lower star formation activity in high-latitude regions. For the GDS, the concentrations could be influenced by both the presence of star-forming regions and the patchy sky coverage of the survey. This limitation may lead to uneven sampling of both high-latitude and Galactic plane regions, potentially over- or under-representing the observed numbers in specific areas. Therefore, these peaks should be interpreted with this limitation in mind.

\section{Machine Learning Approaches}
\label{sec:ML}
In this section, inspired by the goal of separating Galactic sources from extragalactic ones in our H$\alpha$ excess list, we applied machine learning approaches. Our list of H$\alpha$ excess sources selected in the MS of S-PLUS naturally includes extragalactic compact objects with redshifted lines detected in the $J0660$ filter. To classify the sources in our H$\alpha$ excess list, we utilized the multi-band coverage provided by S-PLUS optical photometry. To achieve this, we employed two unsupervised machine learning algorithms: UMAP and HDBSCAN. UMAP is used to reduce the dimensions of our data and perform a feature extraction, while HDBSCAN classifies the data based on the results from UMAP. We conducted two experiments: one using the 66 colours generated from the 12 S-PLUS filters, and a second one by adding filters from the Wide-Field Infrared Survey Explorer \citep[WISE][]{Wright:2010}. This classification helps identify specific types of objects for subsequent spectroscopic follow-up. Additionally, we used a Random Forest algorithm to identify important features and construct colour-colour diagrams to separate the classes of objects identified by HDBSCAN. This methodology is applied to the list of H{$\alpha$} excess sources obtained from the MS of S-PLUS. The classification results also provide the basis for defining tentative colour criteria, which can be used to refine the separation between different classes of H$\alpha$ excess sources, based on the new colour-colour diagrams proposed here.

\subsection{Dimensionality Reduction and Clustering}
\label{sec:dimen}

\subsubsection{UMAP}
\label{sec:umap}

Uniform Manifold Approximation and Projection (UMAP; \citealp{Becht:2018, mcinnes:2020}) is a dimensionality reduction algorithm designed to handle high-dimensional data while preserving its underlying structure. Unlike some other techniques, UMAP is based on a mathematical framework that combines aspects of Riemannian geometry and algebraic topology. This enables UMAP to capture both local and global relationships within the data. UMAP aims to create a low-dimensional representation that retains the intricate nonlinear relationships present in the original high-dimensional features. This process involves constructing a high-dimensional graph representation of the data and then optimizing a low-dimensional graph to match it. By doing so, UMAP effectively preserves the essential information and structure encoded in the data. This makes UMAP particularly well-suited for datasets where parameters exhibit complex nonlinear behaviour. In our analysis, we use UMAP to reduce the dimensionality of our input space, consisting of 66 colours and additional WISE bands, while retaining essential information encoded in the data

For the implementation of the algorithm, we used the Python package \texttt{umap}\footnote{For more details, see \url{https://umap-learn.readthedocs.io/en/latest/index.html}}. UMAP has three key hyperparameters: \texttt{n\_neighbors}, \texttt{n\_components}, and \texttt{min\_dist}.

The \texttt{n\_neighbors} parameter balances local versus global structures in the data by setting the number of neighbouring points UMAP considers for each data point when learning the manifold structure. Low values of \texttt{n\_neighbors} cause UMAP to focus on very local structures, while higher values make UMAP look at larger neighbourhoods, potentially losing fine details in favour of capturing broader patterns.

The \texttt{n\_components} parameter, similar to the parameter used in standard dimension reduction algorithms in the \texttt{scikit-learn} package \citep{scikit-learn}, allows us to set the number of dimensions in the reduced space into which we will embed the data. \texttt{scikit-learn} is a widely used Python library for machine learning, built on top of SciPy, and distributed under the 3-Clause BSD license. It provides implementations for many state-of-the-art machine learning techniques, making it a versatile tool for data analysis and modelling. 

The \texttt{min\_dist} parameter controls how closely UMAP can pack points together in the low-dimensional representation. Lower values result in clumpier embeddings, which are useful for clustering and capturing fine topological structures, while higher values focus on preserving broader topological structures.

\subsubsection{HDBSCAN}
\label{sec:hdbscan}

After obtaining a new system of reduced variables that condenses all the information from the original variables, we utilized HDBSCAN to identify clusters within the data. This clustering approach complements the reduction achieved by UMAP, allowing for a comprehensive understanding of the underlying structure of the dataset.

\begin{figure*}
        \includegraphics[width=\linewidth]{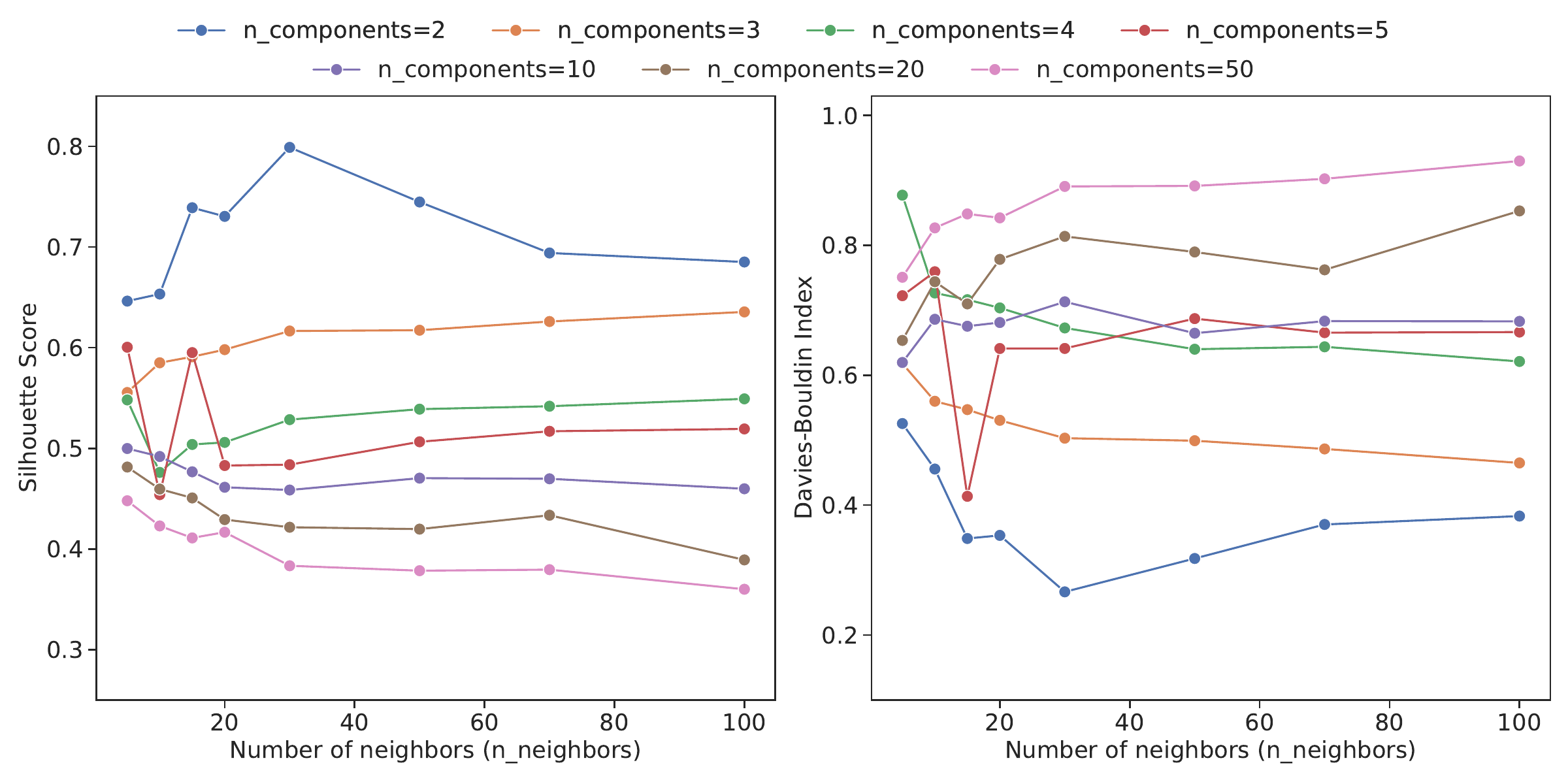}
       \caption{Silhouette Score (left panel) and Davies-Bouldin Index (right panel) as functions of the number of neighbours (\textit{n\_neighbors}) for different values of UMAP components (\textit{n\_components}). Higher Silhouette Score values and lower Davies-Bouldin Index values indicate better clustering performance.}
        \label{fig:Silhouette-Score-Davies-Bouldin}
    \end{figure*}

Hierarchical Density-Based Spatial Clustering of Applications with Noise (HDBSCAN; \citealp{Campello:2013}) is an unsupervised machine learning algorithm for clustering. It builds on the density-based spatial clustering of applications with noise (DBSCAN; \citealp{Ester:1996}) by introducing a hierarchy to the clustering process, which allows for the extraction of "persistent" clusters from the hierarchical tree. HDBSCAN's main advantage over DBSCAN is its ability to find clusters of varying densities and shapes.

For this task, we adopted the Python implementation of \texttt{HDBSCAN}\footnote{\url{https://hdbscan.readthedocs.io/en/latest/}} \citep{McInnes:2017}. The two most critical parameters are the "minimum cluster size" (\texttt{min\_cluster\_size}) and "minimum number of samples" (\texttt{min\_samples}). The "minimum cluster size" refers to the smallest group size that is considered a cluster. The "minimum number of samples" determines how conservative the clustering will be; larger values result in more points being classified as noise, restricting clusters to denser areas. 

HDBSCAN can also classify sources as noise if they do not fit well into any cluster based on these parameters. Additionally, the algorithm relies on a distance metric, such as Euclidean distance, to measure the distance between points and determine their density. The choice of metric can significantly affect the clustering results, as it influences how distances are computed and, consequently, how clusters are formed.

    \subsection{Classification Results}
    \label{sec:results-umap-hd}

    \begin{figure*}
    \centering
    \begin{tabular}{l l}
        \includegraphics[width=0.5\linewidth, trim=10 10 5 8, clip]{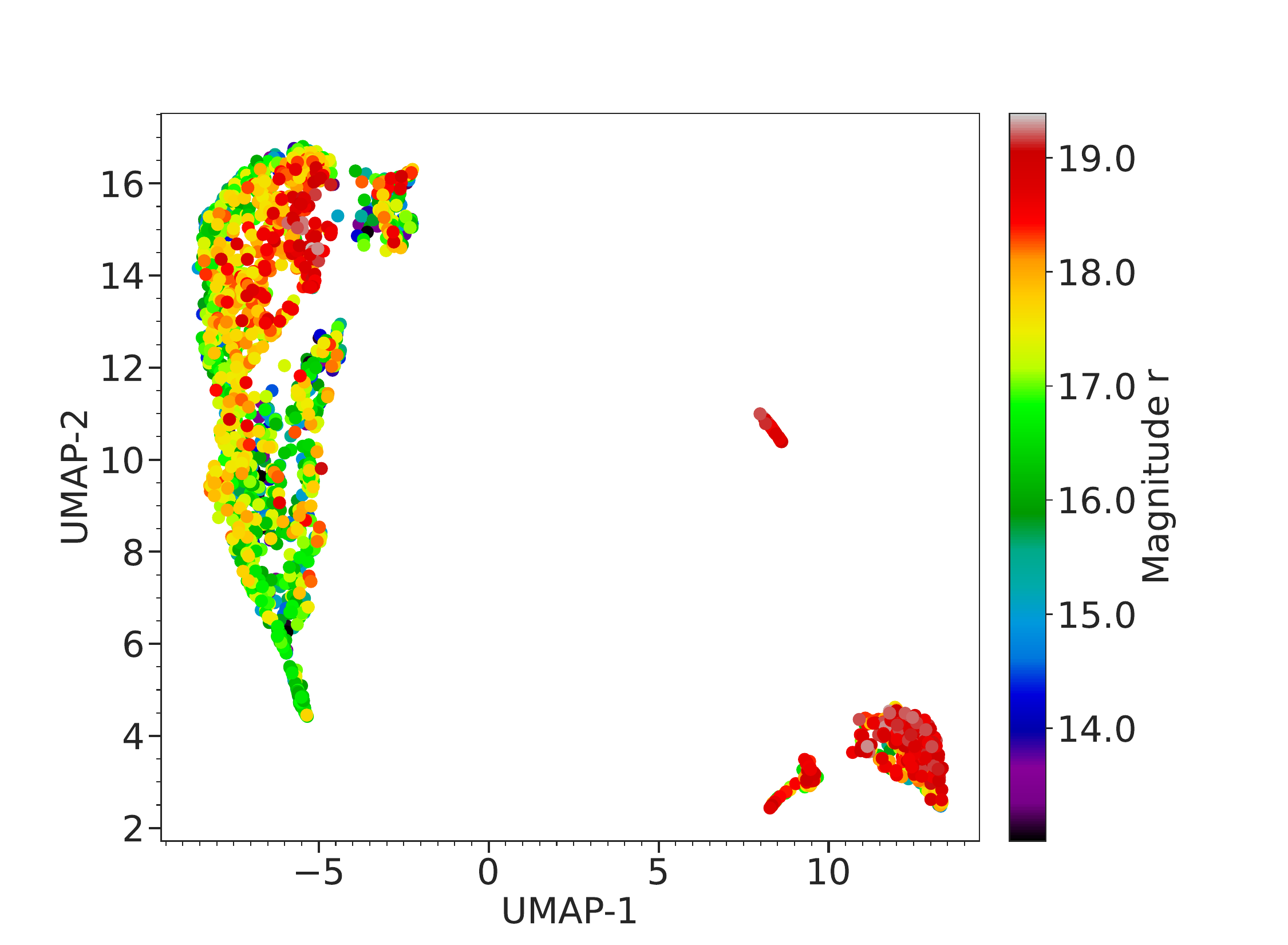} 
         \includegraphics[width=0.45\linewidth, trim=10 10 5 8, clip]{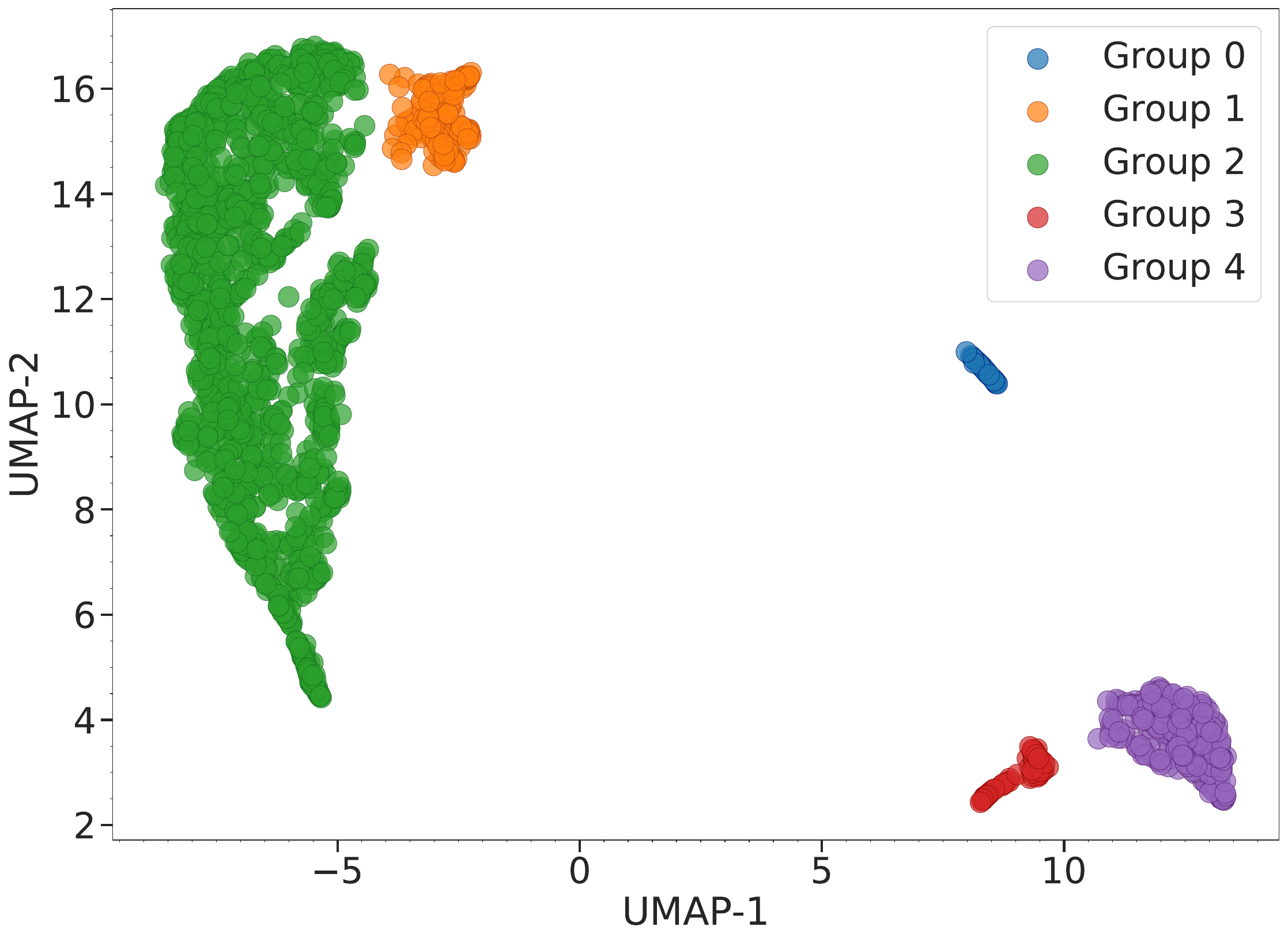} 
        \end{tabular}  
        \caption{UMAP dimension reduction applied to the MS from S-PLUS data.
The left panel shows the UMAP result using only the S-PLUS
colours as input parameters, with the colour bar indicating the $r$ magnitude. The right panel displays the result after applying HDBSCAN clustering, revealing five distinct groups.}
        \label{fig:umap}
    \end{figure*}
    
    \begin{figure*}
    \centering
    \begin{tabular}{l l}
        \includegraphics[width=0.5\linewidth, trim=10 10 5 8, clip]{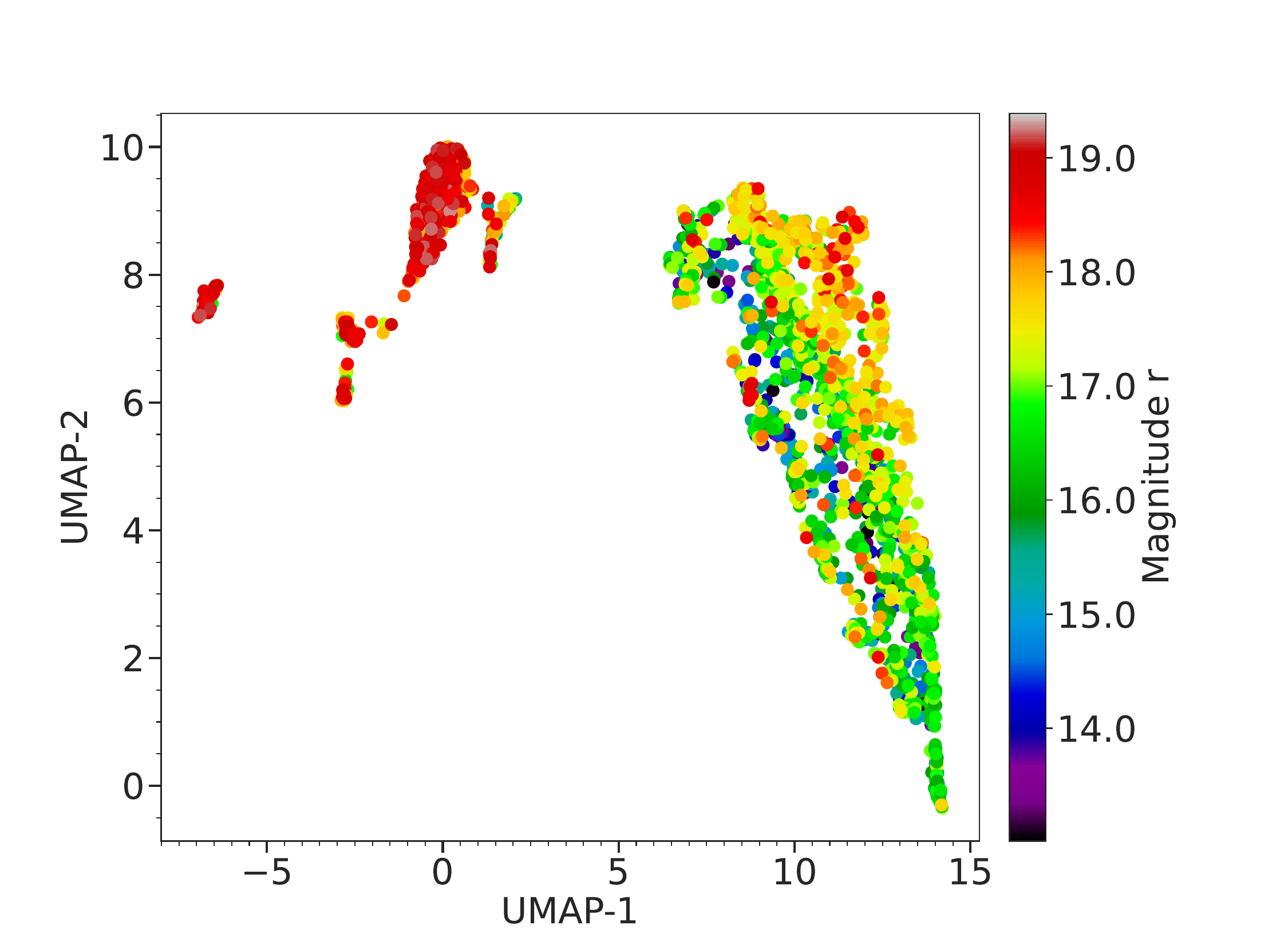}
        \includegraphics[width=0.45\linewidth, trim=10 10 5 8, clip]{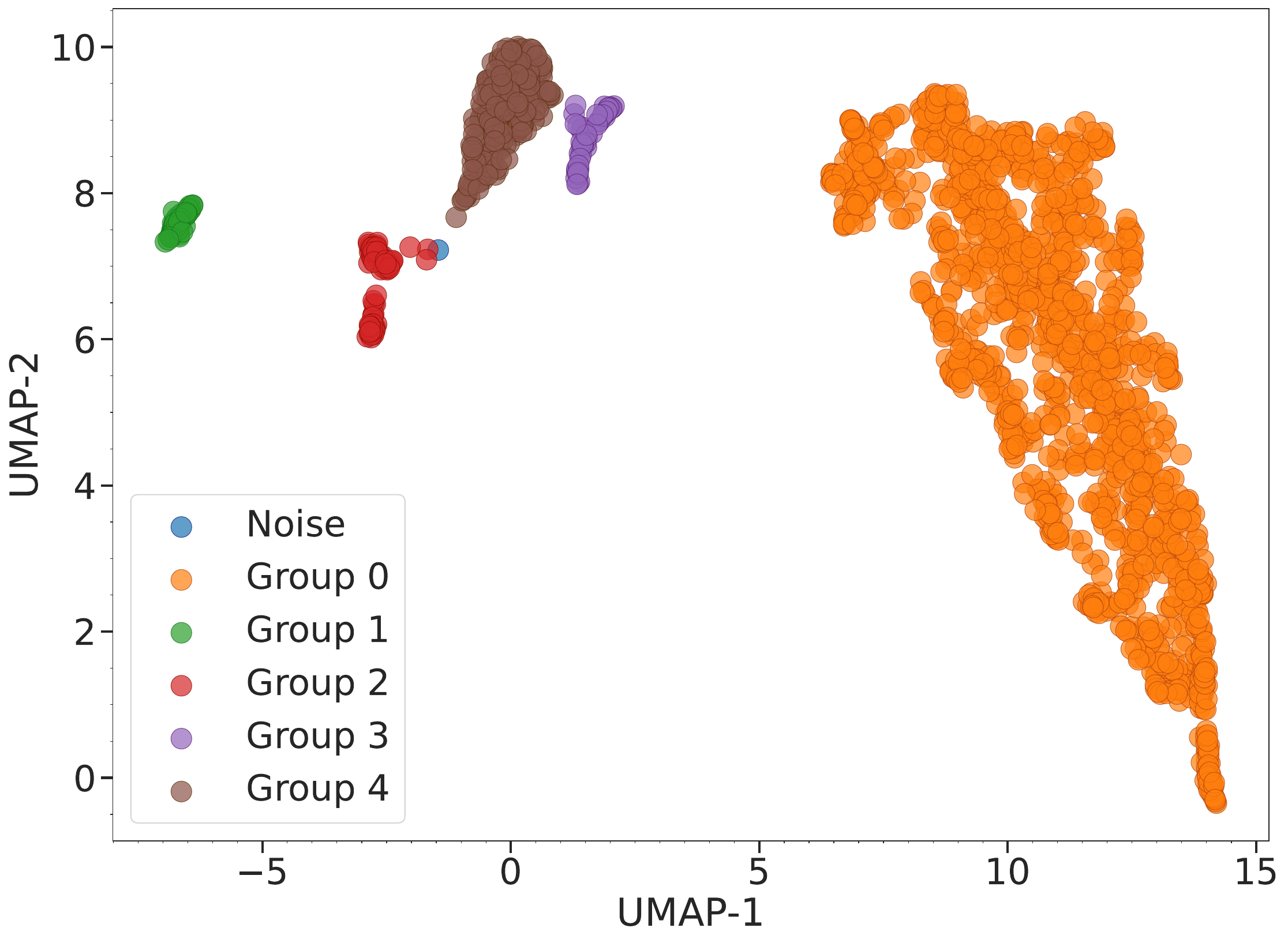}
        \end{tabular}  
        \caption{Similar to Fig. \ref{fig:umap}, but with additional features created using the W1 and W2 bands from WISE. The left panel shows the UMAP result, with the colour bar indicating the $r$ magnitude, and the right panel shows the HDBSCAN clustering result, identifying five distinct groups.}
    
        \label{fig:umap-wise}
    \end{figure*}

    Our unsupervised UMAP model projects the data, and HDBSCAN identifies the clusters. To ensure high-quality photometry, we required errors below 0.2 mag in all filters, reducing the sample to 2181 MS objects. This step minimizes the impact of noisy measurements, improving the performance of UMAP and HDBSCAN. By focusing on reliable photometric data, we enhance the accuracy and robustness of the clustering, ensuring more reliable classifications of H$\alpha$ excess sources.

    To perform cross-validation for selecting the optimal \texttt{n\_neighbors} and \texttt{n\_components} parameters in UMAP, we systematically explored a range of values for these parameters. The selection of parameters n\_neighbors and n\_components in UMAP is critical as it directly influences the quality of the reduced-dimensional representation. Initially, we conducted exploratory data analysis to visualize the dataset in reduced dimensions using various combinations of n\_neighbors and n\_components. This allowed us to qualitatively assess how well UMAP preserved the underlying structure of the data. 

    To objectively evaluate the performance of different parameter combinations, we used two quantitative metrics: the Silhouette Score \citep{ROUSSEEUW198753} and the Davies-Bouldin Index \citep{Davies:1979}. The Silhouette Score measures how well-defined the clusters are in the reduced space, assessing both cohesion (how similar an object is to its own cluster) and separation (how different it is from other clusters). Higher scores indicate better separation, meaning that objects are well-matched to their own cluster and poorly matched to others. The Davies-Bouldin Index measures the average similarity between each cluster and its most similar cluster. A lower value indicates better-defined clusters. Both metrics were used to identify the optimal combination of parameters for UMAP.

    A grid of tests was constructed over a range of \texttt{n\_neighbors} (5, 10, 15, 20, 30, 50, 70, 100) and \texttt{n\_components} (2, 3, 4, 5, 10, 20, 50) values. For each combination, UMAP was applied, followed by clustering using KMeans \citep{Lloyd:1982}, and the metrics were computed to determine the optimal parameter set. The KMeans algorithm clusters data by attempting to separate samples into $n$ groups of equal variance by minimizing a criterion known as inertia, or the within-group sum of squares. It requires the number of clusters (\texttt{n\_clusters}) to be specified, which was set equal to the \texttt{n\_components} from each UMAP computation. This choice assumes that the dimensionality of the reduced space corresponds to the natural clustering structure of the data, making it a reasonable and useful strategy for exploratory data analysis. KMeans scales well to large numbers of samples and has been widely used in various fields for clustering tasks. By applying the Silhouette Score and Davies-Bouldin Index to the results, we identified the optimal parameters for both UMAP and KMeans, ensuring well-defined and separable clusters.
    
    \subsubsection{Initial Analysis Using S-PLUS Photometry}

   For the first experiment, we used the 66 S-PLUS colours as input parameters and applied the metric evaluation method described above. These metrics include the silhouette score, which measures cluster cohesion, and the Davies-Bouldin Index, which assesses cluster separation. Lower values of the Davies-Bouldin Index and higher silhouette scores indicate better clustering performance. After evaluating various hyperparameter combinations, we identified that setting \texttt{n\_neighbors} = 30 and \texttt{n\_components} = 2 yielded the highest silhouette score (0.799) and the lowest Davies-Bouldin Index (0.266). These values were subsequently adopted as the optimal hyperparameters for our analysis. For the \texttt{min\_dist} parameter, we used the default value of 0.1. Figure \ref{fig:Silhouette-Score-Davies-Bouldin} illustrates the behaviour of the silhouette score (left panel) and the Davies-Bouldin Index (right panel) as functions of the \texttt{n\_neighbors} for each \texttt{n\_components}. In the left panel, we observe that the silhouette score varies with \texttt{n\_neighbors}, with the best performance achieved at \texttt{n\_neighbors} = 30 and \texttt{n\_components} = 2. The right panel shows the Davies-Bouldin Index, which decreases consistently for these hyperparameters, confirming their optimality. 
    
    Following dimensionality reduction with UMAP, the resultant variables were utilized to construct HDBSCAN models. We experimented various combination for the "minimum cluster size" and "minimum number of samples" parameters. We ended up with the optimal value of 2 and 50, respectively. Euclidean metric was employed for distance calculations throughout.
    
    The left panel of Fig. \ref{fig:umap} shows the distribution of the new variables in UMAP space, resulting from applying it to the 66 S-PLUS colours of the H$\alpha$ excess objects for the MS. The colour bar indicates the $r$ magnitude, highlighting the bright and faint sources. Visually, it is possible to distinguish at least four groups, with the small clusters located in the upper left of the diagram tending to be fainter. The right panel of the Fig. \ref{fig:umap} shows the same plot but with the results of applying HDBSCAN using the parameters mentioned above. HDBSCAN identified four groups. Table \ref{tab:ML-groups} provides the number of objects in each group. To further understand the nature of each group, we examined their SIMBAD counterparts, which are also detailed in the Table \ref{tab:ML-groups}.
    
    \textbf{Group 0} contains 58 objects, 22 of which are matched in SIMBAD. The majority are QSOs (19), with the remaining objects including one galaxy, one radio source, and one QSO candidate. This composition suggests that Group 0 primarily consists of extragalactic sources, with a redshift distribution peak around 2.45. It is worth noting that the spectral characteristics of QSOs (a type of AGN) differ significantly from typical galaxies; while starburst galaxies show strong extinction at blue wavelengths, the QSO spectrum rises sharply toward the blue, indicative of the high-energy processes associated with active galactic nuclei.
    
    \textbf{Group 1} contains 166 objects, 149 of which have entries in the database. This group is predominantly composed of RR Lyrae stars (107), followed by eclipsing binaries (19), various types of pulsating variables (9), and a few other stellar objects, including 2 QSOs. This group appears to represent objects with H$\alpha$ in absorption, as it is well known that RR Lyrae stars exhibit H$\alpha$ absorption lines.

    \textbf{Group 2} includes 1539 objects, 323 of which are catalogued in SIMBAD. The majority are eclipsing binaries (275), followed by a few stars (10), QSOs (9), and a small number of cataclysmic variables and RR Lyrae stars. In this context, the QSOs are AGNs without detectable H$\alpha$ emission within the S-PLUS wavelength range (or where the SDSS spectra do not cover the H$\alpha$ line). This group is thus characterized by the significant presence of binary star systems and various types of variable stars.

    \textbf{Group 3} consists of 93 objects, 42 of which are matched in the database. According to SIMBAD, the majority are labeled as QSOs (17), along with Seyfert 1 galaxies (10) and other classifications such as AGN candidates, radio sources, and a few galaxies. However, upon inspecting the spectra of several of these QSOs, we find that the line observed within the S-PLUS $J$0660 filter corresponds to [O III] and/or H$\beta$, rather than H$\alpha$. Given the narrow redshift range (0.31 to 0.37), these characteristics suggest that the objects in this group are better classified as AGNs, not QSOs. This may indicate a misclassification in SIMBAD, where sources labelled as QSOs in this group likely correspond to AGNs.
    
    \textbf{Group 4} includes 325 objects, 143 of which are recorded in the database. This group has a high concentration of QSOs (78) and cataclysmic variables (25). Additionally, it features a mix of blue stars, AGNs, radio sources, and white dwarf candidates. The extragalactic objects in this group show a peak in the redshift distribution around 1.35. It is expected that CVs are located closer to the QSOs than to Galactic sources in the UMAP variable space due to their photometric characteristics, which can resemble those of QSOs in certain features, despite the spectral differences \citep{Scaringi:2013}.
    
    In summary, our application of UMAP and HDBSCAN to the H$\alpha$ excess sources has effectively identified distinct groups with varying astrophysical characteristics using S-PLUS photometry. The classification successfully differentiates extragalactic sources, such as QSOs and AGNs, from galactic sources, including variable stars and binary systems. However, distinguishing Galactic cataclysmic variables from QSOs with redshifts around 1.35 remains challenging. Importantly, our results suggest that objects with ($J$0660 - $r$) colour excess due to emission lines can be distinguished from those with excess caused by H$\alpha$ absorption lines, mainly RR Lyrae stars.
    
    \subsubsection{Integration of S-PLUS and WISE Photometry}
    
        The second experiment incorporated the W1 and W2 filters from the WISE survey. These filters were selected because they provide the best sensitivity and reliability for detecting sources with infrared excess or thermal emission \citep{Nakazono:2021}. To include these data, we crossmatched the H$\alpha$ sources from the Main Survey of S-PLUS with the ALLWISE catalogue \citep{Cutri:2013} using a search radius of 2\,arcsec. This radius was chosen considering the broader point-spread function (PSF) of WISE compared to S-PLUS, as discussed in \citet{Nakazono:2021}. This process initially yielded 3173 matches, which were reduced to 1910 after applying photometric quality cuts, including errors smaller than 0.5 magnitudes in W1 and W2 and equivalent constraints for S-PLUS filters.

    Additional colours were constructed by combining WISE bands (W1 and W2) with S-PLUS broadband filters, such as W1 - W2, W1 - $u$, W2 - $u$, W1 - $g$, and so on. This expanded the parameter space from 66 to 77 variables, enriching the dataset and enhancing the performance of machine learning models in characterizing the physical properties of the H$\alpha$ sources. We identified optimal parameters for UMAP as \texttt{n\_neighbors} = 50 and \texttt{n\_components} = 2, based on the silhouette score and the Davies-Bouldin Index. For HDBSCAN, we employed \texttt{min\_cluster\_size} = 50 and \texttt{min\_samples} = 5.
    
    Figure \ref{fig:umap-wise} shows the results of the reduction in dimensionality and the groups identified by applying UMAP followed by HDBSCAN, using the input parameters described in the previous paragraph. On this occasion, HDBSCAN found five groups and one objects that were classified as noise. Table \ref{tab:ML-groups} summarizes these results:
    
    \textbf{Group 0} contains 1,437 objects, 424 of which correspond to entries in the SIMBAD database. Among these, 262 are eclipsing binaries (EB*), followed by 98 RR Lyrae stars (RRLyr). Other objects include EB* candidates, stars, pulsating variables, and a few QSOs. This group predominantly consists of variable stars and a small number of extragalactic sources.
    
    \textbf{Group 1} includes 59 objects, 23 of which are matched with SIMBAD. The majority are QSOs (20), with a few other objects like a galaxy, a radio source, and a QSO candidate. This group mainly represents extragalactic sources, particularly active galactic nuclei. The redshift distribution has a peak around 2.45. This group resembles \textbf{Group 0} from the previous case (without WISE analysis), albeit with one additional QSO. 
    
    \textbf{Group 2} consists of 93 objects, with 43 identified in the database. The group is primarily composed of QSOs (18), Seyfert 1 galaxies (10), and AGN candidates, with some galaxies and radio sources. This indicates a strong presence of active galactic nuclei and other extragalactic objects. The redshift distribution for extragalactic objects in this group ranges approximately from 0.31 to 0.37. This group is analogous to \textbf{Group 3} from the previous analysis (without WISE data).
    
    \textbf{Group 3} includes 51 objects, with 36 matches in SIMBAD. The majority are cataclysmic variables (24), with a few CV candidates, hot subdwarf candidates, and white dwarf candidates. This group is largely composed of cataclysmic variables and related stellar objects.
    
   \textbf{Group 4} contains 269 objects, 100 of which are matched with the database. The majority are QSOs (83), with a mix of blue stars, AGNs, radio sources, stars, and galaxies. This group shows a variety of astrophysical phenomena, both stellar and extragalactic, with a redshift distribution peaking around 1.35. It is similar to \textbf{Group 4} from the previous analysis using only S-PLUS data. In the S-PLUS-only group, the majority of objects are QSOs and cataclysmic variables, while the S-PLUS + WISE group contains more QSOs but no cataclysmic variables. The photometric characteristics of CVs in S-PLUS resemble those of QSOs, explaining why they cluster closer to QSOs than to Galactic sources in the UMAP variable space. The inclusion of WISE data likely contributed to the increase in QSOs by providing infrared information that helps differentiate extragalactic objects.

    In summary, the inclusion of WISE filters in our analysis has significantly enhanced the clustering of H$\alpha$ excess sources. The integration of WISE data has allowed for a more precise differentiation between galactic and extragalactic sources, enriching our understanding of the objects in our dataset. Notably, it has facilitated the separation of cataclysmic variables from QSOs with redshifts around 1.35. For detailed insights, refer to Sect. \ref{sec:results} where the redshifted emission lines of extragalactic objects are highlighted in the $J$0660 filter. However, it is important to note that the addition of WISE data has introduced challenges in identifying the group of RR Lyrae stars using HDBSCAN. 
    
    Uncertainties in photometric colours of variable stars based on single, random observations are inherently biased due to the stars' intrinsic variability. This effect is more pronounced with an increase in the amplitude of variability, as seen in classes of stars like RR Lyrae and Mira variables, which typically exhibit amplitudes higher than 0.3 to 2 magnitudes for RR Lyrae stars Chandra X-ray Observatory. The S-PLUS survey offers a significant advantage in this regard, as its 12 photometric wavebands are observed nearly simultaneously within approximately 1.5 hours SPLUS. Consequently, these observations are closely spaced in phase for variable stars.

For instance, RR Lyrae stars (RRab subtype), which have periods of approximately 0.5 days Chandra X-ray Observatory, will have all 12 S-PLUS wavebands captured within a phase range of $\simeq$0.1. This minimizes the variability effects on the observed photometric colours and ensures a more precise and reliable measurement of stellar parameters derived from these data. In contrast, random or non-simultaneous observations are likely to result in larger uncertainties due to phase mismatches, particularly for variable stars with significant amplitude changes over short timescales.

When S-PLUS wavebands are combined with external data, such as WISE wavebands, the phase mismatch becomes a critical source of uncertainty. WISE observations, which are not time-synchronized with S-PLUS, can introduce errors because the observed phases of variable stars in the combined dataset will be random. As a result, the derived colours and parameters will suffer from increased scatter and reduced precision. Therefore, we attribute the improved performance of models relying solely on S-PLUS observations to the reduced uncertainties in colours achieved by observing all wavebands in a near-simultaneous manner. This emphasizes the importance of phase-coherent photometric observations for the precise characterization of variable stars, especially those with significant amplitude variability.

    \begin{table*}[h]
    \centering
    \caption{Summary of clustering outcomes achieved using the UMAP and HDBSCAN unsupervised machine learning methods applied to H$\alpha$ excess sources of the MS. Clustering is performed using S-PLUS and S-PLUS + WISE filter combinations for the MS. The table displays the number of objects allocated to each cluster, providing insights into the distribution of sources identified through the clustering process.}
    \label{tab:ML-groups}
    \begin{adjustbox}{max width=\textwidth}
    \begin{tabular}{lcccc}
    \toprule
    \textbf{} & \textbf{Group} & \textbf{Number of Objects} & \textbf{Number with SIMBAD Match} & \textbf{Comments about SIMBAD Match} \\
    \midrule
    \multicolumn{5}{c}{\textbf{Main Survey}} \\
    \midrule
    \multicolumn{5}{c}{Only S-PLUS Filters} \\
    \midrule
    & Group 0 & 58 & 22 & \parbox[t]{7cm}{QSO (19), QSO\_Candidate (1), Galaxy (1), Radio (1)} \\
    \cmidrule{2-5}
    & Group 1 & 166 & 149 & \parbox[t]{7cm}{RRLyr (107), EB* (19), EB*\_Candidate (1), PulsV* (9), PulsVdelSct (6), Star (2), QSO (2), RotV* (1), SB*\_Candidate (1), BlueStraggler (1)} \\
    \cmidrule{2-5}
    & Group 2 & 1539 & 323 & \parbox[t]{7cm}{EB* (275), EB*\_Candidate (11), Star (10), QSO (9), CataclyV* (1), CV*\_Candidate (3), V* (3), RotV* (1), Pec* (2), low-mass* (2), RRLyr (2), AGB* (1), PulsV* (1), PulsVdelSct (1), RSCVn (1)} \\
    \cmidrule{2-5}
    & Group 3 & 93 & 42 & \parbox[t]{7cm}{QSO (17), Seyfert\_1 (10), AGN (3), AGN\_Candidate (6), Galaxy (3), Radio (2), RadioG (1)} \\
    \cmidrule{2-5}
    & Group 4 & 325 & 143 & \parbox[t]{7cm}{QSO (78), CataclyV* (25), CV*\_Candidate (6), Blue (7), Star (6), Hsd\_Candidate (4), AGN (3), Radio (3), WD* (2), WD*\_Candidate (3), RRLyr (2), Galaxy (2), EB* (1), Seyfert\_1 (1)} \\
    \midrule
    \textbf{Total} & & \textbf{2181} & \textbf{679} & \\
    \midrule
    \multicolumn{5}{c}{S-PLUS + WISE Filters} \\
    \midrule
    & Group 0 & 1437 & 424 & \parbox[t]{7cm}{EB* (262), EB*\_Candidate (23), RRLyr (98), Star (13), PulsV* (8), V* (4), RotV* (3), QSO (3), PulsVdelSct (2), low-mass* (2), Pec* (2), CataclyV* (1), CV*\_Candidate (1), AGB* (1), SB*\_Candidate (1)} \\
    \cmidrule{2-5}
    & Group 1 & 59 & 23 & \parbox[t]{7cm}{QSO (20), QSO\_Candidate (1). Galaxy (1), Radio (1)} \\
    \cmidrule{2-5}
    & Group 2 & 93 & 43 & \parbox[t]{7cm}{QSO (18), Seyfert\_1 (10),  AGN (3), AGN\_Candidate (6), Galaxy (3), Radio (2), RadioG (1)} \\
    \cmidrule{2-5}
    & Group 3 & 51 & 36 & \parbox[t]{7cm}{CataclyV* (24), CV*\_Candidate (3), Hsd\_Candidate (3), WD*\_Candidate (3), RRLyr (1), Seyfert\_1 (1), Star (1)} \\
    \cmidrule{2-5}
    & Group 4 & 269 & 100 & \parbox[t]{7cm}{QSO (83), AGN (3), Blue (7), Radio (3), Star (2), Galaxy (2)} \\
    \cmidrule{2-5}
    & Noise & 1 & -- & -- \\
    \midrule
    \textbf{Total} & & \textbf{1910} & \textbf{626} & \\
    \bottomrule
    \end{tabular}
    \end{adjustbox}
    \end{table*}

    \subsection{Extracting Main Features: Colour Analysis}
    \label{sec:more-colors}
    
    In this section, we focus on the colours derived from the S-PLUS and WISE filters, which are effective in distinguishing the different groups of H$\alpha$-excess objects identified by the combined UMAP and HDBSCAN analysis of the MS S-PLUS data.
    
    In the MS H$\alpha$-excess list, we identified extragalactic sources with higher redshifts, where blueward emission lines are redshifted to wavelengths near H$\alpha$, resulting in an apparent H$\alpha$ excess in the $J$0660 filter. By incorporating the WISE filters to create additional colours for the unsupervised machine learning models, we achieved better separation of extragalactic sources from Galactic sources (see Sect. \ref{sec:results-umap-hd} for more details).
    
    We used the classifications made by combining UMAP and HDBSCAN to create Random Forest \citep{Breiman:2001} models and identified the most important features, specifically the colours that contribute to the separation or classification of the classes of objects. The Random Forest algorithm is an ensemble learning method that builds multiple decision trees during the training phase. Each tree is trained on a random subset of the data and a random subset of features, which helps reduce overfitting and improves model generalization. During prediction, the results from all trees are aggregated by voting (for classification) or averaging (for regression), providing more stable and accurate predictions. Random forests are widely used for classification and regression tasks, known for their ability to handle complex data and offer reliable results. We implemented Random Forest algorithm, using 66 S-PLUS colours plus 11 additional colours generated with the W1 and W2 filters as input parameters, and labels generated by HDBSCAN. 
    
    The dataset used in this study exhibited a class imbalance: cluster 0 (1437 points), cluster 1 (59 points), cluster 2 (93 points), cluster 3 (51 points), and cluster 4 (269 points). To address this imbalance, we used the \texttt{class\_weight=\textquotesingle balanced\textquotesingle} parameter in the Random Forest algorithm. The classifier achieved an F1 Macro Average of 0.95 ($\pm$0.08) during 5-fold cross-validation. This high score, along with low variability, indicates that the model effectively handles the imbalance and consistently classifies the different clusters. The Macro F1 score is the average of the F1 scores calculated for each class, where each class is given equal weight, regardless of its frequency. The F1 score itself is the harmonic mean of precision and recall, providing a single metric that balances both. This metric is particularly useful in the presence of class imbalance, as it ensures that each class contributes equally to the overall score. For more details, see \citet{Sokolova:2009}. The Random Forest algorithm and Macro F1 score were implemented using the \texttt{scikit-learn} package.
    
\begin{figure}
        \includegraphics[width=\linewidth]{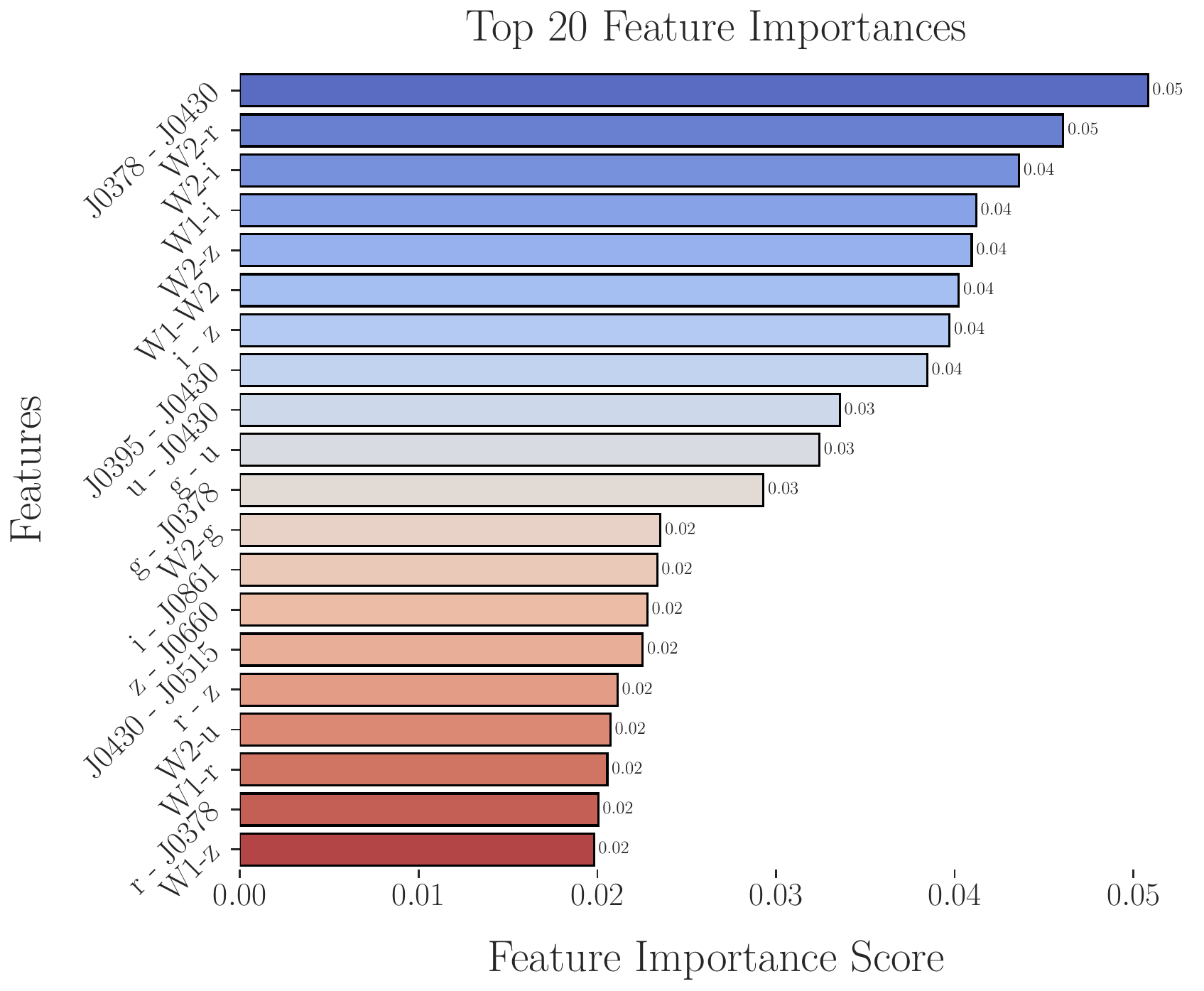}
        \caption{Top 20 feature importances identified by the Random Forest model, showing the colours that contributed most significantly to the clustering of H$\alpha$-excess objects using UMAP + HDBSCAN. The importance scores indicate the relative impact of each colour on the classification of different object classes.}
        \label{fig:importnace-features}
    \end{figure}

    After performing the model, we accessed the feature importances using \texttt{feature\_importances} from the Random Forest package. Figure \ref{fig:importnace-features} shows the top 20 feature importances and their respective scores, indicating the colours that contributed most to clustering the different classes of objects identified by UMAP + HDBSCAN.

    Now that we have identified the important colours, we used the \texttt{pairplot} routine in the \texttt{seaborn} package \citep{Waskom:2021} to generate all possible colour-colour diagrams using the top 20 features. Seaborn is a Python library designed to simplify the creation of statistical graphics. It extends Matplotlib and integrates seamlessly with pandas, making it particularly effective for handling and visualizing structured datasets. The \texttt{pairplot} function is especially useful for creating scatterplot matrices, allowing the simultaneous visualization of pairwise relationships between multiple features in the data. This allowed us to identify the colour-colour diagrams that best separate the different classes of objects and choose the most effective ones. 
    
\begin{figure*}
        \includegraphics[width=\linewidth]{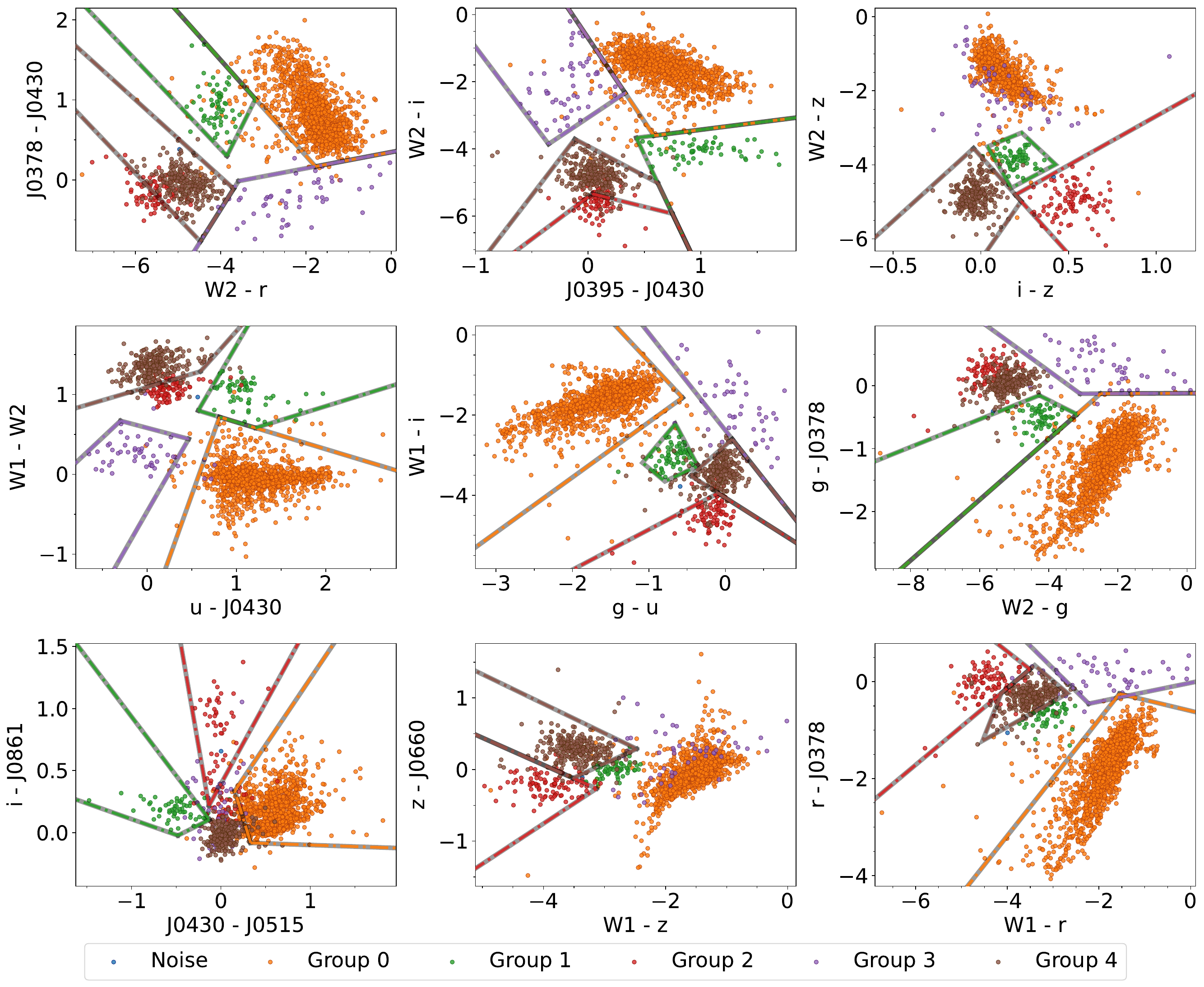}
        \caption{Examples of colour-colour diagrams using the top 20 features identified by the Random Forest model. These diagrams show the separation of different classes of objects from the H$\alpha$-excess sources list. The selected diagrams illustrate effective clustering achieved through UMAP + HDBSCAN, highlighting key colours that contribute to classification. The coloured lines represent tentative colour cuts for separating different classes of H$\alpha$-excess sources.}
        \label{fig:alternative-color-diagram}
    \end{figure*}

    Figure \ref{fig:alternative-color-diagram} shows nine  colour-colour diagrams that we selected for their ability to better separate the groups found in our H$\alpha$-excess sources list. These diagrams are based on the key features of importance, and we aimed to utilize nearly all of the 20 colours. Tentative colour cuts are presented in the Fig. \ref{fig:alternative-color-diagram} to differentiate the various classes of H$\alpha$ sources.
    
    This exercise demonstrates that using specific colour-colour diagrams with selected filters can effectively classify objects. By relying on a few key colours instead of all 12 S-PLUS filters and 2 WISE filters required for the machine learning in Sect. \ref{sec:results-umap-hd}, we can reduce the number of necessary observations. This approach is advantageous because not all objects have complete photometry in all filters, and some magnitudes may not meet the clean criteria, reducing the number of objects available for classification. Consequently, using a few specific colour criteria enables the classification of more objects, as it circumvents the need for complete data across all filters. 
    
   This analysis provides a practical framework for classifying H$\alpha$-excess sources without relying on complex algorithms. The proposed colour-colour diagrams enable the direct application of selection criteria, offering an effective method to distinguish between different classes based on the key features identified. Building on the methodology of \citet{Corradi:2008} for the $r-i$ vs. $r$-H$\alpha$ diagram, we extend it with new and effective colour combinations. These criteria provide a valuable tool for identifying distinct classes of H$\alpha$-excess sources, as shown in Table \ref{tab:ML-groups}, and can aid in targeted spectroscopic follow-ups.

\section{Conclusions}
\label{sec:conclu}

In this study, we have leveraged the S-PLUS project to analyze and classify H$\alpha$-excess sources in the Southern Sky, resulting in the following key conclusions:

\begin{enumerate}

    \item We identified 6\,956 H$\alpha$-excess candidates by using the narrow $J$0660 filter in combination with the broad $r$ and $i$ filters from S-PLUS. This included 3637 candidates from the high-latitude MS and 3319 from the GDS.

  \item Cross-referencing with the SIMBAD database enabled us to explore the types of objects in our list, identifying various emission line objects such as EM stars, YSOs, Be stars, CVs, PNe, and others. We also identified QSOs, non-local galaxies, and objects with H$\alpha$ in absorption, including RR Lyrae stars, primarily within the MS. The higher detection of RR Lyrae stars (111) in the MS compared to the GDS (8), based on SIMBAD, aligns with their expected distribution in older stellar populations.

    \item Validation with spectroscopic data from LAMOST and SDSS showed that approximately 60\% of the spectra exhibit H$\alpha$ emission lines, while around 30\% show H$\alpha$ in absorption in the MS. This comparison indicates the general accuracy of our classifications and supports the reliability of our H$\alpha$-excess source identifications. Furthermore, the VPHAS+ data for the GDS are consistent with our findings.

    \item The S-PLUS 12-filter system facilitates the detection of RR Lyrae stars and eclipsing binaries as H$\alpha$-excess sources, capturing their short-period variability through the sequential exposures of specific filters, such as $r$ and $J0660$. This makes S-PLUS particularly suited for identifying and studying such variables, enabling detailed analysis of their photometric behaviour and potential H$\alpha$ features.

    \item The use of machine learning techniques, specifically UMAP for dimensionality reduction and HDBSCAN for clustering, significantly enhanced our analysis of H$\alpha$-excess sources. The 12 S-PLUS filters allowed for effective differentiation between Galactic H$\alpha$-emission objects and extragalactic sources, as well as those with H$\alpha$ in absorption, such as RR Lyrae stars. However, the classification of CVs versus QSOs or AGNs, particularly with redshifts around 1.35, remained challenging. The similarity in the photometric characteristics of these objects made the boundaries between them less distinct, highlighting the inherent complexity in separating these sources, even when applying  machine learning techniques with S-PLUS colours.

     \item  The integration of WISE filter data significantly improved the clustering process, leading to a more accurate separation between extragalactic and Galactic sources. In particular, it facilitated the differentiation between CVs and QSOs, especially for objects with redshifts around 1.35, where their photometric characteristics previously overlapped. This enhancement enabled the clear separation of CVs from QSOs, refining the classification of these sources. Additionally, the infrared data allowed the identification of specific groups corresponding to AGNs or QSOs at particular redshifts, clearly separating them from Galactic sources. Certain groups identified in the clustering process were distinctly linked to AGNs or QSOs at specific redshifts, underscoring the value of combining optical and infrared data to resolve subtle differences in photometric signatures. However, the integration of WISE data also introduced challenges in classifying RR Lyrae stars, as the combination of optical and infrared data introduced noise in the clustering algorithm, complicating their precise classification.

     \item Using data from the S-PLUS Main Survey (MS) and WISE, we constructed new, effective colour-colour diagrams. By applying a Random Forest model to the results of clustering with UMAP and HDBSCAN, we identified key photometric features that differentiate the various classes of H$\alpha$-excess sources. Additionally, tentative colour criteria are proposed within these colour-colour diagrams, enabling a preliminary classification of sources without the need for complex algorithms.

\end{enumerate}

Our study used observational and analytical techniques to gain valuable insights into H$\alpha$-excess sources. Although challenges were encountered, particularly with RR Lyrae stars and certain extragalactic objects, our methods provided a robust framework for understanding H$\alpha$-excess phenomena. Our findings underscore the versatility of S-PLUS for identifying and understanding H$\alpha$-excess phenomena, demonstrating its potential for future applications in different sky regions and astrophysical environments. Future research should focus on expanding sample sizes and incorporating additional spectroscopic data to further refine classifications. Applying these methods to other sky regions or wavelengths could enhance our understanding of H$\alpha$-excess sources and their astrophysical contexts.

\section*{Acknowledgements}

LAG-S acknowledges funding for this work from CONICET and FAPESP grants 2019/26412-0.
RLO acknowledges financial support from the Brazilian institutions CNPq (PQ-312705/2020-4 and 445047/2024-0) and FAPESP (\#2020/00457-4). 
DRG acknowledges grants from FAPERJ (E-26/211.527/2023) and CNPq (315307/2023-4).
LLN thanks Funda\c{c}\~ao de Amparo \`a Pesquisa do Estado do Rio de Janeiro (FAPERJ) for granting the postdoctoral research fellowship E-40/2021(280692). M.B.F. acknowledges financial support from the National Council for Scientific and Tech-
nological Development (CNPq) Brazil (grant number: 307711/2022-6). S.D. acknowledges CNPq/MCTI for grant 306859/2022-0.
PKH gratefully acknowledges the Fundação de Amparo à Pesquisa do Estado de São Paulo (FAPESP) for the support grant 2023/14272-4. CL-D acknowledges a grant from the ESO Comite Mixto 2022.
SP is supported by the international Gemini Observatory, a program of NSF NOIRLab, which is managed by the Association of Universities for Research in Astronomy (AURA) under a cooperative agreement with the U.S. National Science Foundation, on behalf of the Gemini partnership of Argentina, Brazil, Canada, Chile, the Republic of Korea, and the United States of America.
This work is sponsored (in part) by the Chinese Academy of Sciences (CAS), through a grant to the CAS South 
America Center for Astronomy (CASSACA). We acknowledge the science research grants from the China Manned 
Space Project with NO. CMS-CSST-2021-A05.
AAC acknowledges financial support from the Severo Ochoa grant CEX2021- 001131-S funded by MCIN/AEI/10.13039/501100011033

The S-PLUS project, including the T80-South robotic telescope and
the S-PLUS scientific survey, was founded as a partnership between the
Fundação de Amparo à Pesquisa do Estado de S\~{a}o Paulo
(FAPESP), the Observatório Nacional (ON), the Federal University of
Sergipe (UFS), and the Federal University of Santa Catarina
(UFSC), with important financial and practical contributions from
other collaborating institutes in Brazil, Chile (Universidad de La
Serena), and Spain (Centro de Estudios de Física del Cosmos de
Aragón, CEFCA). We further acknowledge financial support from
the São Paulo Research Foundation (FAPESP), the Brazilian National
Research Council (CNPq), the Coordination for the Improvement of
Higher Education Personnel (CAPES), the Carlos Chagas Filho Rio
de Janeiro State Research Foundation (FAPERJ), and the Brazilian
Innovation Agency (FINEP).

Funding for the SDSS and SDSS-II has been provided by the Alfred P.
Sloan Foundation, the Participating Institutions, the National Science
Foundation, the U.S. Department of Energy, the National Aeronautics
and Space Administration, the Japanese Monbukagakusho, the Max
Planck Society, and the Higher Education Funding Council for England.
The SDSS Web Site is \url{http://www.sdss.org/}.

The SDSS is managed by the Astrophysical Research Consortium for
the Participating Institutions. The Participating Institutions
are the American Museum of Natural History, Astrophysical Institute Potsdam,
University of Basel, University of Cambridge, Case Western Reserve University,
University of Chicago, Drexel University, Fermilab, the Institute for Advanced
Study, the Japan Participation Group, Johns Hopkins University, the Joint
Institute for Nuclear Astrophysics, the Kavli Institute for Particle Astrophysics
and Cosmology, the Korean Scientist Group, the Chinese Academy of Sciences (LAMOST),
Los Alamos National Laboratory, the Max-Planck-Institute for Astronomy (MPIA),
the Max-Planck-Institute for Astrophysics (MPA), New Mexico State University,
Ohio State University, University of Pittsburgh, University of Portsmouth,
Princeton University, the United States Naval Observatory, and the University
of Washington.

Guoshoujing Telescope (the Large Sky Area Multi-Object Fiber Spectroscopic
Telescope LAMOST) is a National Major Scientific Project built by the Chinese
Academy of Sciences. Funding for the project has been provided by the National
Development and Reform Commission. LAMOST is operated and managed by the
National Astronomical Observatories, Chinese Academy of Sciences. 

Scientific software
and databases used in this work include 
TOPCAT\footnote{\url{http://www.star.bristol.ac.uk/~mbt/topcat/}} \citep{Taylor:2005}, 
simbad and vizier from Strasbourg Astronomical Data Center (CDS)\footnote{\url{https://cds.u-strasbg.fr/}} 
and the following  python packages: \texttt{numpy}, \texttt{astropy}, \texttt{matplotlib}, 
\texttt{seaborn}, \texttt{pandas}, \texttt{scikit-learn}, \texttt{hdbscan}, \texttt{umap}.

%
\bibliographystyle{aa} 
\bibliography{ref} 

\begin{thebibliography}{87}
\expandafter\ifx\csname natexlab\endcsname\relax\def\natexlab#1{#1}\fi

\bibitem[{{Abril} {et~al.}(2020){Abril}, {Schmidtobreick}, {Ederoclite}, \&
  {L{\'o}pez-Sanjuan}}]{Abril:2020}
{Abril}, J., {Schmidtobreick}, L., {Ederoclite}, A., \& {L{\'o}pez-Sanjuan}, C.
  2020, \mnras, 492, L40

\bibitem[{{Ahumada} {et~al.}(2020){Ahumada}, {Prieto}, {Almeida}, {Anders},
  {Anderson}, {Andrews}, {Anguiano}, {Arcodia}, {Armengaud}, {Aubert}, {Avila},
  {Avila-Reese}, {Badenes}, {Balland}, {Barger}, {Barrera-Ballesteros}, {Basu},
  {Bautista}, {Beaton}, {Beers}, {Benavides}, {Bender}, {Bernardi}, {Bershady},
  {Beutler}, {Bidin}, {Bird}, {Bizyaev}, {Blanc}, {Blanton}, {Boquien},
  {Borissova}, {Bovy}, {Brandt}, {Brinkmann}, {Brownstein}, {Bundy}, {Bureau},
  {Burgasser}, {Burtin}, {Cano-D{\'\i}az}, {Capasso}, {Cappellari}, {Carrera},
  {Chabanier}, {Chaplin}, {Chapman}, {Cherinka}, {Chiappini}, {Doohyun Choi},
  {Chojnowski}, {Chung}, {Clerc}, {Coffey}, {Comerford}, {Comparat}, {da
  Costa}, {Cousinou}, {Covey}, {Crane}, {Cunha}, {Ilha}, {Dai}, {Damsted},
  {Darling}, {Davidson}, {Davies}, {Dawson}, {De}, {de la Macorra}, {De Lee},
  {Queiroz}, {Deconto Machado}, {de la Torre}, {Dell'Agli}, {du Mas des
  Bourboux}, {Diamond-Stanic}, {Dillon}, {Donor}, {Drory}, {Duckworth},
  {Dwelly}, {Ebelke}, {Eftekharzadeh}, {Davis Eigenbrot}, {Elsworth},
  {Eracleous}, {Erfanianfar}, {Escoffier}, {Fan}, {Farr},
  {Fern{\'a}ndez-Trincado}, {Feuillet}, {Finoguenov}, {Fofie},
  {Fraser-McKelvie}, {Frinchaboy}, {Fromenteau}, {Fu}, {Galbany}, {Garcia},
  {Garc{\'\i}a-Hern{\'a}ndez}, {Oehmichen}, {Ge}, {Maia}, {Geisler}, {Gelfand},
  {Goddy}, {Gonzalez-Perez}, {Grabowski}, {Green}, {Grier}, {Guo}, {Guy},
  {Harding}, {Hasselquist}, {Hawken}, {Hayes}, {Hearty}, {Hekker}, {Hogg},
  {Holtzman}, {Horta}, {Hou}, {Hsieh}, {Huber}, {Hunt}, {Chitham}, {Imig},
  {Jaber}, {Angel}, {Johnson}, {Jones}, {J{\"o}nsson}, {Jullo}, {Kim},
  {Kinemuchi}, {Kirkpatrick}, {Kite}, {Klaene}, {Kneib}, {Kollmeier}, {Kong},
  {Kounkel}, {Krishnarao}, {Lacerna}, {Lan}, {Lane}, {Law}, {Le Goff}, {Leung},
  {Lewis}, {Li}, {Lian}, {Lin}, {Long}, {Longa-Pe{\~n}a}, {Lundgren}, {Lyke},
  {Ted Mackereth}, {MacLeod}, {Majewski}, {Manchado}, {Maraston}, {Martini},
  {Masseron}, {Masters}, {Mathur}, {McDermid}, {Merloni}, {Merrifield},
  {M{\'e}sz{\'a}ros}, {Miglio}, {Minniti}, {Minsley}, {Miyaji}, {Mohammad},
  {Mosser}, {Mueller}, {Muna}, {Mu{\~n}oz-Guti{\'e}rrez}, {Myers}, {Nadathur},
  {Nair}, {Nandra}, {do Nascimento}, {Nevin}, {Newman}, {Nidever}, {Nitschelm},
  {Noterdaeme}, {O'Connell}, {Olmstead}, {Oravetz}, {Oravetz}, {Osorio},
  {Pace}, {Padilla}, {Palanque-Delabrouille}, {Palicio}, {Pan}, {Pan},
  {Parker}, {Paviot}, {Peirani}, {Ram{\'r}ez}, {Penny}, {Percival},
  {Perez-Fournon}, {P{\'e}rez-R{\`a}fols}, {Petitjean}, {Pieri},
  {Pinsonneault}, {Poovelil}, {Povick}, {Prakash}, {Price-Whelan}, {Raddick},
  {Raichoor}, {Ray}, {Rembold}, {Rezaie}, {Riffel}, {Riffel}, {Rix}, {Robin},
  {Roman-Lopes}, {Rom{\'a}n-Z{\'u}{\~n}iga}, {Rose}, {Ross}, {Rossi},
  {Rowlands}, {Rubin}, {Salvato}, {S{\'a}nchez}, {S{\'a}nchez-Menguiano},
  {S{\'a}nchez-Gallego}, {Sayres}, {Schaefer}, {Schiavon}, {Schimoia},
  {Schlafly}, {Schlegel}, {Schneider}, {Schultheis}, {Schwope}, {Seo},
  {Serenelli}, {Shafieloo}, {Shamsi}, {Shao}, {Shen}, {Shetrone}, {Shirley},
  {Aguirre}, {Simon}, {Skrutskie}, {Slosar}, {Smethurst}, {Sobeck}, {Sodi},
  {Souto}, {Stark}, {Stassun}, {Steinmetz}, {Stello}, {Stermer},
  {Storchi-Bergmann}, {Streblyanska}, {Stringfellow}, {Stutz}, {Su{\'a}rez},
  {Sun}, {Taghizadeh-Popp}, {Talbot}, {Tayar}, {Thakar}, {Theriault}, {Thomas},
  {Thomas}, {Tinker}, {Tojeiro}, {Toledo}, {Tremonti}, {Troup}, {Tuttle},
  {Unda-Sanzana}, {Valentini}, {Vargas-Gonz{\'a}lez}, {Vargas-Maga{\~n}a},
  {V{\'a}zquez-Mata}, {Vivek}, {Wake}, {Wang}, {Weaver}, {Weijmans}, {Wild},
  {Wilson}, {Wilson}, {Wolthuis}, {Wood-Vasey}, {Yan}, {Yang}, {Y{\`e}che},
  {Zamora}, {Zarrouk}, {Zasowski}, {Zhang}, {Zhao}, {Zhao}, {Zheng}, {Zheng},
  {Zhu}, \& {Zou}}]{Ahumada:2020}
{Ahumada}, R., {Prieto}, C.~A., {Almeida}, A., {et~al.} 2020, \apjs, 249, 3

\bibitem[{{Akras}(2023)}]{Akras:2023}
{Akras}, S. 2023, \mnras, 519, 6044

\bibitem[{{Akras} {et~al.}(2021){Akras}, {Gon{\c{c}}alves}, {Alvarez-Candal},
  \& {Pereira}}]{Akras:2021}
{Akras}, S., {Gon{\c{c}}alves}, D.~R., {Alvarez-Candal}, A., \& {Pereira},
  C.~B. 2021, \mnras, 502, 2513

\bibitem[{{Akras} {et~al.}(2019{\natexlab{a}}){Akras}, {Guzman-Ramirez}, \&
  {Gon{\c{c}}alves}}]{Akras:2019c}
{Akras}, S., {Guzman-Ramirez}, L., \& {Gon{\c{c}}alves}, D.~R.
  2019{\natexlab{a}}, \mnras, 488, 3238

\bibitem[{{Akras} {et~al.}(2019{\natexlab{b}}){Akras}, {Guzman-Ramirez},
  {Leal-Ferreira}, \& {Ramos-Larios}}]{Akras:2019a}
{Akras}, S., {Guzman-Ramirez}, L., {Leal-Ferreira}, M.~L., \& {Ramos-Larios},
  G. 2019{\natexlab{b}}, \apjs, 240, 21

\bibitem[{{Akras} {et~al.}(2019{\natexlab{c}}){Akras}, {Leal-Ferreira},
  {Guzman-Ramirez}, \& {Ramos-Larios}}]{Akras:2019b}
{Akras}, S., {Leal-Ferreira}, M.~L., {Guzman-Ramirez}, L., \& {Ramos-Larios},
  G. 2019{\natexlab{c}}, \mnras, 483, 5077

\bibitem[{{Almeida-Fernandes} {et~al.}(2022){Almeida-Fernandes}, {SamPedro},
  {Herpich}, {Molino}, {Barbosa}, {Buzzo}, {Overzier}, {de Lima}, {Nakazono},
  {Oliveira Schwarz}, {Perottoni}, {Bolutavicius}, {Guti{\'e}rrez-Soto},
  {Santos-Silva}, {Vitorelli}, {Werle}, {Whitten}, {Costa Duarte}, {Bom},
  {Coelho}, {Sodr{\'e}}, {Placco}, {Teixeira}, {Alonso-Garc{\'\i}a}, {Barbosa},
  {Beers}, {Bonatto}, {Chies-Santos}, {Hartmann}, {Lopes de Oliveira},
  {Navarete}, {Kanaan}, {Ribeiro}, {Schoenell}, \& {Mendes de
  Oliveira}}]{Fernandes:2022}
{Almeida-Fernandes}, F., {SamPedro}, L., {Herpich}, F.~R., {et~al.} 2022,
  \mnras, 511, 4590

\bibitem[{{Barentsen} {et~al.}(2014){Barentsen}, {Farnhill}, {Drew},
  {Gonz{\'a}lez-Solares}, {Greimel}, {Irwin}, {Miszalski}, {Ruhland}, {Groot},
  {Mampaso}, {Sale}, {Henden}, {Aungwerojwit}, {Barlow}, {Carter}, {Corradi},
  {Drake}, {Eisl{\"o}ffel}, {Fabregat}, {G{\"a}nsicke}, {Gentile Fusillo},
  {Greiss}, {Hales}, {Hodgkin}, {Huckvale}, {Irwin}, {King}, {Knigge},
  {Kupfer}, {Lagadec}, {Lennon}, {Lewis}, {Mohr-Smith}, {Morris}, {Naylor},
  {Parker}, {Phillipps}, {Pyrzas}, {Raddi}, {Roelofs}, {Rodr{\'{\i}}guez-Gil},
  {Sabin}, {Scaringi}, {Steeghs}, {Suso}, {Tata}, {Unruh}, {van Roestel},
  {Viironen}, {Vink}, {Walton}, {Wright}, \& {Zijlstra}}]{Barentsen:2014}
{Barentsen}, G., {Farnhill}, H.~J., {Drew}, J.~E., {et~al.} 2014, \mnras, 444,
  3230

\bibitem[{{Barentsen} {et~al.}(2011){Barentsen}, {Vink}, {Drew}, {Greimel},
  {Wright}, {Drake}, {Martin}, {Valdivielso}, \& {Corradi}}]{Barentsen:2011}
{Barentsen}, G., {Vink}, J.~S., {Drew}, J.~E., {et~al.} 2011, \mnras, 415, 103

\bibitem[{Becht {et~al.}(2018)Becht, McInnes, Healy, Dutertre, Kwok, Ng,
  Ginhoux, \& Newell}]{Becht:2018}
Becht, E., McInnes, L., Healy, J., {et~al.} 2018, Nature biotechnology

\bibitem[{{Benitez} {et~al.}(2014){Benitez}, {Dupke}, {Moles}, {Sodre},
  {Cenarro}, {Marin-Franch}, {Taylor}, {Cristobal}, {Fernandez-Soto}, {Mendes
  de Oliveira}, {Cepa-Nogue}, {Abramo}, {Alcaniz}, {Overzier},
  {Hernandez-Monteagudo}, {Alfaro}, {Kanaan}, {Carvano}, {Reis}, {Martinez
  Gonzalez}, {Ascaso}, {Ballesteros}, {Xavier}, {Varela}, {Ederoclite},
  {Vazquez Ramio}, {Broadhurst}, {Cypriano}, {Angulo}, {Diego}, {Zandivarez},
  {Diaz}, {Melchior}, {Umetsu}, {Spinelli}, {Zitrin}, {Coe}, {Yepes}, {Vielva},
  {Sahni}, {Marcos-Caballero}, {Shu Kitaura}, {Maroto}, {Masip}, {Tsujikawa},
  {Carneiro}, {Gonzalez Nuevo}, {Carvalho}, {Reboucas}, {Carvalho}, {Abdalla},
  {Bernui}, {Pigozzo}, {Ferreira}, {Chandrachani Devi}, {Bengaly}, {Campista},
  {Amorim}, {Asari}, {Bongiovanni}, {Bonoli}, {Bruzual}, {Cardiel}, {Cava},
  {Cid Fernandes}, {Coelho}, {Cortesi}, {Delgado}, {Diaz Garcia}, {Espinosa},
  {Galliano}, {Gonzalez-Serrano}, {Falcon-Barroso}, {Fritz}, {Fernandes},
  {Gorgas}, {Hoyos}, {Jimenez-Teja}, {Lopez-Aguerri}, {Lopez-San Juan},
  {Mateus}, {Molino}, {Novais}, {OMill}, {Oteo}, {Perez-Gonzalez}, {Poggianti},
  {Proctor}, {Ricciardelli}, {Sanchez-Blazquez}, {Storchi-Bergmann}, {Telles},
  {Schoennell}, {Trujillo}, {Vazdekis}, {Viironen}, {Daflon},
  {Aparicio-Villegas}, {Rocha}, {Ribeiro}, {Borges}, {Martins}, {Marcolino},
  {Martinez-Delgado}, {Perez-Torres}, {Siffert}, {Calvao}, {Sako}, {Kessler},
  {Alvarez-Candal}, {De Pra}, {Roig}, {Lazzaro}, {Gorosabel}, {Lopes de
  Oliveira}, {Lima-Neto}, {Irwin}, {Liu}, {Alvarez}, {Balmes}, {Chueca},
  {Costa-Duarte}, {da Costa}, {Dantas}, {Diaz}, {Fabregat}, {Ferrari},
  {Gavela}, {Gracia}, {Gruel}, {Gutierrez}, {Guzman}, {Hernandez-Fernandez},
  {Herranz}, {Hurtado-Gil}, {Jablonsky}, {Laporte}, {Le Tiran}, {Licandro},
  {Lima}, {Martin}, {Martinez}, {Montero}, {Penteado}, {Pereira}, {Peris},
  {Quilis}, {Sanchez-Portal}, {Soja}, {Solano}, {Torra}, \&
  {Valdivielso}}]{Benitez:2014}
{Benitez}, N., {Dupke}, R., {Moles}, M., {et~al.} 2014, arXiv e-prints,
  arXiv:1403.5237

\bibitem[{{Bertin}(2011)}]{Psfex:2011}
{Bertin}, E. 2011, in Astronomical Society of the Pacific Conference Series,
  Vol. 442, Astronomical Data Analysis Software and Systems XX, ed. I.~N.
  {Evans}, A.~{Accomazzi}, D.~J. {Mink}, \& A.~H. {Rots}, 435

\bibitem[{{Bertin} \& {Arnouts}(1996)}]{Sextractor:1996}
{Bertin}, E. \& {Arnouts}, S. 1996, \aaps, 117, 393

\bibitem[{{Blair} \& {Long}(2004)}]{Blair:2004}
{Blair}, W.~P. \& {Long}, K.~S. 2004, \apjs, 155, 101

\bibitem[{{Bom} {et~al.}(2021){Bom}, {Cortesi}, {Lucatelli}, {Dias},
  {Schubert}, {Oliveira Schwarz}, {Cardoso}, {Lima}, {Mendes de Oliveira},
  {Sodre}, {Smith Castelli}, {Ferrari}, {Damke}, {Overzier}, {Kanaan},
  {Ribeiro}, \& {Schoenell}}]{Bom:2021}
{Bom}, C.~R., {Cortesi}, A., {Lucatelli}, G., {et~al.} 2021, \mnras, 507, 1937

\bibitem[{{Bonoli} {et~al.}(2021){Bonoli}, {Mar{\'\i}n-Franch}, {Varela},
  {V{\'a}zquez Rami{\'o}}, {Abramo}, {Cenarro}, {Dupke}, {V{\'\i}lchez},
  {Crist{\'o}bal-Hornillos}, {Gonz{\'a}lez Delgado},
  {Hern{\'a}ndez-Monteagudo}, {L{\'o}pez-Sanjuan}, {Muniesa}, {Civera},
  {Ederoclite}, {Hern{\'a}n-Caballero}, {Marra}, {Baqui}, {Cortesi},
  {Cypriano}, {Daflon}, {de Amorim}, {D{\'\i}az-Garc{\'\i}a}, {Diego},
  {Mart{\'\i}nez-Solaeche}, {P{\'e}rez}, {Placco}, {Prada}, {Queiroz},
  {Alcaniz}, {Alvarez-Candal}, {Cepa}, {Maroto}, {Roig}, {Siffert}, {Taylor},
  {Benitez}, {Moles}, {Sodr{\'e}}, {Carneiro}, {Mendes de Oliveira}, {Abdalla},
  {Angulo}, {Aparicio Resco}, {Balaguera-Antol{\'\i}nez}, {Ballesteros},
  {Brito-Silva}, {Broadhurst}, {Carrasco}, {Castro}, {Cid Fernandes}, {Coelho},
  {de Melo}, {Doubrawa}, {Fernandez-Soto}, {Ferrari}, {Finoguenov},
  {Garc{\'\i}a-Benito}, {Iglesias-P{\'a}ramo}, {Jim{\'e}nez-Teja}, {Kitaura},
  {Laur}, {Lopes}, {Lucatelli}, {Mart{\'\i}nez}, {Maturi}, {Overzier},
  {Pigozzo}, {Quartin}, {Rodr{\'\i}guez-Mart{\'\i}n}, {Salzano}, {Tamm},
  {Tempel}, {Umetsu}, {Valdivielso}, {von Marttens}, {Zitrin},
  {D{\'\i}az-Mart{\'\i}n}, {L{\'o}pez-Alegre}, {L{\'o}pez-Sainz},
  {Yanes-D{\'\i}az}, {Rueda-Teruel}, {Rueda-Teruel}, {Abril Iba{\~n}ez}, {L
  Ant{\'o}n Bravo}, {Bello Ferrer}, {Bielsa}, {Casino}, {Castillo}, {Chueca},
  {Cuesta}, {Garzar{\'a}n Calderaro}, {Iglesias-Marzoa}, {{\'I}niguez},
  {Lamadrid Gutierrez}, {Lopez-Martinez}, {Lozano-P{\'e}rez}, {Ma{\'\i}cas
  Sacrist{\'a}n}, {Molina-Ib{\'a}{\~n}ez}, {Moreno-Signes}, {Rodr{\'\i}guez
  Llano}, {Royo Navarro}, {Tilve Rua}, {Andrade}, {Alfaro}, {Akras},
  {Arnalte-Mur}, {Ascaso}, {Barbosa}, {Beltr{\'a}n Jim{\'e}nez}, {Benetti},
  {Bengaly}, {Bernui}, {Blanco-Pillado}, {Borges Fernandes}, {Bregman},
  {Bruzual}, {Calderone}, {Carvano}, {Casarini}, {Chaves-Montero},
  {Chies-Santos}, {Coutinho de Carvalho}, {Dimauro}, {Duarte Puertas},
  {Figueruelo}, {Gonz{\'a}lez-Serrano}, {Guerrero}, {Gurung-L{\'o}pez},
  {Herranz}, {Huertas-Company}, {Irwin}, {Izquierdo-Villalba}, {Kanaan},
  {Kehrig}, {Kirkpatrick}, {Lim}, {Lopes}, {Lopes de Oliveira},
  {Marcos-Caballero}, {Mart{\'\i}nez-Delgado}, {Mart{\'\i}nez-Gonz{\'a}lez},
  {Mart{\'\i}nez-Somonte}, {Oliveira}, {Orsi}, {Penna-Lima}, {Reis}, {Spinoso},
  {Tsujikawa}, {Vielva}, {Vitorelli}, {Xia}, {Yuan}, {Arroyo-Polonio},
  {Dantas}, {Galarza}, {Gon{\c{c}}alves}, {Gon{\c{c}}alves}, {Gonzalez},
  {Gonzalez}, {Greisel}, {Jim{\'e}nez-Esteban}, {Landim}, {Lazzaro}, {Magris},
  {Monteiro-Oliveira}, {Pereira}, {Rebou{\c{c}}as}, {Rodriguez-Espinosa},
  {Santos da Costa}, \& {Telles}}]{Bonoli:2021}
{Bonoli}, S., {Mar{\'\i}n-Franch}, A., {Varela}, J., {et~al.} 2021, \aap, 653,
  A31

\bibitem[{Breiman(2001)}]{Breiman:2001}
Breiman, L. 2001, Machine Learning, 45, 5

\bibitem[{Campello {et~al.}(2013)Campello, Moulavi, \& Sander}]{Campello:2013}
Campello, R. J. G.~B., Moulavi, D., \& Sander, J. 2013, in Advances in
  Knowledge Discovery and Data Mining, ed. J.~Pei, V.~S. Tseng, L.~Cao,
  H.~Motoda, \& G.~Xu (Berlin, Heidelberg: Springer Berlin Heidelberg),
  160--172

\bibitem[{{Cenarro} {et~al.}(2019){Cenarro}, {Moles},
  {Crist{\'o}bal-Hornillos}, {Mar{\'\i}n-Franch}, {Ederoclite}, {Varela},
  {L{\'o}pez-Sanjuan}, {Hern{\'a}ndez-Monteagudo}, {Angulo}, {V{\'a}zquez
  Rami{\'o}}, {Viironen}, {Bonoli}, {Orsi}, {Hurier}, {San Roman}, {Greisel},
  {Vilella-Rojo}, {D{\'\i}az-Garc{\'\i}a}, {Logro{\~n}o-Garc{\'\i}a},
  {Gurung-L{\'o}pez}, {Spinoso}, {Izquierdo-Villalba}, {Aguerri}, {Allende
  Prieto}, {Bonatto}, {Carvano}, {Chies-Santos}, {Daflon}, {Dupke},
  {Falc{\'o}n-Barroso}, {Gon{\c{c}}alves}, {Jim{\'e}nez-Teja}, {Molino},
  {Placco}, {Solano}, {Whitten}, {Abril}, {Ant{\'o}n}, {Bello}, {Bielsa de
  Toledo}, {Castillo-Ram{\'\i}rez}, {Chueca}, {Civera},
  {D{\'\i}az-Mart{\'\i}n}, {Dom{\'\i}nguez-Mart{\'\i}nez},
  {Garzar{\'a}n-Calderaro}, {Hern{\'a}ndez-Fuertes}, {Iglesias-Marzoa},
  {I{\~n}iguez}, {Jim{\'e}nez Ruiz}, {Kruuse}, {Lamadrid}, {Lasso-Cabrera},
  {L{\'o}pez-Alegre}, {L{\'o}pez-Sainz}, {Ma{\'\i}cas}, {Moreno-Signes},
  {Muniesa}, {Rodr{\'\i}guez-Llano}, {Rueda-Teruel}, {Rueda-Teruel},
  {Soriano-Lagu{\'\i}a}, {Tilve}, {Valdivielso}, {Yanes-D{\'\i}az}, {Alcaniz},
  {Mendes de Oliveira}, {Sodr{\'e}}, {Coelho}, {Lopes de Oliveira}, {Tamm},
  {Xavier}, {Abramo}, {Akras}, {Alfaro}, {Alvarez-Candal}, {Ascaso}, {Beasley},
  {Beers}, {Borges Fernandes}, {Bruzual}, {Buzzo}, {Carrasco}, {Cepa},
  {Cortesi}, {Costa-Duarte}, {De Pr{\'a}}, {Favole}, {Galarza}, {Galbany},
  {Garcia}, {Gonz{\'a}lez Delgado}, {Gonz{\'a}lez-Serrano},
  {Guti{\'e}rrez-Soto}, {Hernandez-Jimenez}, {Kanaan}, {Kuncarayakti},
  {Landim}, {Laur}, {Licandro}, {Lima Neto}, {Lyman}, {Ma{\'\i}z
  Apell{\'a}niz}, {Miralda-Escud{\'e}}, {Morate}, {Nogueira-Cavalcante},
  {Novais}, {Oncins}, {Oteo}, {Overzier}, {Pereira}, {Rebassa-Mansergas},
  {Reis}, {Roig}, {Sako}, {Salvador-Rusi{\~n}ol}, {Sampedro},
  {S{\'a}nchez-Bl{\'a}zquez}, {Santos}, {Schmidtobreick}, {Siffert}, {Telles},
  \& {Vilchez}}]{Cenarro:2019}
{Cenarro}, A.~J., {Moles}, M., {Crist{\'o}bal-Hornillos}, D., {et~al.} 2019,
  \aap, 622, A176

\bibitem[{{Coelho}(2014)}]{Coelho:2014}
{Coelho}, P.~R.~T. 2014, \mnras, 440, 1027

\bibitem[{{Cook} {et~al.}(2019){Cook}, {Kasliwal}, {Van Sistine}, {Kaplan},
  {Sutter}, {Kupfer}, {Shupe}, {Laher}, {Masci}, {Dale}, {Sesar}, {Brady},
  {Yan}, {Ofek}, {Reitze}, \& {Kulkarni}}]{Cook:2019}
{Cook}, D.~O., {Kasliwal}, M.~M., {Van Sistine}, A., {et~al.} 2019, \apj, 880,
  7

\bibitem[{{Corradi} \& {Giammanco}(2010)}]{Corradi:2010}
{Corradi}, R.~L.~M. \& {Giammanco}, C. 2010, \aap, 520, A99

\bibitem[{{Corradi} {et~al.}(2008){Corradi}, {Rodr{\'\i}guez-Flores},
  {Mampaso}, {Greimel}, {Viironen}, {Drew}, {Lennon}, {Mikolajewska}, {Sabin},
  \& {Sokoloski}}]{Corradi:2008}
{Corradi}, R.~L.~M., {Rodr{\'\i}guez-Flores}, E.~R., {Mampaso}, A., {et~al.}
  2008, \aap, 480, 409

\bibitem[{{Corradi} {et~al.}(2011){Corradi}, {Sabin}, {Munari}, {Cetrulo},
  {Englaro}, {Angeloni}, {Greimel}, \& {Mampaso}}]{Corradi:2011}
{Corradi}, R.~L.~M., {Sabin}, L., {Munari}, U., {et~al.} 2011, \aap, 529, A56

\bibitem[{{Cutri} {et~al.}(2013){Cutri}, {Wright}, {Conrow}, {Fowler},
  {Eisenhardt}, {Grillmair}, {Kirkpatrick}, {Masci}, {McCallon}, {Wheelock},
  {Fajardo-Acosta}, {Yan}, {Benford}, {Harbut}, {Jarrett}, {Lake}, {Leisawitz},
  {Ressler}, {Stanford}, {Tsai}, {Liu}, {Helou}, {Mainzer}, {Gettings},
  {Gonzalez}, {Hoffman}, {Marsh}, {Padgett}, {Skrutskie}, {Beck}, {Papin}, \&
  {Wittman}}]{Cutri:2013}
{Cutri}, R.~M., {Wright}, E.~L., {Conrow}, T., {et~al.} 2013, {Explanatory
  Supplement to the AllWISE Data Release Products}, Explanatory Supplement to
  the AllWISE Data Release Products, by R. M. Cutri et al.

\bibitem[{Davies \& Bouldin(1979)}]{Davies:1979}
Davies, D.~L. \& Bouldin, D.~W. 1979, IEEE Transactions on Pattern Analysis and
  Machine Intelligence, PAMI-1, 224

\bibitem[{{Davies} {et~al.}(1976){Davies}, {Elliott}, \&
  {Meaburn}}]{1976MmRAS..81...89D}
{Davies}, R.~D., {Elliott}, K.~H., \& {Meaburn}, J. 1976, \memras, 81, 89

\bibitem[{{Drew} {et~al.}(2014){Drew}, {Gonzalez-Solares}, {Greimel}, {Irwin},
  {K{\"u}pc{\"u} Yoldas}, {Lewis}, {Barentsen}, {Eisl{\"o}ffel}, {Farnhill},
  {Martin}, {Walsh}, {Walton}, {Mohr-Smith}, {Raddi}, {Sale}, {Wright},
  {Groot}, {Barlow}, {Corradi}, {Drake}, {Fabregat}, {Frew}, {G{\"a}nsicke},
  {Knigge}, {Mampaso}, {Morris}, {Naylor}, {Parker}, {Phillipps}, {Ruhland},
  {Steeghs}, {Unruh}, {Vink}, {Wesson}, \& {Zijlstra}}]{Drew:2014}
{Drew}, J.~E., {Gonzalez-Solares}, E., {Greimel}, R., {et~al.} 2014, \mnras,
  440, 2036

\bibitem[{{Drew} {et~al.}(2005){Drew}, {Greimel}, {Irwin}, {Aungwerojwit},
  {Barlow}, {Corradi}, {Drake}, {G{\"a}nsicke}, {Groot}, {Hales}, {Hopewell},
  {Irwin}, {Knigge}, {Leisy}, {Lennon}, {Mampaso}, {Masheder}, {Matsuura},
  {Morales-Rueda}, {Morris}, {Parker}, {Phillipps}, {Rodriguez-Gil}, {Roelofs},
  {Skillen}, {Sokoloski}, {Steeghs}, {Unruh}, {Viironen}, {Vink}, {Walton},
  {Witham}, {Wright}, {Zijlstra}, \& {Zurita}}]{Drew:2005}
{Drew}, J.~E., {Greimel}, R., {Irwin}, M.~J., {et~al.} 2005, \mnras, 362, 753

\bibitem[{{Drew} {et~al.}(2008){Drew}, {Greimel}, {Irwin}, \&
  {Sale}}]{Drew:2008}
{Drew}, J.~E., {Greimel}, R., {Irwin}, M.~J., \& {Sale}, S.~E. 2008, \mnras,
  386, 1761

\bibitem[{Ester {et~al.}(1996)Ester, Kriegel, Sander, \& Xu}]{Ester:1996}
Ester, M., Kriegel, H.-P., Sander, J., \& Xu, X. 1996, in Proc. of 2nd
  International Conference on Knowledge Discovery and Data Mining (KDD-96),
  226--231

\bibitem[{{Fratta} {et~al.}(2021){Fratta}, {Scaringi}, {Drew}, {Mongui{\'o}},
  {Knigge}, {Maccarone}, {Court}, {I{\l}kiewicz}, {Pala}, {Gandhi}, \&
  {G{\"a}nsicke}}]{Fratta:2021}
{Fratta}, M., {Scaringi}, S., {Drew}, J.~E., {et~al.} 2021, \mnras, 505, 1135

\bibitem[{{Frew}(2008)}]{Frew:2008}
{Frew}, D.~J. 2008, PhD thesis, Department of Physics, Macquarie University,
  NSW 2109, Australia

\bibitem[{{Fukugita} {et~al.}(1996){Fukugita}, {Ichikawa}, {Gunn}, {Doi},
  {Shimasaku}, \& {Schneider}}]{Fukugita:1996}
{Fukugita}, M., {Ichikawa}, T., {Gunn}, J.~E., {et~al.} 1996, \aj, 111, 1748

\bibitem[{{Gonz{\'a}lez-L{\'o}pezlira}
  {et~al.}(2017){Gonz{\'a}lez-L{\'o}pezlira}, {Lomel{\'{\i}}-N{\'u}{\~n}ez},
  {{\'A}lamo-Mart{\'{\i}}nez}, {{\'O}rdenes-Brice{\~n}o}, {Loinard},
  {Georgiev}, {Mu{\~n}oz}, {Puzia}, {Bruzual A.}, \& {Gwyn}}]{Lomeli:2017}
{Gonz{\'a}lez-L{\'o}pezlira}, R.~A., {Lomel{\'{\i}}-N{\'u}{\~n}ez}, L.,
  {{\'A}lamo-Mart{\'{\i}}nez}, K., {et~al.} 2017, \apj, 835, 184

\bibitem[{{Gonz{\'a}lez-L{\'o}pezlira}
  {et~al.}(2022){Gonz{\'a}lez-L{\'o}pezlira}, {Lomel{\'\i}-N{\'u}{\~n}ez},
  {Ordenes-Brice{\~n}o}, {Loinard}, {Gwyn}, {Alamo-Mart{\'\i}nez}, {Bruzual},
  {Lan{\c{c}}on}, \& {Puzia}}]{Lomeli:2022cfht}
{Gonz{\'a}lez-L{\'o}pezlira}, R.~A., {Lomel{\'\i}-N{\'u}{\~n}ez}, L.,
  {Ordenes-Brice{\~n}o}, Y., {et~al.} 2022, \apj, 941, 53

\bibitem[{{Greer} {et~al.}(2017){Greer}, {Payne}, {Norton}, {Maxted},
  {Smalley}, {West}, {Wheatley}, \& {Kolb}}]{Greer:2017A}
{Greer}, P.~A., {Payne}, S.~G., {Norton}, A.~J., {et~al.} 2017, \aap, 607, A11

\bibitem[{{Guti{\'e}rrez-Soto} {et~al.}(2020){Guti{\'e}rrez-Soto},
  {Gon{\c{c}}alves}, {Akras}, {Cortesi}, {L{\'o}pez-Sanjuan}, {Guerrero},
  {Daflon}, {Borges Fernandes}, {Mendes de Oliveira}, {Ederoclite},
  {Sodr{\'e}}, {Pereira}, {Kanaan}, {Werle}, {V{\'a}zquez Rami{\'o}},
  {Alcaniz}, {Angulo}, {Cenarro}, {Crist{\'o}bal-Hornillos}, {Dupke},
  {Hern{\'a}ndez-Monteagudo}, {Mar{\'\i}n-Franch}, {Moles}, {Varela},
  {Ribeiro}, {Schoenell}, {Alvarez-Candal}, {Galbany}, {Jim{\'e}nez-Esteban},
  {Logro{\~n}o-Garc{\'\i}a}, \& {Sobral}}]{Gutierrez:2020}
{Guti{\'e}rrez-Soto}, L.~A., {Gon{\c{c}}alves}, D.~R., {Akras}, S., {et~al.}
  2020, \aap, 633, A123

\bibitem[{{Guti{\'e}rrez-Soto} {et~al.}(2024){Guti{\'e}rrez-Soto}, {Mari},
  {Weidmann}, \& {Faifer}}]{Gutierrez-Soto:2024}
{Guti{\'e}rrez-Soto}, L.~A., {Mari}, M.~B., {Weidmann}, W.~A., \& {Faifer},
  F.~R. 2024, \na, 109, 102207

\bibitem[{{Herpich} {et~al.}(2024){Herpich}, {Almeida-Fernandes}, {Oliveira
  Schwarz}, {Lima}, {Nakazono}, {Alonso-Garc{\'\i}a}, {Fonseca-Faria},
  {Sartori}, {Bolutavicius}, {Fabiano de Souza}, {Hartmann}, {Li}, {Espinosa},
  {Kanaan}, {Schoenell}, {Werle}, {Machado-Pereira}, {Guti{\'e}rrez-Soto},
  {Santos-Silva}, {Smith Castelli}, {Lacerda}, {Barbosa}, {Perottoni},
  {Ferreira Lopes}, {Valen{\c{c}}a}, {Re Martho}, {Bom}, {Bonatto}, {Carvalho},
  {Cernic}, {Cid Fernandes}, {Coelho}, {Cortesi}, {Cubillos Palma}, {Doubrawa},
  {Ferreira Alberice}, {Quispe-Huaynasi}, {Jacob Perin}, {Jaque Arancibia},
  {Krabbe}, {Lima-Dias}, {Lomel{\'\i}-N{\'u}{\~n}ez}, {Lopes de Oliveira},
  {Lopes}, {Luiz Figueiredo}, {L{\"o}sch}, {Navarete}, {Oliveira}, {Overzier},
  {Placco}, {Roig}, {Rubet}, {Santos}, {Sasse}, {Thain{\'a}-Batista},
  {Torres-Flores}, {Beers}, {Alvarez-Candal}, {Akras}, {Panda}, {Limberg},
  {Nilo Castell{\'o}n}, {Telles}, {Lopes}, {Pardo Montaguth}, {Beraldo e
  Silva}, {Humire}, {Borges Fernandes}, {Cordeiro}, {Ribeiro}, \& {Mendes de
  Oliveira}}]{Herpich:2024}
{Herpich}, F.~R., {Almeida-Fernandes}, F., {Oliveira Schwarz}, G.~B., {et~al.}
  2024, \aap, 689, A249

\bibitem[{{Jacoby} {et~al.}(2010){Jacoby}, {Kronberger}, {Patchick}, {Teutsch},
  {Saloranta}, {Howell}, {Crisp}, {Riddle}, {Acker}, {Frew}, \&
  {Parker}}]{Jacoby:2010}
{Jacoby}, G.~H., {Kronberger}, M., {Patchick}, D., {et~al.} 2010, \pasa, 27,
  156

\bibitem[{{Jaiswal} \& {Omar}(2016)}]{Jaiswal:2016}
{Jaiswal}, S. \& {Omar}, A. 2016, \mnras, 462, 92

\bibitem[{{Kalari} {et~al.}(2015){Kalari}, {Vink}, {Drew}, {Barentsen},
  {Drake}, {Eisl{\"o}ffel}, {Mart{\'\i}n}, {Parker}, {Unruh}, {Walton}, \&
  {Wright}}]{Kalari:2015}
{Kalari}, V.~M., {Vink}, J.~S., {Drew}, J.~E., {et~al.} 2015, \mnras, 453, 1026

\bibitem[{{Kron}(1980)}]{Kron:1980}
{Kron}, R.~G. 1980, \apjs, 43, 305

\bibitem[{Lloyd(1982)}]{Lloyd:1982}
Lloyd, S. 1982, IEEE Transactions on Information Theory, 28, 129

\bibitem[{{Lomel{\'\i}-N{\'u}{\~n}ez}
  {et~al.}(2022){Lomel{\'\i}-N{\'u}{\~n}ez}, {Mayya}, {Rodr{\'\i}guez-Merino},
  {Ovando}, \& {Rosa-Gonz{\'a}lez}}]{Lomeli:2022}
{Lomel{\'\i}-N{\'u}{\~n}ez}, L., {Mayya}, Y.~D., {Rodr{\'\i}guez-Merino},
  L.~H., {Ovando}, P.~A., \& {Rosa-Gonz{\'a}lez}, D. 2022, \mnras, 509, 180

\bibitem[{{Mar{\'\i}n-Franch} {et~al.}(2012){Mar{\'\i}n-Franch}, {Chueca},
  {Moles}, {Benitez}, {Taylor}, {Cepa}, {Cenarro}, {Cristobal-Hornillos},
  {Ederoclite}, {Gruel}, {Hern{\'a}ndez-Fuertes}, {L{\'o}pez-Sainz},
  {Luis-Simoes}, {Rueda-Teruel}, {Rueda-Teruel}, {Varela}, {Yanes-D{\'\i}az},
  {Brauneck}, {Danielou}, {Dupke}, {Fern{\'a}ndez-Soto}, {Mendes de Oliveira},
  \& {Sodr{\'e}}}]{Marin-Franch:2012}
{Mar{\'\i}n-Franch}, A., {Chueca}, S., {Moles}, M., {et~al.} 2012, in Society
  of Photo-Optical Instrumentation Engineers (SPIE) Conference Series, Vol.
  8450, Modern Technologies in Space- and Ground-based Telescopes and
  Instrumentation II, ed. R.~{Navarro}, C.~R. {Cunningham}, \& E.~{Prieto},
  84503S

\bibitem[{McInnes {et~al.}(2017)McInnes, Healy, \& Astels}]{McInnes:2017}
McInnes, L., Healy, J., \& Astels, S. 2017, The Journal of Open Source
  Software, 2

\bibitem[{McInnes {et~al.}(2020)McInnes, Healy, \& Melville}]{mcinnes:2020}
McInnes, L., Healy, J., \& Melville, J. 2020, UMAP: Uniform Manifold
  Approximation and Projection for Dimension Reduction

\bibitem[{{Mendes de Oliveira} {et~al.}(2019){Mendes de Oliveira}, {Ribeiro},
  {Schoenell}, {Kanaan}, {Overzier}, {Molino}, {Sampedro}, {Coelho}, {Barbosa},
  {Cortesi}, {Costa-Duarte}, {Herpich}, {Hernand ez-Jimenez}, {Placco},
  {Xavier}, {Abramo}, {Saito}, {Chies-Santos}, {Ederoclite}, {Lopes de
  Oliveira}, {Gon{\c{c}}alves}, {Akras}, {Almeida}, {Almeida-Fernandes},
  {Beers}, {Bonatto}, {Bonoli}, {Cypriano}, {Vinicius-Lima}, {de Souza},
  {Fabiano de Souza}, {Ferrari}, {Gon{\c{c}}alves}, {Gonzalez},
  {Guti{\'e}rrez-Soto}, {Hartmann}, {Jaffe}, {Kerber}, {Lima-Dias}, {Lopes},
  {Menendez-Delmestre}, {Nakazono}, {Novais}, {Ortega-Minakata}, {Pereira},
  {Perottoni}, {Queiroz}, {Reis}, {Santos}, {Santos-Silva}, {Santucci},
  {Barbosa}, {Siffert}, {Sodr{\'e}}, {Torres-Flores}, {Westera}, {Whitten},
  {Alcaniz}, {Alonso-Garc{\'\i}a}, {Alencar}, {Alvarez-Cand al}, {Amram},
  {Azanha}, {Barb{\'a}}, {Bernardinelli}, {Borges Fernandes}, {Branco},
  {Brito-Silva}, {Buzzo}, {Caffer}, {Campillay}, {Cano}, {Carvano}, {Castejon},
  {Cid Fernandes}, {Dantas}, {Daflon}, {Damke}, {de la Reza}, {de Melo de
  Azevedo}, {De Paula}, {Diem}, {Donnerstein}, {Dors}, {Dupke}, {Eikenberry},
  {Escudero}, {Faifer}, {Far{\'\i}as}, {Fernandes}, {Fernandes}, {Fontes},
  {Galarza}, {Hirata}, {Katena}, {Gregorio-Hetem},
  {Hern{\'a}ndez-Fern{\'a}ndez}, {Izzo}, {Jaque Arancibia}, {Jatenco-Pereira},
  {Jim{\'e}nez-Teja}, {Kann}, {Krabbe}, {Labayru}, {Lazzaro}, {Lima Neto},
  {Lopes}, {Magalh{\~a}es}, {Makler}, {de Menezes}, {Miralda-Escud{\'e}},
  {Monteiro-Oliveira}, {Montero-Dorta}, {Mu{\~n}oz-Elgueta}, {Nemmen}, {Nilo
  Castell{\'o}n}, {Oliveira}, {Ort{\'\i}z}, {Pattaro}, {Pereira}, {Quint},
  {Riguccini}, {Rocha Pinto}, {Rodrigues}, {Roig}, {Rossi}, {Saha}, {Santos},
  {Schnorr M{\"u}ller}, {Sesto}, {Silva}, {Smith Castelli}, {Teixeira},
  {Telles}, {Thom de Souza}, {Th{\"o}ne}, {Trevisan}, {de Ugarte Postigo},
  {Urrutia-Viscarra}, {Veiga}, {Vika}, {Vitorelli}, {Werle}, {Werner}, \&
  {Zaritsky}}]{Mendes:2019}
{Mendes de Oliveira}, C., {Ribeiro}, T., {Schoenell}, W., {et~al.} 2019,
  \mnras, 489, 241

\bibitem[{{Merc} {et~al.}(2019){Merc}, {G{\`a}lis}, \& {Wolf}}]{Merc:2019}
{Merc}, J., {G{\`a}lis}, R., \& {Wolf}, M. 2019, Eruptive Stars Information
  Letter, 41, 78

\bibitem[{{Merc} {et~al.}(2022){Merc}, {G{\'a}lis}, {Wolf}, {Velez}, {Bohlsen},
  \& {Barlow}}]{Merc:2022}
{Merc}, J., {G{\'a}lis}, R., {Wolf}, M., {et~al.} 2022, \mnras, 510, 1404

\bibitem[{{Merc} {et~al.}(2021){Merc}, {G{\'a}lis}, {Wolf}, {Velez}, {Buil},
  {Sims}, {Bohlsen}, {Vra{\v{s}}{\v{t}}{\'a}k}, {Boussin}, {Boussier},
  {Cazzato}, {Diarrasouba}, \& {Teyssier}}]{Merc:2021}
{Merc}, J., {G{\'a}lis}, R., {Wolf}, M., {et~al.} 2021, \mnras, 506, 4151

\bibitem[{{Merc} {et~al.}(2020){Merc}, {Miko{\l}ajewska}, {Gromadzki},
  {Ga{\l}an}, {I{\l}kiewicz}, {Skowron}, {Wyrzykowski}, {Hodgkin}, {Rybicki},
  {Zieli{\'n}ski}, {Kruszy{\'n}ska}, {Godunova}, {Simon}, {Reshetnyk}, {Lewis},
  {Kolb}, {Morrell}, {Norton}, {Awiphan}, {Poshyachinda}, {Reichart}, {Greet},
  \& {Kolgjini}}]{Merc:2020}
{Merc}, J., {Miko{\l}ajewska}, J., {Gromadzki}, M., {et~al.} 2020, \aap, 644,
  A49

\bibitem[{{Miko{\l}ajewska} {et~al.}(2014){Miko{\l}ajewska}, {Caldwell}, \&
  {Shara}}]{Mikolajewska:2014}
{Miko{\l}ajewska}, J., {Caldwell}, N., \& {Shara}, M.~M. 2014, \mnras, 444, 586

\bibitem[{{Miko{\l}ajewska} {et~al.}(2017){Miko{\l}ajewska}, {Shara},
  {Caldwell}, {I{\l}kiewicz}, \& {Zurek}}]{Mikolajewska:2017}
{Miko{\l}ajewska}, J., {Shara}, M.~M., {Caldwell}, N., {I{\l}kiewicz}, K., \&
  {Zurek}, D. 2017, \mnras, 465, 1699

\bibitem[{{Miszalski} {et~al.}(2009){Miszalski}, {Acker}, {Moffat}, {Parker},
  \& {Udalski}}]{Miszalski:2009}
{Miszalski}, B., {Acker}, A., {Moffat}, A.~F.~J., {Parker}, Q.~A., \&
  {Udalski}, A. 2009, \aap, 496, 813

\bibitem[{{Miszalski} \& {Miko{\l}ajewska}(2014)}]{Miszalski:2014}
{Miszalski}, B. \& {Miko{\l}ajewska}, J. 2014, \mnras, 440, 1410

\bibitem[{{Mongui{\'o}} {et~al.}(2020){Mongui{\'o}}, {Greimel}, {Drew},
  {Barentsen}, {Groot}, {Irwin}, {Casares}, {G{\"a}nsicke}, {Carter},
  {Corral-Santana}, {Gentile-Fusillo}, {Greiss}, {van Haaften}, {Hollands},
  {Jones}, {Kupfer}, {Manser}, {Murphy}, {McLeod}, {Oosting}, {Parker},
  {Pyrzas}, {Rodr{\'\i}guez-Gil}, {van Roestel}, {Scaringi}, {Schellart},
  {Toloza}, {Vaduvescu}, {van Spaandonk}, {Verbeek}, {Wright}, {Eisl{\"o}ffel},
  {Fabregat}, {Harris}, {Morris}, {Phillipps}, {Raddi}, {Sabin}, {Unruh},
  {Vink}, {Wesson}, {Cardwell}, {de Burgos}, {Cochrane}, {Doostmohammadi},
  {Mocnik}, {Stoev}, {Su{\'a}rez-Andr{\'e}s}, {Tudor}, {Wilson}, \&
  {Zegmott}}]{Monguio:2020}
{Mongui{\'o}}, M., {Greimel}, R., {Drew}, J.~E., {et~al.} 2020, \aap, 638, A18

\bibitem[{{Munari} {et~al.}(2022){Munari}, {Alcal{\'a}}, {Frasca}, {Masetti},
  {Traven}, {Akras}, \& {Zampieri}}]{Munari:2022}
{Munari}, U., {Alcal{\'a}}, J.~M., {Frasca}, A., {et~al.} 2022, \aap, 661, A124

\bibitem[{{Munari} {et~al.}(2021){Munari}, {Traven}, {Masetti}, {Valisa},
  {Righetti}, {Hambsch}, {Frigo}, {{\v{C}}otar}, {De Silva}, {Freeman},
  {Lewis}, {Martell}, {Sharma}, {Simpson}, {Ting}, {Wittenmyer}, \&
  {Zucker}}]{Munari:2021}
{Munari}, U., {Traven}, G., {Masetti}, N., {et~al.} 2021, \mnras, 505, 6121

\bibitem[{{Nakazono} {et~al.}(2021){Nakazono}, {Mendes de Oliveira}, {Hirata},
  {Jeram}, {Queiroz}, {Eikenberry}, {Gonzalez}, {Abramo}, {Overzier},
  {Espadoto}, {Martinazzo}, {Sampedro}, {Herpich}, {Almeida-Fernandes},
  {Werle}, {Barbosa}, {Sodr{\'e}}, {Lima}, {Buzzo}, {Cortesi},
  {Men{\'e}ndez-Delmestre}, {Akras}, {Alvarez-Candal}, {Lopes}, {Telles},
  {Schoenell}, {Kanaan}, \& {Ribeiro}}]{Nakazono:2021}
{Nakazono}, L., {Mendes de Oliveira}, C., {Hirata}, N.~S.~T., {et~al.} 2021,
  \mnras, 507, 5847

\bibitem[{{Oke} \& {Gunn}(1983)}]{Oke:1983}
{Oke}, J.~B. \& {Gunn}, J.~E. 1983, \apj, 266, 713

\bibitem[{{Parker} {et~al.}(2016){Parker}, {Boji{\v c}i{\'c}}, \&
  {Frew}}]{Parker:2016}
{Parker}, Q.~A., {Boji{\v c}i{\'c}}, I.~S., \& {Frew}, D.~J. 2016, in Journal
  of Physics Conference Series, Vol. 728, Journal of Physics Conference Series,
  032008

\bibitem[{{Parker} {et~al.}(2005){Parker}, {Phillipps}, {Pierce}, {Hartley},
  {Hambly}, {Read}, {MacGillivray}, {Tritton}, {Cass}, {Cannon}, {Cohen},
  {Drew}, {Frew}, {Hopewell}, {Mader}, {Malin}, {Masheder}, {Morgan}, {Morris},
  {Russeil}, {Russell}, \& {Walker}}]{2005MNRAS.362..689P}
{Parker}, Q.~A., {Phillipps}, S., {Pierce}, M.~J., {et~al.} 2005, \mnras, 362,
  689

\bibitem[{Pedregosa {et~al.}(2011)Pedregosa, Varoquaux, Gramfort, Michel,
  Thirion, Grisel, Blondel, Prettenhofer, Weiss, Dubourg, Vanderplas, Passos,
  Cournapeau, Brucher, Perrot, \& Duchesnay}]{scikit-learn}
Pedregosa, F., Varoquaux, G., Gramfort, A., {et~al.} 2011, Journal of Machine
  Learning Research, 12, 2825

\bibitem[{{Peters} {et~al.}(2015){Peters}, {Richards}, {Myers}, {Strauss},
  {Schmidt}, {Ivezi{\'c}}, {Ross}, {MacLeod}, \& {Riegel}}]{Peters:2015}
{Peters}, C.~M., {Richards}, G.~T., {Myers}, A.~D., {et~al.} 2015, \apj, 811,
  95

\bibitem[{{Pickles}(1998)}]{Pickles:1998}
{Pickles}, A.~J. 1998, \pasp, 110, 863

\bibitem[{{Pollmann} {et~al.}(2018){Pollmann}, {Bennett}, {Vollmann}, \&
  {Somogyi}}]{Pollmann:2018}
{Pollmann}, E., {Bennett}, P.~D., {Vollmann}, W., \& {Somogyi}, P. 2018,
  Information Bulletin on Variable Stars, 6249, 1

\bibitem[{{Raddi} {et~al.}(2015){Raddi}, {Drew}, {Steeghs}, {Wright}, {Drake},
  {Barentsen}, {Fabregat}, \& {Sale}}]{Raddi:2015}
{Raddi}, R., {Drew}, J.~E., {Steeghs}, D., {et~al.} 2015, \mnras, 446, 274

\bibitem[{Rousseeuw(1987)}]{ROUSSEEUW198753}
Rousseeuw, P.~J. 1987, Journal of Computational and Applied Mathematics, 20, 53

\bibitem[{{Sabin} {et~al.}(2010){Sabin}, {Zijlstra}, {Wareing}, {Corradi},
  {Mampaso}, {Viironen}, {Wright}, \& {Parker}}]{Sabin:2010}
{Sabin}, L., {Zijlstra}, A.~A., {Wareing}, C., {et~al.} 2010, \pasa, 27, 166

\bibitem[{{Scaringi} {et~al.}(2013){Scaringi}, {Groot}, {Verbeek}, {Greiss},
  {Knigge}, \& {K{\"o}rding}}]{Scaringi:2013}
{Scaringi}, S., {Groot}, P.~J., {Verbeek}, K., {et~al.} 2013, \mnras, 428, 2207

\bibitem[{Sokolova \& Lapalme(2009)}]{Sokolova:2009}
Sokolova, M. \& Lapalme, G. 2009, Information Processing \& Management, 45, 427

\bibitem[{{Taylor}(2005)}]{Taylor:2005}
{Taylor}, M.~B. 2005, in Astronomical Society of the Pacific Conference Series,
  Vol. 347, Astronomical Data Analysis Software and Systems XIV, ed.
  P.~{Shopbell}, M.~{Britton}, \& R.~{Ebert}, 29

\bibitem[{{Viironen} {et~al.}(2009){Viironen}, {Mampaso}, {Corradi},
  {Rodr{\'\i}guez}, {Greimel}, {Sabin}, {Sale}, {Unruh}, {Delgado-Inglada},
  {Drew}, {Giammanco}, {Groot}, {Parker}, {Sokoloski}, \&
  {Zijlstra}}]{Viirone:2009}
{Viironen}, K., {Mampaso}, A., {Corradi}, R.~L.~M., {et~al.} 2009, \aap, 502,
  113

\bibitem[{{Vink} {et~al.}(2008){Vink}, {Drew}, {Steeghs}, {Wright}, {Martin},
  {G{\"a}nsicke}, {Greimel}, \& {Drake}}]{Vink:2008}
{Vink}, J.~S., {Drew}, J.~E., {Steeghs}, D., {et~al.} 2008, \mnras, 387, 308

\bibitem[{Waskom(2021)}]{Waskom:2021}
Waskom, M.~L. 2021, Journal of Open Source Software, 6, 3021

\bibitem[{{Wevers} {et~al.}(2017){Wevers}, {Jonker}, {Nelemans}, {Torres},
  {Groot}, {Steeghs}, {Maccarone}, {Hynes}, {Heinke}, \& {Britt}}]{Wevers:2017}
{Wevers}, T., {Jonker}, P.~G., {Nelemans}, G., {et~al.} 2017, \mnras, 466, 163

\bibitem[{{Witham} {et~al.}(2007){Witham}, {Knigge}, {Aungwerojwit}, {Drew},
  {G{\"a}nsicke}, {Greimel}, {Groot}, {Roelofs}, {Steeghs}, \&
  {Woudt}}]{Witham:2007}
{Witham}, A.~R., {Knigge}, C., {Aungwerojwit}, A., {et~al.} 2007, \mnras, 382,
  1158

\bibitem[{{Witham} {et~al.}(2008){Witham}, {Knigge}, {Drew}, {Greimel},
  {Steeghs}, {G{\"a}nsicke}, {Groot}, \& {Mampaso}}]{Witham:2008}
{Witham}, A.~R., {Knigge}, C., {Drew}, J.~E., {et~al.} 2008, \mnras, 384, 1277

\bibitem[{{Witham} {et~al.}(2006){Witham}, {Knigge}, {G{\"a}nsicke},
  {Aungwerojwit}, {Corradi}, {Drew}, {Greimel}, {Groot}, {Morales-Rueda},
  {Rodriguez-Flores}, {Rodriguez-Gil}, \& {Steeghs}}]{Witham:2006}
{Witham}, A.~R., {Knigge}, C., {G{\"a}nsicke}, B.~T., {et~al.} 2006, \mnras,
  369, 581

\bibitem[{{Wright} {et~al.}(2010){Wright}, {Eisenhardt}, {Mainzer}, {Ressler},
  {Cutri}, {Jarrett}, {Kirkpatrick}, {Padgett}, {McMillan}, {Skrutskie},
  {Stanford}, {Cohen}, {Walker}, {Mather}, {Leisawitz}, {Gautier}, {McLean},
  {Benford}, {Lonsdale}, {Blain}, {Mendez}, {Irace}, {Duval}, {Liu}, {Royer},
  {Heinrichsen}, {Howard}, {Shannon}, {Kendall}, {Walsh}, {Larsen}, {Cardon},
  {Schick}, {Schwalm}, {Abid}, {Fabinsky}, {Naes}, \& {Tsai}}]{Wright:2010}
{Wright}, E.~L., {Eisenhardt}, P. R.~M., {Mainzer}, A.~K., {et~al.} 2010, \aj,
  140, 1868

\bibitem[{{Wu} {et~al.}(2011){Wu}, {Luo}, {Li}, {Shi}, {Prugniel}, {Liang},
  {Zhao}, {Zhang}, {Bai}, {Wei}, {Dong}, {Zhang}, \& {Chen}}]{Wu:2011}
{Wu}, Y., {Luo}, A.~L., {Li}, H.-N., {et~al.} 2011, Research in Astronomy and
  Astrophysics, 11, 924

\bibitem[{{Yang} {et~al.}(2022){Yang}, {Yuan}, {Xiang}, {Duan}, {Huang}, {Liu},
  {Beers}, {Galarza}, {Daflon}, {Fern{\'a}ndez-Ontiveros}, {Cenarro},
  {Crist{\'o}bal-Hornillos}, {Hern{\'a}ndez-Monteagudo}, {L{\'o}pez-Sanjuan},
  {Mar{\'\i}n-Franch}, {Moles}, {Varela}, {V{\'a}zquez Rami{\'o}}, {Alcaniz},
  {Dupke}, {Ederoclite}, {Sodr{\'e}}, \& {Angulo}}]{Yang:2022}
{Yang}, L., {Yuan}, H., {Xiang}, M., {et~al.} 2022, \aap, 659, A181

\bibitem[{{York} {et~al.}(2000){York}, {Adelman}, {Anderson}, {Anderson},
  {Annis}, {Bahcall}, {Bakken}, {Barkhouser}, {Bastian}, {Berman}, {Boroski},
  {Bracker}, {Briegel}, {Briggs}, {Brinkmann}, {Brunner}, {Burles}, {Carey},
  {Carr}, {Castander}, {Chen}, {Colestock}, {Connolly}, {Crocker}, {Csabai},
  {Czarapata}, {Davis}, {Doi}, {Dombeck}, {Eisenstein}, {Ellman}, {Elms},
  {Evans}, {Fan}, {Federwitz}, {Fiscelli}, {Friedman}, {Frieman}, {Fukugita},
  {Gillespie}, {Gunn}, {Gurbani}, {de Haas}, {Haldeman}, {Harris}, {Hayes},
  {Heckman}, {Hennessy}, {Hindsley}, {Holm}, {Holmgren}, {Huang}, {Hull},
  {Husby}, {Ichikawa}, {Ichikawa}, {Ivezi{\'c}}, {Kent}, {Kim}, {Kinney},
  {Klaene}, {Kleinman}, {Kleinman}, {Knapp}, {Korienek}, {Kron}, {Kunszt},
  {Lamb}, {Lee}, {Leger}, {Limmongkol}, {Lindenmeyer}, {Long}, {Loomis},
  {Loveday}, {Lucinio}, {Lupton}, {MacKinnon}, {Mannery}, {Mantsch}, {Margon},
  {McGehee}, {McKay}, {Meiksin}, {Merelli}, {Monet}, {Munn}, {Narayanan},
  {Nash}, {Neilsen}, {Neswold}, {Newberg}, {Nichol}, {Nicinski}, {Nonino},
  {Okada}, {Okamura}, {Ostriker}, {Owen}, {Pauls}, {Peoples}, {Peterson},
  {Petravick}, {Pier}, {Pope}, {Pordes}, {Prosapio}, {Rechenmacher}, {Quinn},
  {Richards}, {Richmond}, {Rivetta}, {Rockosi}, {Ruthmansdorfer}, {Sandford},
  {Schlegel}, {Schneider}, {Sekiguchi}, {Sergey}, {Shimasaku}, {Siegmund},
  {Smee}, {Smith}, {Snedden}, {Stone}, {Stoughton}, {Strauss}, {Stubbs},
  {SubbaRao}, {Szalay}, {Szapudi}, {Szokoly}, {Thakar}, {Tremonti}, {Tucker},
  {Uomoto}, {Vanden Berk}, {Vogeley}, {Waddell}, {Wang}, {Watanabe},
  {Weinberg}, {Yanny}, {Yasuda}, \& {SDSS Collaboration}}]{York:2000}
{York}, D.~G., {Adelman}, J., {Anderson}, John~E., J., {et~al.} 2000, \aj, 120,
  1579

\end{thebibliography}
%


\begin{appendix}

\section{RR Lyrae Stars in the $(r - J0660)$ versus $(r - i)$ Diagram}
\label{sec:Appenx-A}

\begin{figure}
    \includegraphics[width=\linewidth, clip]{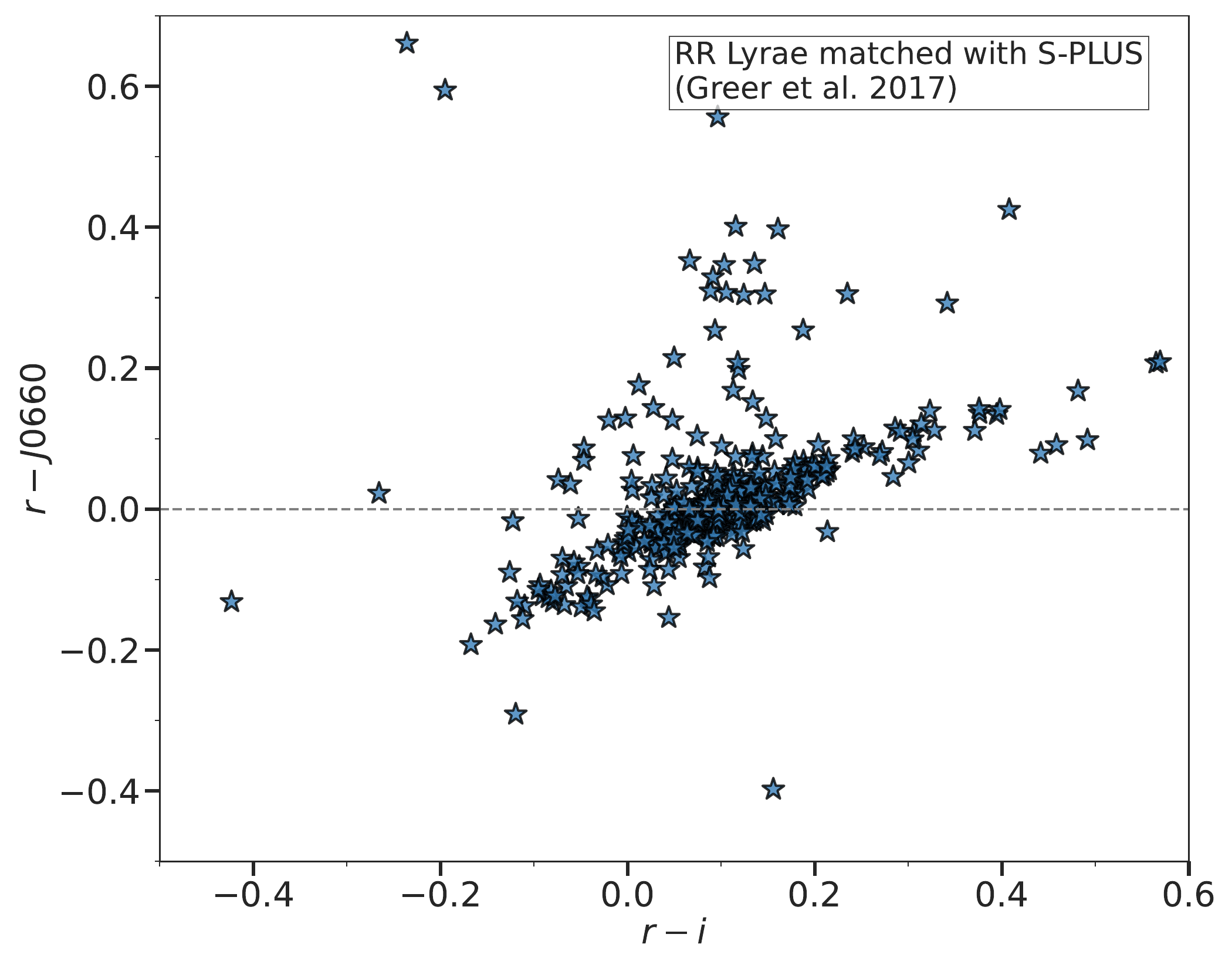} 
    \caption{Distribution of RR Lyrae stars from the catalog of \citet{Greer:2017A} matched with S-PLUS data, shown in the $(r - J0660)$ versus $(r - i)$ diagram. The horizontal dashed line indicates the $(r - J0660) = 0$ value.}
\label{fig:Diagram-RRlyra}
\end{figure}

We cross-matched the RR Lyrae catalog from \citet{Greer:2017A}, which contains 4\,963 objects, and found 375 matches with S-PLUS data. Figure \ref{fig:Diagram-RRlyra} shows the distribution of these RR Lyrae stars in the $(r - J0660)$ versus $(r - i)$ diagram.

\end{appendix}
\end{document}